\documentclass[11pt,headings=big,tightenlines,numbers=noenddot,DIV=14,a4paper]{article}%

\pdfoutput=1

\usepackage[margin=0.7 in]{geometry}
\usepackage{comment}
\usepackage{graphicx}  
\usepackage{url}
\usepackage{dcolumn} 
\usepackage{bm,relsize}   
\usepackage{amsfonts,amsmath,amssymb,amsthm,mathtools}
\usepackage{mathtools}
\usepackage{physics}
\usepackage{subfig}
\usepackage{slashed}
\usepackage{placeins}
\usepackage{adjustbox}
\usepackage[normalem]{ulem}
\usepackage{lscape}
\usepackage{multirow}
\usepackage{lipsum, color}
\usepackage{tabularray}
\usepackage{multirow}
\usepackage{geometry}
\usepackage{array}
\usepackage{booktabs}
\newcolumntype{C}[1]{>{\centering\arraybackslash}p{#1}}
\def\beq{\begin{equation}}
\def\eeq{\end{equation}}
\def\bea{\begin{eqnarray}}
\def\eea{\end{eqnarray}}
\usepackage{chngpage}
\usepackage[usenames,dvipsnames,svgnames,table]{xcolor}
\usepackage[linktoc=page,bookmarks=false,colorlinks=true,linkbordercolor=RoyalBlue,citebordercolor=ForestGreen,urlbordercolor=CornflowerBlue]{hyperref}
\hypersetup{
    colorlinks=true,
    linkcolor=ForestGreen,
    filecolor=ForestGreen,      
    urlcolor=ForestGreen,
    citecolor=ForestGreen,
}

\usepackage[sort&compress,numbers,merge]{natbib}
\usepackage{breakcites}
\usepackage{array}
\usepackage{makecell}

\usepackage{tikz}
\usepackage{amsmath}
\usetikzlibrary{arrows.meta, positioning}
\usepackage[compat=1.1.0]{tikz-feynman}
\usepackage{tabularray}
\UseTblrLibrary{booktabs} 

\usepackage{tabularray}
\usepackage[dvipsnames,table]{xcolor}
\usetikzlibrary{decorations.pathmorphing}

\tikzfeynmanset{
  graviton/.style={
    decoration={
      zigzag,
      segment length=2pt,
      amplitude=1pt,
      pre=lineto,
      post=lineto
    },
    decorate,
    draw,
    double
  }
}

\usepackage{tikz-feynman}
\usetikzlibrary{decorations.pathmorphing,decorations.markings,arrows.meta}

\usepackage{empheq}
\usepackage[most]{tcolorbox}
\newtcbox{\mymath}[1][]{%
    nobeforeafter, math upper, tcbox raise base,
    enhanced, colframe=blue!20!black,
    colback=blue!15, boxrule=1pt,
    #1}

\tikzset{
  photon/.style={line width=0.9pt, decorate, decoration={snake, amplitude=1.6pt, segment length=6pt}},
  fermion/.style={line width=0.9pt, postaction={decorate},
                  decoration={markings, mark=at position 0.55 with {\arrow{>}}}},
  scalar/.style={line width=0.9pt, densely dashed},
}

\usepackage[
    starfontserif 
    ]{starfont}
\usepackage{amsmath}
\usepackage{relsize}

\DeclareSymbolFont{starfontsym}{OT1}{sts}{m}{n}
\DeclareMathSymbol{\mathSun}{\mathord}{starfontsym}{115}
\DeclareMathSymbol{\mathMercury}{\mathord}{starfontsym}{102}
\DeclareMathSymbol{\mathVenus}{\mathord}{starfontsym}{103}
\DeclareMathSymbol{\mathTerra}{\mathord}{starfontsym}{76}
\DeclareMathSymbol{\mathvarTerra}{\mathord}{starfontsym}{108}
\DeclareMathSymbol{\mathMoon}{\mathord}{starfontsym}{100}
\DeclareMathSymbol{\mathvarMoon}{\mathord}{starfontsym}{97}
\DeclareMathSymbol{\mathMars}{\mathord}{starfontsym}{104}
\DeclareMathSymbol{\mathJupiter}{\mathord}{starfontsym}{106}
\DeclareMathSymbol{\mathSaturn}{\mathord}{starfontsym}{83}
\DeclareMathSymbol{\mathUranus}{\mathord}{starfontsym}{70}
\DeclareMathSymbol{\mathvarUranus}{\mathord}{starfontsym}{65}
\DeclareMathSymbol{\mathNeptune}{\mathord}{starfontsym}{71}
\DeclareMathSymbol{\mathPluto}{\mathord}{starfontsym}{74}
\DeclareMathSymbol{\mathvarPluto}{\mathord}{starfontsym}{72}

\begin{document}

\noindent \makebox[16cm][r]{\footnotesize 
CERN-TH-2025-169, MIT-CTP/5904, MITP-25-056}

\begin{flushright}
\footnotesize
\end{flushright}
\color{black}
\begin{center}



{\huge \bf Implications for Pulsar Timing Arrays}\\[3mm]
{\huge \bf of Sub-solar Black Hole Detections:}\\[5mm]
{\Large \bf From LVK to Einstein Telescope and Cosmic Explorer}

\medskip
\bigskip\color{black}\vspace{0.5cm}

{
{\large Yann Gouttenoire}$^{a}$,
{\large Sokratis Trifinopoulos}$^{b,c,d}$,
{\large Miguel Vanvlasselaer}$^{e}$
}
\\[7mm]

{\it \small $^a$ PRISMA+ Cluster of Excellence $\&$ MITP, Johannes Gutenberg University, 55099 Mainz, Germany}\\
{\it \small $^b$ Center  for  Theoretical  Physics -- a Leinweber Institute,  Massachusetts  Institute  of  Technology,  Cambridge,  MA  02139,  USA}\\
{\it \small $^c$ Theoretical Physics Department, CERN, Geneva, Switzerland}\\
{\it \small $^d$ Physik-Institut, Universit\"at Z\"urich, 8057 Z\"urich, Switzerland}\\
{\it \small $^e$ Theoretische Natuurkunde and IIHE/ELEM, Vrije Universiteit Brussel, \& The  International Solvay Institutes, Pleinlaan 2, B-1050 Brussels, Belgium}\\
\end{center}

\bigskip

\centerline{\bf Abstract}
\begin{quote}

The detection of compact binary mergers with sub-solar masses at gravitational-wave observatories could mark the groundbreaking discovery of primordial black holes (PBHs). Concurrently, evidence for a nHz stochastic gravitational wave background observed by pulsar timing arrays (PTAs) could suggest a non-astrophysical origin, potentially arising from scalar-induced gravitational waves (SIGW). In this work, we analyze the connection between the two phenomena in the case where they share a common origin: the collapse of large primordial curvature perturbations in the early universe. We focus on sub-solar PBH populations within reach of upcoming experiments, including the current and future runs of LIGO–Virgo–KAGRA as well as the third generation observatories such as the Einstein Telescope and Cosmic Explorer. Using a Bayesian framework with physically motivated priors, we perform a consistent model comparison that incorporates existing astrophysical bounds together with the discovery potential of future detectors. Our analysis shows that if PBHs are discovered then the SIGW interpretation --- especially in the presence of primordial non-Gaussianities --- could become favored over the astrophysical one, as the narrowed priors place greater weight on the region of highest likelihood. Ultimately, we illustrate that combining PTA data with interferometer searches can deliver correlated evidence for new physics across multiple gravitational-wave bands.

\end{quote}

\clearpage
\noindent\makebox[\linewidth]{\rule{\textwidth}{1pt}} 
\setcounter{tocdepth}{2}
\tableofcontents
\noindent\makebox[\linewidth]{\rule{\textwidth}{1pt}} 	

\newpage

\section{Introduction}
\label{sec:intro}

Gravitational wave (GW) astronomy has opened an entirely new observational window onto the universe, unveiling phenomena that were previously inaccessible and spanning a wide range of frequency bands. In the Hz/kHz range, the LIGO-Virgo-KAGRA (LVK) collaboration~\cite{LIGOScientific:2007fwp,LIGOScientific:2016aoc,LIGOScientific:2018glc,LIGOScientific:2019kan,LIGOScientific:2020ibl,LIGOScientific:2021job,KAGRA:2021vkt,LIGOScientific:2025slb}  has registered GW signals originating from the coalescence of black hole and neutron star binaries for 218 events to date. In the nHz range, the NANOGrav~\cite{NANOGrav:2020bcs,NANOGrav:2023gor}, in conjunction with other pulsar timing array (PTA) collaborations~\cite{Chen:2021rqp,Antoniadis:2023rey,Goncharov:2021oub,Reardon:2023gzh,Xu:2023wog,Antoniadis:2022pcn}, has presented evidence for the presence of a \emph{stochastic} gravitational wave background (SGWB). In the future, many more experiments are envisioned in a wide range of frequency bands~\cite{Caprini:2019egz,Caprini:2015zlo,Hild:2010id,Sathyaprakash:2012jk,Maggiore:2019uih,Evans:2021gyd, Srivastava:2022slt,Seto:2001qf, Yagi:2011wg, Isoyama:2018rjb,Corbin:2005ny, Crowder:2005nr,Badurina:2019hst,Bertoldi:2019tck,Babak:2024yhu,Foster:2025nzf,Blas:2025lzc} and their complementarity is expected to facilitate the identification of the GW sources~\cite{Ellis:2023oxs}.

The leading interpretation of the PTA signal is that it originates from astrophysical supermassive black hole (SMBH) binaries~\cite{Sesana:2004sp,Volonteri:2010wz,Middleton:2017nbg,Thrane:2018qnx,Taylor:2021yjx,NANOGrav:2023hfp,Antoniadis:2023zhi,Ellis:2023dgf,Bi:2023tib,Ellis:2024wdh,Raidal:2024odr,Raidal:2024tui,Goncharov:2024htb}.
However, there remains a noticeable degree of uncertainty regarding the 
GW amplitude expected from SMBH binaries. In order to explain the PTA observations, current 
astrophysical models must be pushed toward the upper end of their expected GW signal~\cite{NANOGrav:2023hvm,NANOGrav:2023hfp,Antoniadis:2023zhi} as well as to rely on optimistic assumptions about binary environments~\cite{Ellis:2023dgf} or eccentricity~\cite{Raidal:2024tui}, see however \cite{Goncharov:2024htb}.
This has, in turn, prompted the exploration of alternative 
explanations involving a primordial origin, see Refs.~\cite{Ratzinger:2020koh,Madge:2023dxc,NANOGrav:2023hvm,EPTA:2023xxk,Bian:2023dnv,Figueroa:2023zhu,Ellis:2023oxs,Esmyol:2025ket} for reviews. In such scenarios, the observed peak frequency is typically given by an order one
fraction of the Hubble horizon at the time of production, $f_\star = \mathcal{O}(H_\star/2\pi)$, redshifted up to today

\begin{equation}
\label{eq:f0_Hstar}
f_{\rm 0}= \left(\frac{a_\star}{a_0}\right)\left(\frac{2\pi}{H_\star/f_\star} \right)\frac{H_\star}{2\pi} \simeq 2.3~{\rm nHz}\left(\frac{g_\star(T_\star)}{50}\right)\left(\frac{T_\star}{0.1~\rm GeV}\right)\,.
\end{equation}
where $g_\star$ and $T_\star$ are the number of relativistic degrees of freedom and the temperature at the time of production. Among the viable options are \emph{scalar-induced} gravitational waves (SIGW)~\cite{Ananda:2006af,Baumann:2007zm,Espinosa:2018eve,Kohri:2018awv,Domenech:2021and,Liu:2023pau, Dandoy:2023jot, Franciolini:2023pbf,Vagnozzi:2023lwo,Franciolini:2023wjm,Wang:2023ost,Liu:2023ymk,Firouzjahi:2023lzg,Balaji:2023ehk,Harigaya:2023pmw,Tagliazucchi:2023dai,Cai:2023dls,Chen:2023uiz,Frosina:2023nxu,Chen:2024fir,Chen:2024twp,Domenech:2024rks,Iovino:2024tyg,Clesse:2024epo}, which are naturally motivated by inflationary dynamics. In this mechanism, GWs are sourced at second order in perturbation theory by gradient of the curvature perturbation.

To fit the PTA signal, the amplitude of the primordial SGWB must be around $\Omega_{\rm GW} \sim 10^{-8}$, which is significantly larger than the standard expectation~\cite{Giblin:2014gra} suggesting that primordial sources must be in a regime of large GW production. This is generally associated to violent processes involving significant energy overdensities or metric inhomogeneities, conditions likely to produce \emph{primordial black holes} (PBHs)~\cite{10.1093/mnras/152.1.75,Carr:1974nx,Carr:1975qj}. 
The PBH mass is related to the mass within the Hubble horizon at that time, which is related to the observed GW frequency in Eq.~\eqref{eq:f0_Hstar} by
\begin{equation}
M_{\rm PBH} = \frac{4\pi M_{\rm pl}^2}{H_\star}\simeq M_{\odot}\left( \frac{50}{g_\star(T_\star)}\right)^{1/6} \left(\frac{2\pi}{H_\star/f_\star} \right)^2\left(\frac{6.2~\rm nHz}{f_0} \right)^2\,. \label{eq:PBH_mass}
\end{equation}
Some of those PBHs can then form compact binaries with each other and merge in the late universe, possibly constituting a fraction of today's black hole mergers. According to Eq.~\eqref{eq:PBH_mass}, the masses of such mergers lie in the 
(sub-)solar range, which is currently being probed by LVK~\cite{Sasaki:2016jop,
Bird:2016dcv,Clesse:2016vqa,Sasaki:2018dmp,Raidal:2017mfl,Raidal:2018bbj,
Vaskonen:2019jpv,DeLuca:2020sae}, and this 
sensitivity will be further extended by upcoming ground-based laser interferometers, 
including the Einstein Telescope (ET)~\cite{Hild:2010id,Sathyaprakash:2012jk,Maggiore:2019uih}, and Cosmic Explorer (CE)~\cite{Evans:2021gyd, Srivastava:2022slt}.

As a consequence of Eq.~\eqref{eq:PBH_mass}, a coincidence of cosmic scales arises, that relates the GW signals in experiments that \emph{a priori} probe distinct frequencies: the same primordial curvature peaks that can source SIGW at scales and magnitudes relevant to PTAs, can collapse upon reentering the horizon and form PBH compact binaries that may be detectable through kHz GW observatories.\footnote{Such correlations have been observed also for other mechanisms of primordial origin, e.g. first-order phase transitions~\cite{Gouttenoire:2023bqy,Ellis:2023oxs,Gouttenoire:2023gbn,Lewicki:2024ghw,Franciolini:2025ztf,Addazi:2023jvg,Bringmann:2023opz,Gouttenoire:2023bqy,Ellis:2023oxs,Lewicki:2024ghw}, and domain wall networks~\cite{Ferreira:2022zzo,Kitajima:2023cek,Gouttenoire:2023ftk,Gouttenoire:2025ofv,Lu:2024ngi,Lu:2024szr}.} This observation has been exploited to translate the PTA signal into bounds 
on the PBH parameter space, arising from the associated overproduction of 
GWs~\cite{Saito:2008jc, Nakama:2016gzw,Inomata:2016rbd, Ando:2017veq,Garcia-Bellido:2017aan,Clesse:2018ogk,Chen:2019xse,Cai:2019elf,Kohri:2020qqd,Inomata:2020xad,Vaskonen:2020lbd,DeLuca:2020agl,Domenech:2020ers,Liu:2021jnw,Yuan:2021qgz,Franciolini:2023pbf,Wang:2023ost,Inomata:2023zup,Iovino:2024tyg,Kumar:2025jfi}.  
In turn, in this work we address the following question: \emph{Should transient GW signals in the Hz/kHz regime be discovered and attributed to PBH mergers that originate from large curvature perturbations, what are the implications for the stochastic GW signal of PTAs?} 

To strengthen the case for a primordial origin of the mergers, we will focus on black hole binaries with sub-solar masses. Candidates for such events have been identified in previous LVK datasets~\cite{LIGOScientific:2021job,KAGRA:2021vkt}, but no definite signal has yet been confirmed~\cite{LVK:2022ydq,Morras:2023jvb,Nitz:2022ltl}. Sub-solar black holes are regarded as smoking guns for new physics, since standard astrophysical formation channels are unable to produce black holes below the Chandrasekhar limit of approximately $1.4~M_{\odot}$. Nevertheless, there are still alternative new physics scenarios that can accommodate non-primordial sub-solar binary merger events, such as exotic compact objects~\cite{Coleman:1985ki,Colpi:1986ye,Liebling:2012fv,Lee:1986tr,DelGrosso:2023trq} and stellar transmutation mechanisms~\cite{Goldman:1989nd,Bramante:2013hn,Kouvaris:2013kra,Bramante:2015dfa,Takhistov:2017bpt,Kouvaris:2018wnh,Takhistov:2020vxs,Dasgupta:2020mqg,Garani:2021gvc,Singh:2022wvw,Steigerwald:2022pjo,Bhattacharya:2023stq,Chakraborty:2024eyx}, which we aim to disfavor through the distinct phenomenology~\cite{Dasgupta:2020mqg,Yamamoto:2023tsr,Carr:2023tpt,DeLuca:2023mio,Crescimbeni:2024cwh, Crescimbeni:2024qrq,Golomb:2024mmt,Yuan:2024yyo}. Furthermore, we will explore the impact of primordial non-Gaussianities (NGs) which has been shown to be significant both at PBH formation~\cite{Bugaev:2013vba, Nakama:2016gzw,Byrnes:2012yx,Young:2013oia,Kitajima:2021fpq,Kawasaki:2019mbl,DeLuca:2019qsy,Young:2019yug,Cai:2019elf,Taoso:2021uvl,Meng:2022ixx,Escriva:2022pnz,Ferrante:2022mui,Yoo:2019pma,Riccardi:2021rlf,Young:2022phe} and SIGW~\cite{Unal:2018yaa, Liu:2023ymk,Cai:2018dig, Atal:2019cdz, Cai:2019amo, Ragavendra:2020sop, Yuan:2020iwf, Atal:2021jyo, Adshead:2021hnm, Garcia-Saenz:2022tzu, Perna:2024ehx}. 

\begin{figure*}[t!]
    \centering
    \includegraphics[width=0.9\linewidth]{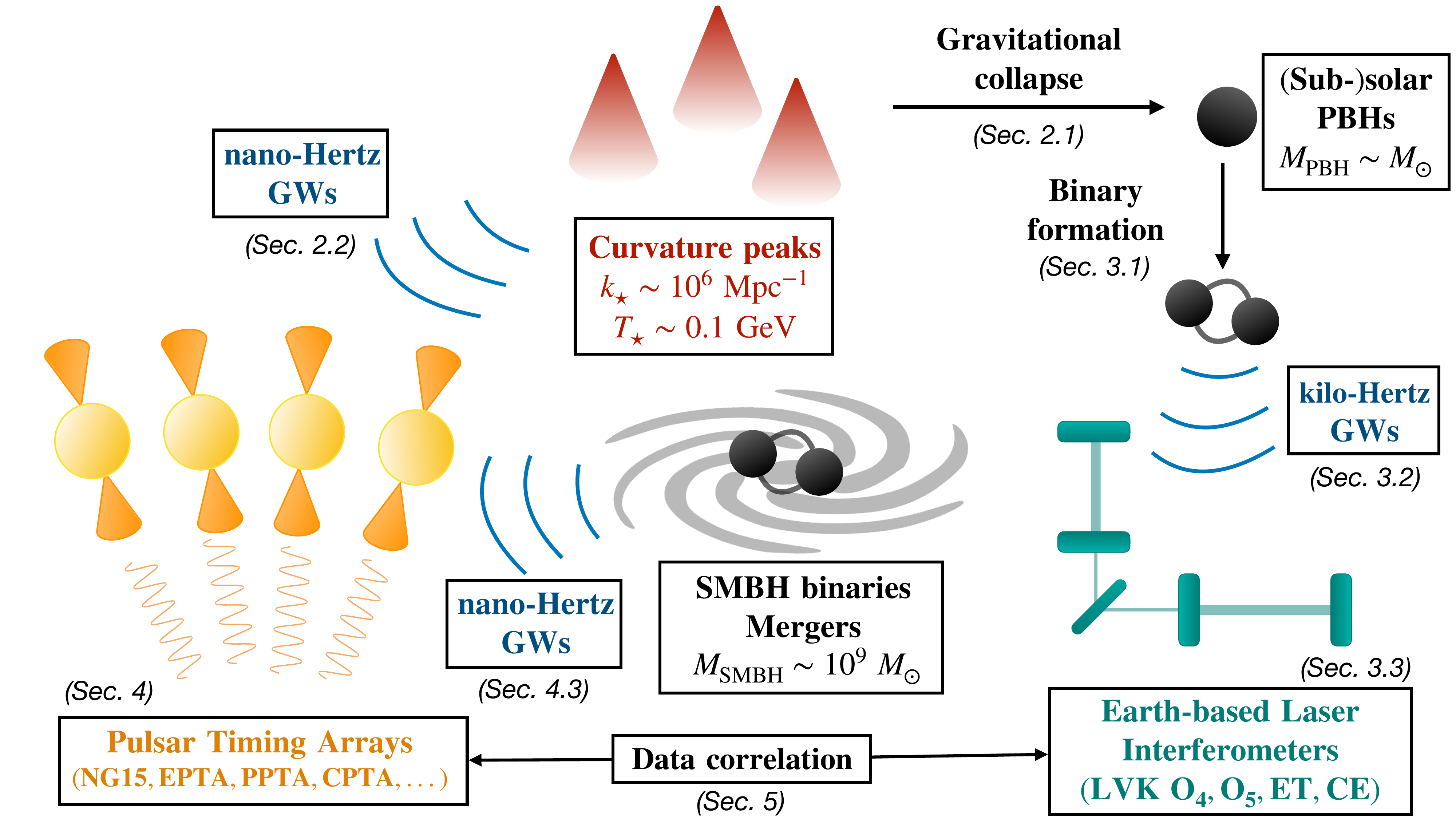}  
    \caption{Roadmap of the paper. \label{fig:drawing_subsolar_paper}}
\end{figure*}

This work is intended to be as self-contained as possible, with the goal of providing a useful reference for future studies. We therefore review in detail the underlying theoretical framework, as well as the details of the statistical analysis. In Sec.~\ref{sec:curvature_peaks} we discuss PBH formation from curvature peaks, relying both on threshold statistics and peak theory prescriptions, and the associated SIGW signal. Sec.~\ref{sec:GW_from_PBH_mergers} computes the distribution of PBH binaries formed in the early universe and their resulting transient GW emission, and estimates the reach of future experiments, such as LIGO O5, ET, and CE, in the sub-solar mass range. In Sec.~\ref{sec:PTAs} we perform Bayesian inference of curvature-power-spectrum parameters using the NANOGrav 15-year dataset (NG15), incorporating astrophysical constraints on PBHs. Importantly, we quantify how prospective detections of sub-solar PBHs by kHz interferometers would update the evidence for a SIGW component in PTA data. Results are presented in Sec.~\ref{sec:results}, and Sec.~\ref{sec:conclusion} summarizes our conclusions and outlines future directions. The above structure is illustrated in Fig.~\ref{fig:drawing_subsolar_paper}.

\section{Large curvature perturbations}
\label{sec:curvature_peaks}

The early Universe's inflationary paradigm is a well-known phase of exponential expansion that provides the initial conditions to density perturbations responsible for all structures observed today. During this inflationary phase, primordial perturbations (scalar, vector, tensor) across a broad wavelength range are produced. The decomposition theorem states that, at the linear level of cosmological perturbation theory, these modes of perturbation evolve independently. The scalar fluctuations manifest as temperature anisotropies in the Cosmic Microwave Background (CMB) radiation and eventually form the seed density perturbations that evolve into large-scale structures later on. This picture has proven to be exceptionally effective in matching the observed features of the CMB and galaxy power spectrum~\cite{Mukhanov:1981xt,Planck:2018jri}. Equivalently, those observations fix the power spectrum of the scalar fluctuations at the CMB scales.

However, at smaller scales, the scalar power spectrum is not constrained, and enhanced curvature peaks could be produced during ultra-slow roll inflation~\cite{Carr:1993aq,Ivanov:1994pa,Yokoyama:1998pt, Kannike:2017bxn, Garcia-Bellido:2017mdw,Byrnes:2018txb,Inomata:2018cht,Karam:2022nym, Chen:2020uhe}, curvaton scenarios~\cite{Enqvist:2001zp,Lyth:2001nq,Moroi:2001ct,Clesse:2015wea, Ando:2017veq,Ando:2018nge, Cheng:2018yyr,Espinosa:2018eve,Chen:2019zza,Palma:2020ejf, Braglia:2020eai}, resonant particle production
during inflation \cite{Chung:1999ve,Romano:2008rr,Barnaby:2009dd,Pearce:2017bdc,Cai:2019bmk,Olea-Romacho:2025mxj}, a turn during slow roll inflation\cite{Fumagalli:2020adf}, first-order phase transitions~\cite{Sasaki:1982fi,Wasserman:1986gi,Turner:1992tz,Liu:2022lvz,Giombi:2023jqq,Elor:2023xbz,Lewicki:2024ghw,Buckley:2024nen,Cai:2024nln,Jinno:2024nwb,Franciolini:2025ztf}, domain wall networks~\cite{Turner:1990uw,Goetz:1990qz,Goetz:1990pj,Lola:1993qu,Lazanu:2015fua,Takahashi:2020tqv,Kitajima:2022jzz,Ramberg:2022irf,Zeng:2023jut,Lu:2024dzj,Gouttenoire:2025wxc}, or come from strong primordial magnetic fields~\cite{Kim:1994zh,Jedamzik:2013gua,Ralegankar:2023pyx,Olea-Romacho:2025qag}.  
The variance of the primordial curvature perturbation $\zeta$ is
\begin{equation} \label{eq:zeta_PS_def}
\langle \zeta(k) \zeta(p) \rangle \equiv (2\pi)^{3}\delta^{(3)}(k+p) \frac{2\pi^2}{k^3} \mathcal{P}_{\zeta}(k)\,.
\end{equation}
where the quantity ${\cal P}_\zeta(k)$ is defined to be dimensionless. 
 In the present work, we will use as benchmark the log-normal distribution, which is given by  
\begin{equation}
\label{eq:log_normal}
\mathcal{P}_\zeta (k)=\frac{\mathcal{A_\zeta}}{\sqrt{2\pi}\Delta}
\exp\left(-\frac{\ln^2(k/k_\star)}{2\Delta^2}\right),
\end{equation}
where $k_\star$ denotes the wavenumber at the peak, $\Delta$ the peak width, and $\mathcal{A_\zeta}$ the amplitude of the distribution. As long as the distribution is narrowly peaked, we assume that the log-normal is a reliable proxy for any family of power spectra where only a narrow range of modes contribute to $\zeta$~\cite{Young:2022phe,Ferrante:2022mui}.

\subsection{Primordial black hole formation}
\label{sec:PBH_formation}

\paragraph{Curvature perturbation and density contrast.}

Let us consider cosmological perturbations at time $t$ with the lengthscale $a(t) \hat{r}_k$, where $\hat{r}_k$ is the comoving wavelength for a specific wavenumber $k$, and we define also the parameter 
\bea
\label{eq:definition_epsilon}
\epsilon(t) = \frac{1}{a(t) H(t)  \hat{r}_k}~.
\eea
In the \emph{long-wavelength limit} $\epsilon(t) \ll 1$, the perturbation is super-horizon, and we may assume that each patch evolves as if it is part of a separate Friedmann–Lemaître–Robertson–Walker (FLRW) universe~\cite{Sasaki:1998ug,Wands:2000dp}. 
We can then describe inhomogeneities by incorporating the curvature perturbation $K(\hat{r})$ and $\zeta(r)$ into the metric as follows~\cite{Polnarev:2006aa,Harada:2015yda,Musco:2018rwt,Young:2019yug}
\begin{equation}
\label{eq:metric_conformal}
    ds^2 = - dt^2 + a^2(t) \left[\frac{d\hat{r}^2}{1-K(\hat{r})\hat{r}^2}+\hat{r}^2 d\Omega^2 \right] = - dt^2 + a^2(t) e^{2\zeta(r)}\left[dr^2+r^2 d\Omega^2 \right] \,, 
\end{equation}
where $d\Omega = d\theta^2 + \sin\theta d\phi^2$ is the induced metric of the 2-sphere. These coordinates are related by\footnote{We use the same notation as Ref. \cite{Ferrante:2022mui} for $\hat{r}$ and $r$ which is inverted $r\leftrightarrow \hat{r}$ with respect to Refs. \cite{Musco:2018rwt,Musco:2020jjb}.}
\begin{equation} \label{eq:transfo}
    \hat{r}=r e^{\zeta(r)}\,, \quad K(\hat{r}) \hat{r}^2 = -r \zeta'(r) (2+r \zeta'(r)) \,.
\end{equation}
In the super-horizon regime, it can be shown that $\zeta$ and $K$ are gauge independent at $\mathcal{O}(\epsilon)$~\cite{Lyth:2004gb,Harada:2015yda}.
We consider next the Einstein equations in the comoving gauge under the assumption of spherical symmetry, which are known as the Misner-Sharp-Hernandez equations~\cite{Misner:1964je}. Solving these equations in the \emph{long-wavelength} approximation $\epsilon(t) \ll 1$, one obtains the density contrast~\cite{Polnarev:2006aa,Harada:2015yda,Musco:2018rwt,Young:2019yug}
\begin{equation} 
\frac{\delta\rho(\hat r,t)}{\rho_b(t)} =  \mathcal{F}(w)  \left(\frac{1}{a(t) H(t)}\right)^2 \left( K(\hat{r}) +\frac{\hat{r}}{3}K'(\hat{r}) \right)\,,
\quad \mathcal{F}(w) = \frac{3(1+w)}{5+3w}\,,
\label{eq:rho_contrast}
 \end{equation}
where $\rho_b$ is the mean background energy density and 
$\mathcal{F}$ is a transfer coefficient that depends on the equation of state parameter $w\equiv p_b/\rho_b$.  In the following, we will denote $\mathcal{F} =2/3$ the value of the transfer coefficient for a radiation-dominated universe ($w=1/3$). Moreover, using the coordinates in Eq. \eqref{eq:metric_conformal} we can write~\cite{Musco:2018rwt} 
\begin{equation} 
\frac{\delta\rho (r,t) }{\rho_b(t)}= - 2\mathcal{F}(w) \left(\frac{1}{a(t) H(t)}\right)^2 e^{-5\zeta(r)/2} \Delta \left(e^{\zeta(r)/2}\right)\,,
\label{eq:rho_contrast_zeta}
\end{equation}
where $\Delta$ is the Laplacian operator.
We observe that the non-linear relation between $\delta \rho$ and $\zeta$ induces NGs in the former even in the Gaussian limit of the latter~\cite{DeLuca:2019qsy,Young:2019yug,Germani:2019zez}.
We introduce the \emph{areal radius} $R(r,t)\equiv \sqrt{A/4\pi}$ which measures the surface area $A$ of a 2-sphere as a function of the radial coordinate $r$ ~\cite{Misner:1964je}.\footnote{For a given spherically-symmetric metric $ds^2 = - e^{2\phi(t,r)} \, dt^2 + e^{2\lambda(t,r)} \, dr^2 + R^2(t,r)\, d\Omega^2 $, the areal radius $R(r,t)$ is the coefficient in front of $d\Omega^2$~\cite{Misner:1964je}.} From Eq.~\eqref{eq:metric_conformal}, we have
\begin{equation}
\label{eq:areal_radius}
    R(r,t) = a(t)\hat{r}=a(t)r e^{\zeta(r)}\,.
\end{equation}
The mean mass excess or \emph{perturbation amplitude} within the volume $V=\frac{4}{3}\pi R^3$ can be calculated as
\begin{equation} 
\label{eq:delta_K_zeta}
\bar \delta(\hat{r},t) \equiv \frac{1}{V} \int_0^R \frac{\delta\rho}{\rho_b}(\tilde{r},t) d\tilde{V} = \frac{3}{\hat{r}^3} \int_0^{\hat{r}}  \frac{\delta\rho}{\rho_b}(\tilde{r},t) \tilde{r}^2 d\tilde{r}=\epsilon(t)^2\mathcal{F}(w) K(\hat{r})\hat{r}^2\,,
\end{equation}
where we see that the curvature K measures directly the mass excess within V.

\paragraph{PBH formation threshold.}

\begin{figure*}[t!]
    \centering
    \includegraphics[width=0.515\linewidth]{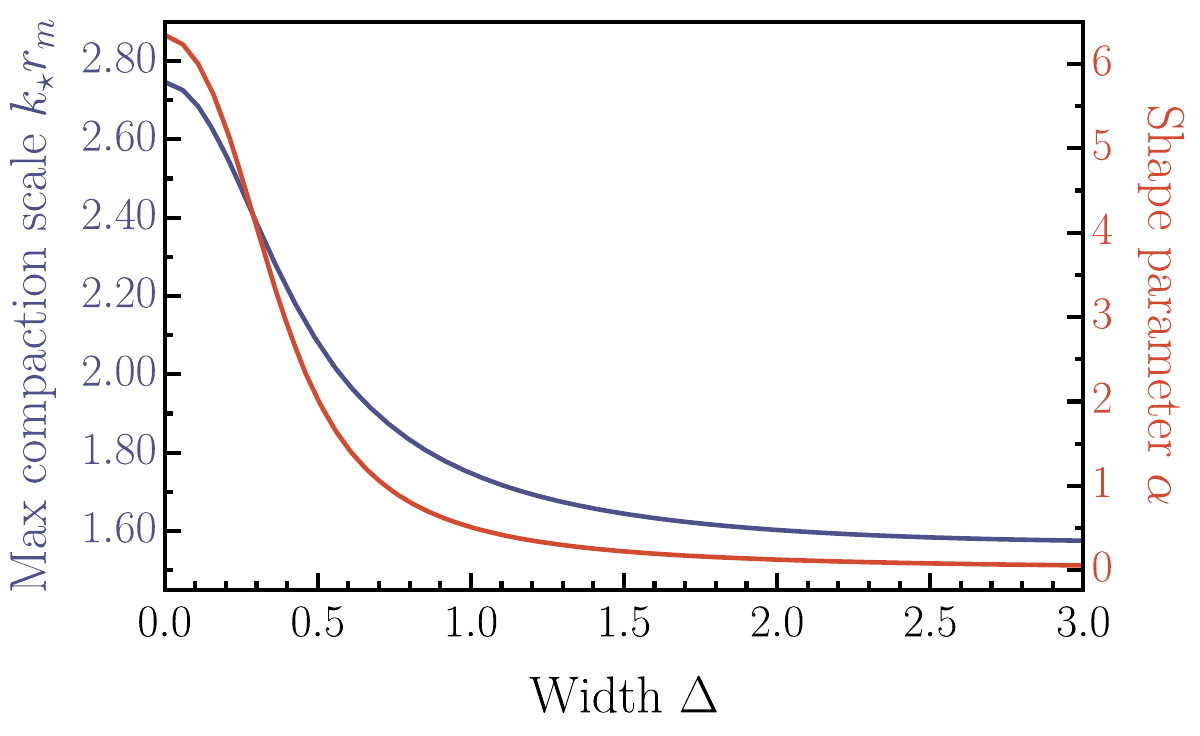}  
    \includegraphics[width=0.475\linewidth]{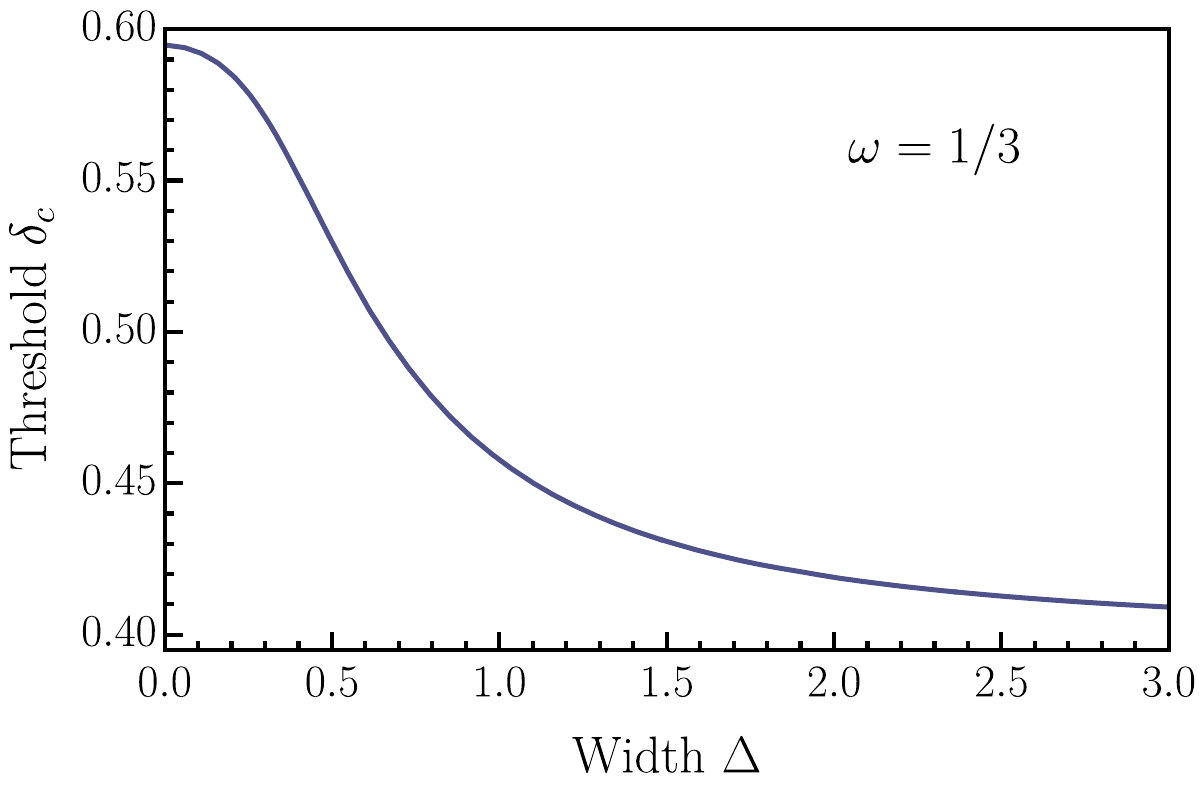}  
    \caption{Determination of the comoving length scale $r_m$ at which the compaction function is maximized $C'(r_m)=0$ (obtained by solving Eq.~\eqref{eq:hat_r_m}), the shape parameter $\alpha$ (obtained by solving Eq.~\eqref{eq:alpha_solve}), and the compaction threshold $\delta_c$ (see Eq.~\eqref{eq:delta_c}). The curvature power spectrum is assumed to be the log-normal distribution in Eq.~\eqref{eq:log_normal}. \label{fig:delta_c_Delta}}
\end{figure*}

A collapsing region forms a black hole once an apparent horizon of radius $R$ encloses a mass $M$ such that $R=2GM$~\cite{Helou:2016xyu}.  
The central quantity that characterizes such an overdensity is the \emph{compaction function}~\cite{Shibata:1999zs,Harada:2015yda,Musco:2012au,Musco:2018rwt,Young:2019yug,Escriva:2019nsa,Escriva:2021aeh,Musco:2020jjb}, which remains conserved on superhorizon scales. It is defined as the mass excess within a sphere of radius $R$,
\begin{equation} \label{eq:compaction_function}
\mathcal{C}(r) \equiv 2G\,\frac{M(r,t)-M_b(r,t)}{R(r,t)}
= \bar\delta(r)\,\frac{r^2}{r_k^2}\,,
\end{equation}
where $M_b=\rho_b V$ is the background FLRW mass, and $\delta(r)$ is the spatial part of Eq.~\eqref{eq:delta_K_zeta}, i.e. $\bar\delta(r,t)=\bar\delta(r)\,\epsilon^2(t)$.
For modes reentering the horizon, $\mathcal{C}(r)$ reduces to the average density contrast in the Hubble patch, $\mathcal{C}(r_k)=\bar\delta(r_k)$. The relevant comoving scale is the radius $r_m$ at which $\mathcal{C}(r)$ attains its maximum,
\begin{equation}
\label{eq:rm_def}
  \mathcal{C}'(r_m)=0 \quad \Rightarrow \quad 
  \begin{cases}
  K(\hat{r}_m) + \tfrac{\hat{r}_m}{2} K'(\hat{r}_m)=0\,,\\[2mm]
  \zeta'(r_m) + r_m\,\zeta''(r_m)=0 \,.
  \end{cases}
\end{equation}
At the time of horizon crossing $t=t_H$ defined by $\epsilon(t_H)=1$, the peak perturbation amplitude is
\begin{equation}
\label{eq:delta_m_def}
\delta_m \equiv \bar\delta(r_m,t_H) = \mathcal{C}(r_m)
= \mathcal{F}(w)\,K(\hat{r}_m)\,\hat{r}_m^2
= 3\,\frac{\delta\rho}{\rho_b}(r_m,t_H)\,,
\end{equation}
where the last equality follows from Eqs.~\eqref{eq:rho_contrast} and \eqref{eq:rm_def}. We observe that Eq.~\eqref{eq:delta_m_def} links an averaged quantity (the mass excess at scale $r_m$) to the local value of the density contrast.  
Introducing the \emph{linear} peak amplitude $\delta_\ell$ via Eq.~\eqref{eq:transfo}, one obtains the non-linear relation between density and curvature perturbation~\cite{DeLuca:2019qsy,
Young:2019yug}
\begin{equation} \label{eq:delta_ell_def}
\delta_m = \delta_\ell - \frac{1}{4\mathcal{F}(w)}\,\delta_\ell^2\,,
\qquad 
\delta_\ell \equiv -2\mathcal{F}(w)\,r_m\,\zeta'(r_m)\,.
\end{equation}
The condition for PBH formation can be written as~\cite{Young:2014ana,Harada:2015yda,Musco:2012au,Musco:2018rwt,Germani:2018jgr,Young:2019yug,Escriva:2019phb,Escriva:2021aeh,Musco:2020jjb}
\begin{equation}
\label{eq:PBH_formation_condition}
\delta_m \geq \delta_c \,,
\end{equation}
where the critical threshold during radiation domination typically lies in the range $\delta_c \in [0.40,0.67]$~\cite{Musco:2018rwt,Escriva:2019phb}.
An analytic expression for $\delta_c$ was obtained in Ref.~\cite{Escriva:2019phb}:
\begin{equation}
\label{eq:delta_c}
\delta_c \simeq 
\frac{4}{15}\,
e^{-1/\alpha}\,
\frac{\alpha^{1-5/(2\alpha)}}
{\Gamma\!\left(\tfrac{5}{2\alpha}\right) -
 \Gamma\!\left(\tfrac{5}{2\alpha},\,1/\alpha\right)}\,,
\qquad 
\alpha \equiv -\frac{\mathcal{C}''(r_m)\,r_m^2}{4\,\mathcal{C}(r_m)} \,,
\end{equation}
where $\alpha$ denotes the \emph{shape parameter}.  
The value of $\alpha$ can be determined by solving~\cite{Musco:2020jjb}
\begin{equation}
\label{eq:alpha_solve}
G(\alpha)\,[1+G(\alpha)]\,\alpha = \alpha_G(r_m) \,,
\end{equation}
where $\alpha_G$ is the Gaussian shape parameter and $G(\alpha)$ is an auxiliary function, 
\begin{equation}
\alpha_G(r_m) = -\frac{1}{2}\left[
1+r_m\,
\frac{\int dk\, k \cos(k r_m)\, \mathcal{P}_\zeta(k)}
{\int dk\, \sin(k r_m)\, \mathcal{P}_\zeta(k)}
\right]\,,
\qquad
G(\alpha) \equiv
\sqrt{
1-\frac{2}{5}\,
\frac{e^{-1/\alpha}\,
\alpha^{1-5/(2\alpha)}}
{\Gamma\!\left(\tfrac{5}{2\alpha}\right) -
 \Gamma\!\left(\tfrac{5}{2\alpha},\,1/\alpha\right)}
}\,.
\end{equation}
The comoving scale $r_m$ entering Eq.~\eqref{eq:alpha_solve} is obtained from the condition that the compaction function is maximized. Fourier transforming Eq.~\eqref{eq:alpha_solve}, this requirement becomes~\cite{Musco:2020jjb}
\begin{equation}
\label{eq:hat_r_m}
\int \frac{dk}{k}\,
\Bigg[
\big(k^2 r_m^2 - 1\big)\,
\frac{\sin(k r_m)}{k r_m}
+ \cos(k r_m)
\Bigg] \mathcal{P}_\zeta(k) = 0 \,.
\end{equation}
In practice, the procedure is as follows:  
(i) One first solves Eq.~\eqref{eq:hat_r_m} to determine $r_m$.  
(ii) This result is then inserted into Eq.~\eqref{eq:alpha_solve} to find $\alpha$.  
(iii) Finally, the value of $\alpha$ is substituted into Eq.~\eqref{eq:delta_c} to obtain the collapse threshold $\delta_c$. The results of these steps for the case of the log-normal distribution are shown in Fig.~\ref{fig:delta_c_Delta}.

\begin{figure*}[t!]
    \centering
    \includegraphics[width=0.495\linewidth]{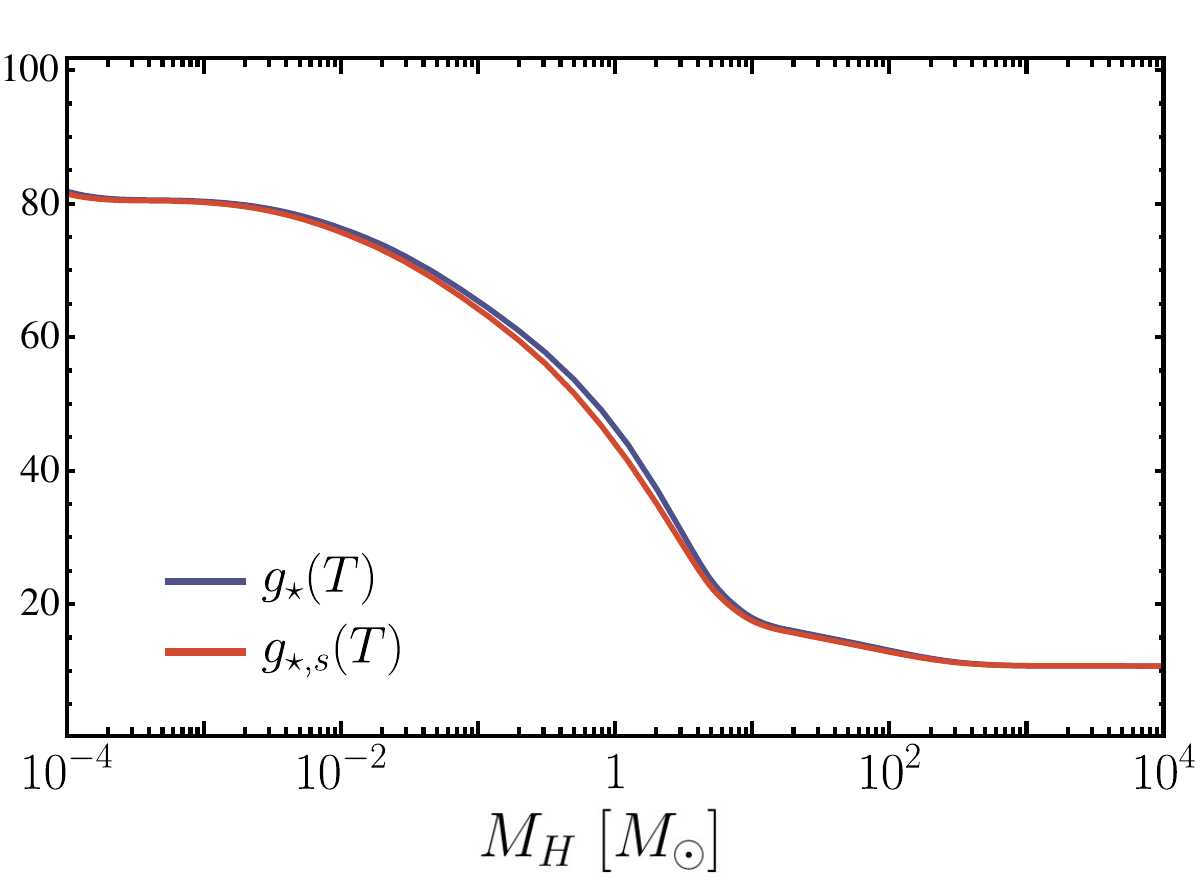}  
    \includegraphics[width=0.495\linewidth]{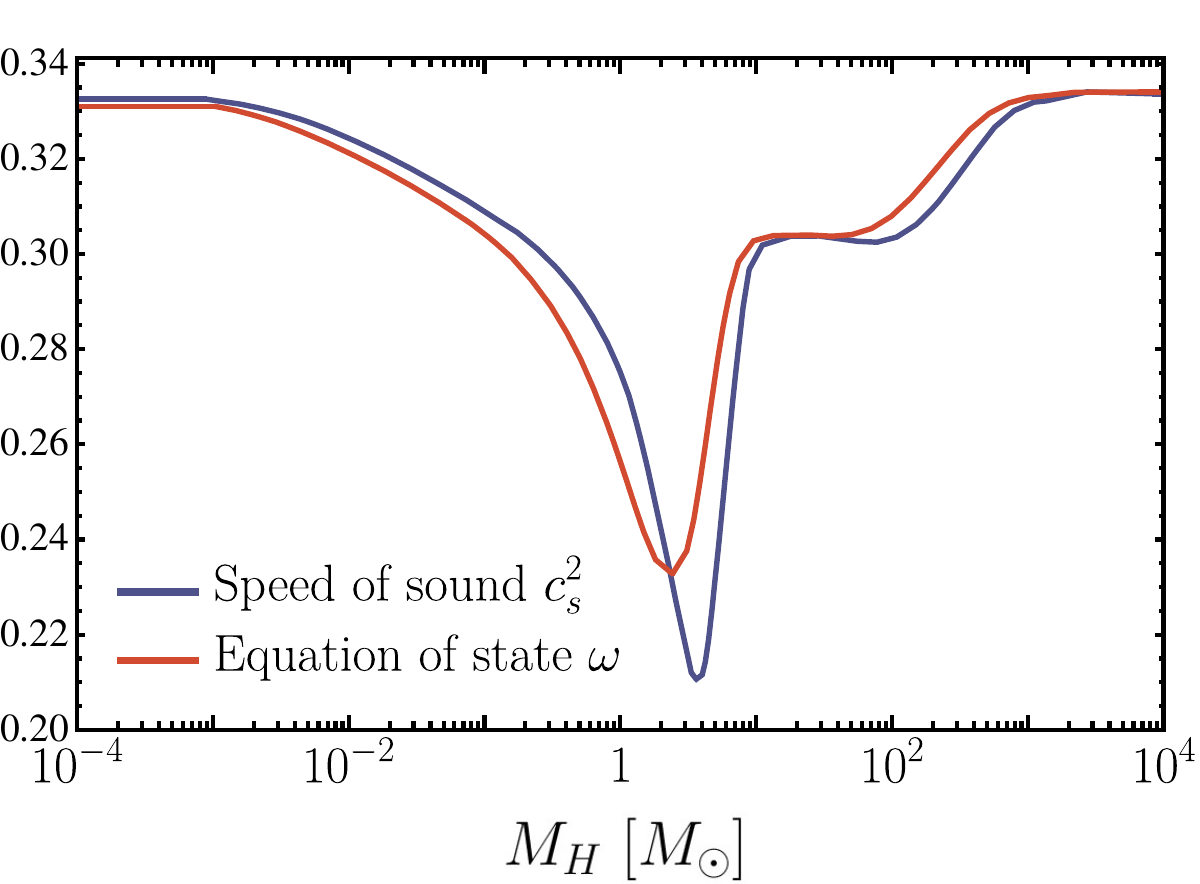}  
    \includegraphics[width=0.495\linewidth]{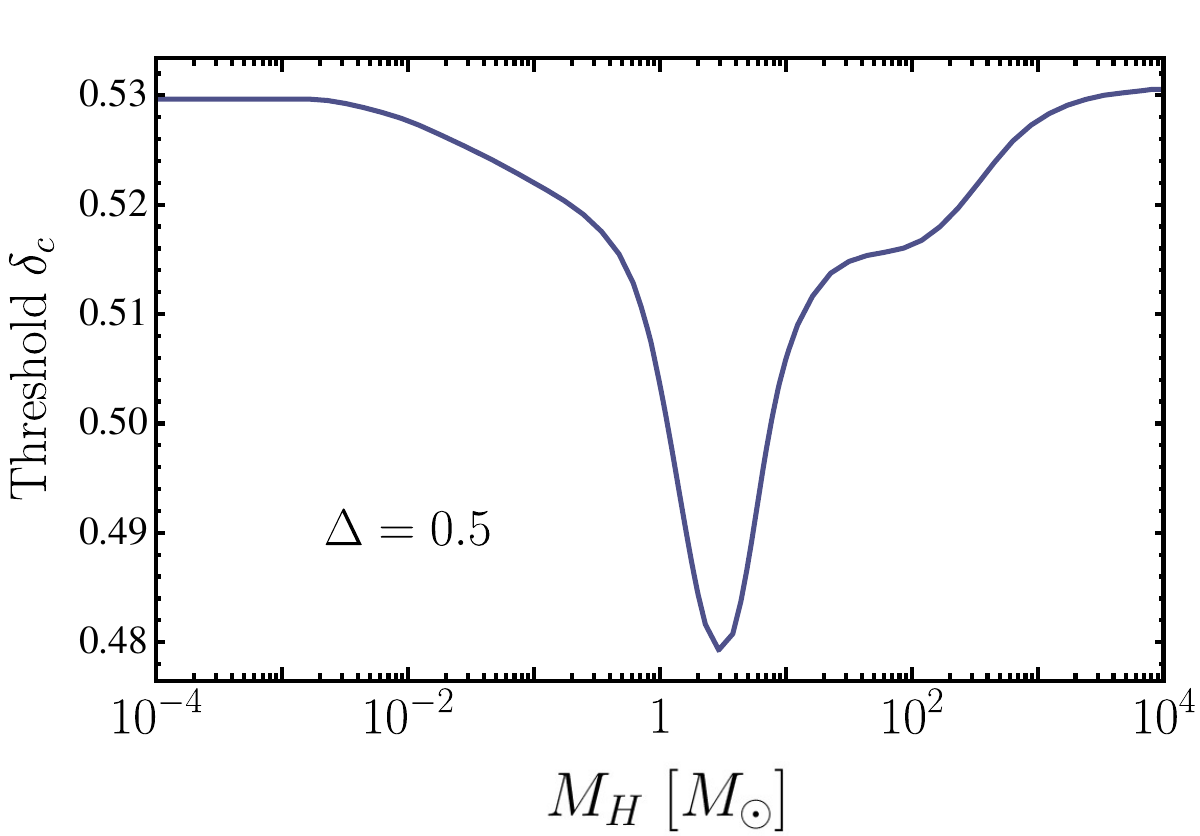}  
    \includegraphics[width=0.495\linewidth]{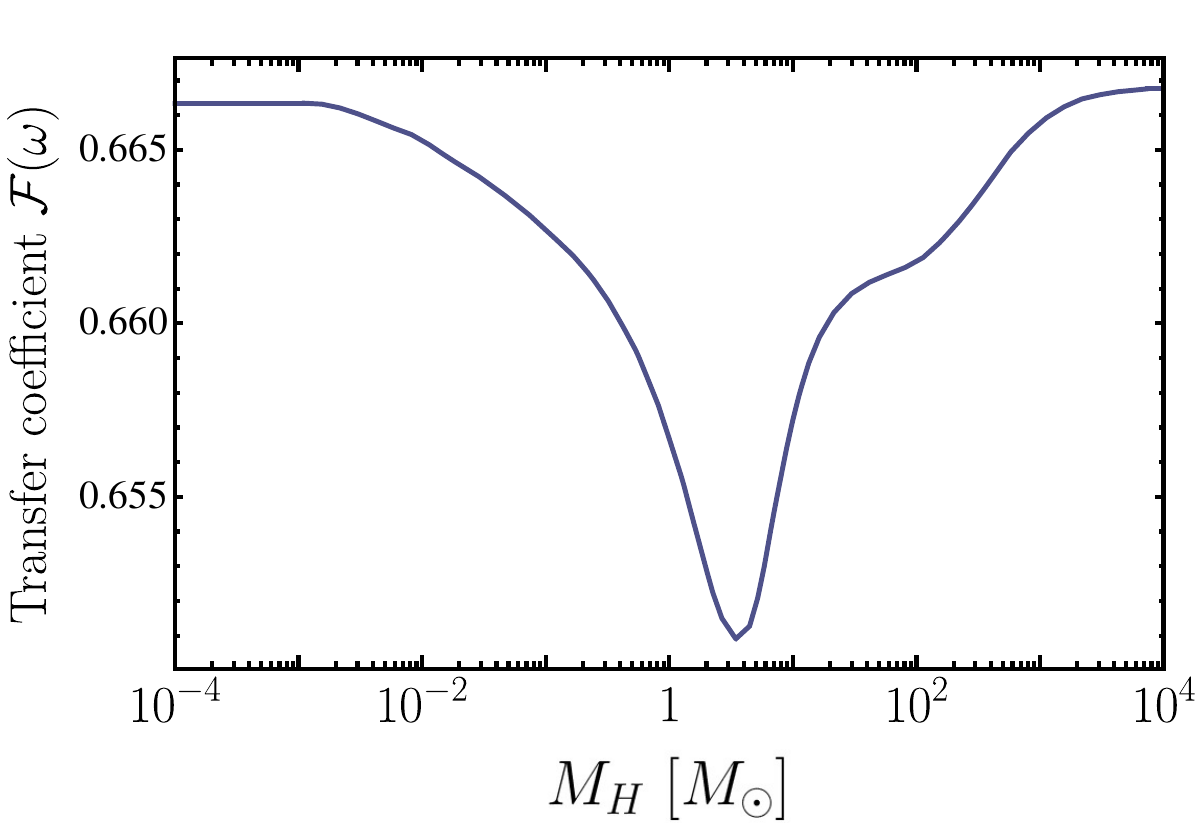}  
    \includegraphics[width=0.495\linewidth]{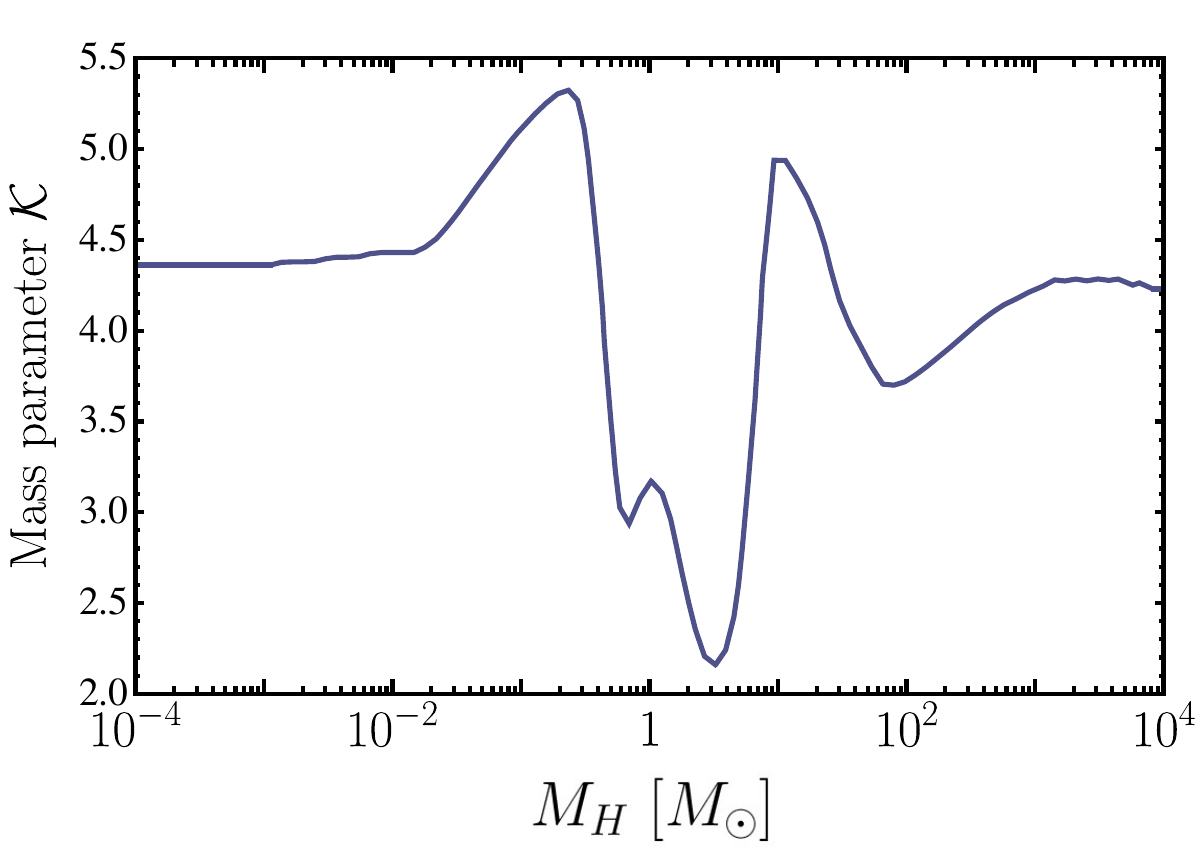}  
    \includegraphics[width=0.495\linewidth]{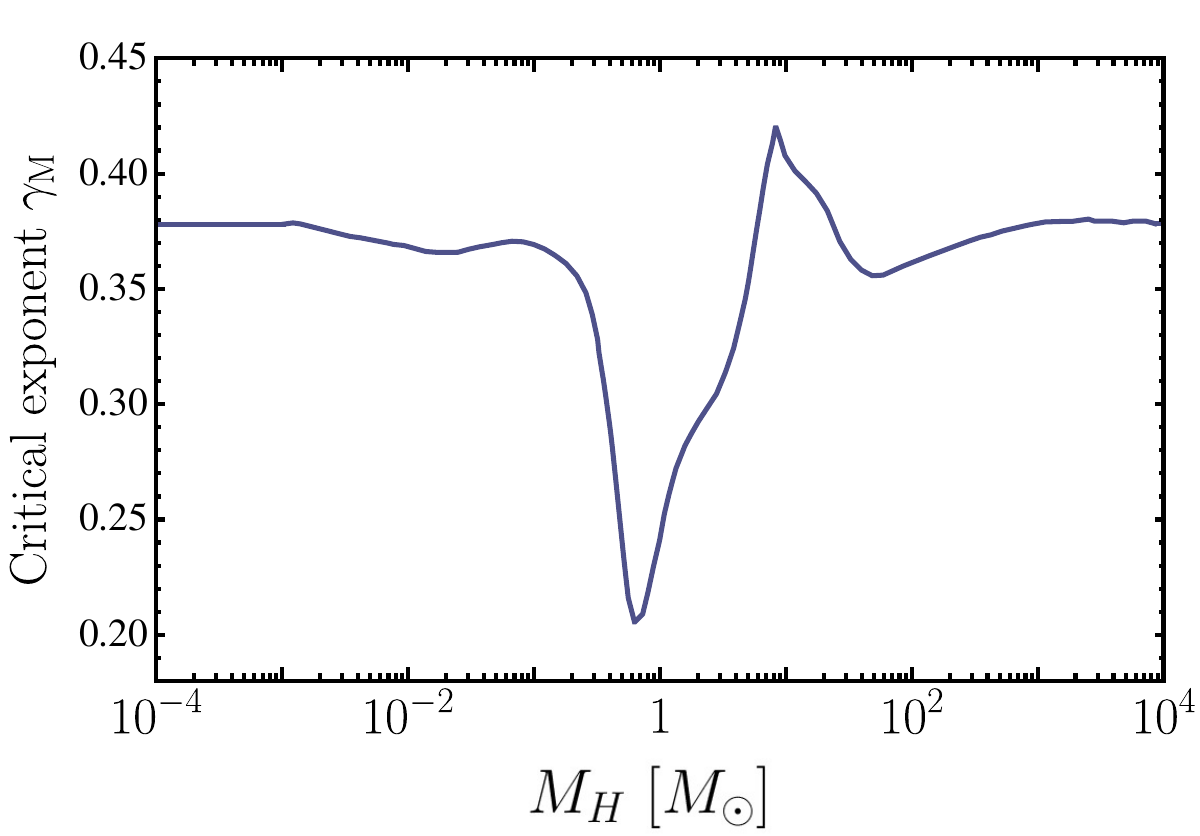}  
    \caption{Variations in the effective numbers of relativistic degrees of freedom, 
$g_\star(T)$ and $g_{\star,s}(T)$, induce corresponding changes in the equation of state $w$, the 
speed of sound $c_s^2$, the transfer 
function $\mathcal{F}(w)$ defined in Eq.~\eqref{eq:rho_contrast}, 
and the mass–scaling parameters $\mathcal{K}$ and $\gamma_{\rm M}$ 
appearing in Eq.~\eqref{eq:choptuik_law}. 
The functions of $g_\star(T)$ and $g_{\star,s}(T)$ are adapted from 
Ref.~\cite{Borsanyi:2016ksw}, while the remaining functions are 
taken from Refs.~\cite{Franciolini:2022tfm,Musco:2023dak}. 
\label{fig:QCD_effects}}
\end{figure*}

\paragraph{Choptuik's scaling law.}
The mass of a PBH follows the critical scaling relation first identified in Ref.~\cite{Choptuik:1992jv} and later applied to cosmological collapse~\cite{Niemeyer:1997mt,Niemeyer:1999ak,Musco:2012au,Musco:2018rwt,Ianniccari:2024ltb},  
\begin{equation}
\label{eq:choptuik_law}
    M_{\rm PBH}(\delta_m) = \mathcal{K}\, M_H \,(\delta_m - \delta_c)^{\gamma_{\rm M}}\,,
\end{equation}
with ${\gamma_{\rm M}} \simeq 0.36$ and $\mathcal{K} \simeq 4$ dimensionless constants. Here $M_H$ denotes the horizon mass at horizon crossing when $\epsilon(t=t_H)=1$ in Eq.~\eqref{eq:definition_epsilon}. Using $aH r_m \simeq 1$, one obtains the horizon mass in terms of the scale $r_m$ associated with the maximum of the compaction function~\cite{Musco:2020jjb},  
\begin{equation}
\label{eq:m_horizon_rm}
    M_H(r_m) \simeq M_\odot
    \left(\frac{10.75}{g_\star(T_H)}\right)^{1/2}
    \left(\frac{120~\rm MeV}{T_H}\right)^{2}
    \simeq M_{\odot}\,
    \left(\frac{10.75}{g_\star(T_H)}\right)^{1/6}
    \left(\frac{4.5\times 10^{6} ~\rm Mpc^{-1}}{r_m^{-1}}\right)^{2},
\end{equation}
where $g_\star(T)$ is the number of relativistic degrees of freedom which in the last equation is assumed to be equal to the entropic ones $g_\star(T)=g_{\star,s}(T)$.  It can also be useful to define the mass $M_\star$ when the peak scale $k_\star$ of the log-normal power spectrum in Eq.~\eqref{eq:log_normal} crosses the horizon  $aH = k_\star$,
\begin{equation}
    M_{\star}\equiv M_H(r_m=k_\star^{-1}).
\end{equation}
The equation of state and sound speed can be written as 
\begin{equation}
    w\equiv\frac{p}{\rho} = \frac{4g_{\star,s}}{3g_\star}-1\,,\qquad \textrm{and}\qquad c_s^2 \equiv \frac{\partial p}{\partial \rho},
\end{equation}
where $\rho$, $p=Ts-\rho$ and $s$ are the energy, pressure and entropy density  of the universe.
Near the QCD transition, variations in equation of state $w(T)$ and the sound speed $c_s^2(T)$ lead to a non-trivial dependence of $\mathcal{F}$, $\delta_c$, $\gamma_{\rm M}$, and $\mathcal{K}$ on $M_H$~\cite{Byrnes:2018clq,Franciolini:2022tfm,Escriva:2022bwe,Musco:2023dak} (see also~\cite{Jedamzik:1996mr,Schmid:1998mx,Carr:2019kxo,Pritchard:2024vix}). We note that NGs could also impact the value of $\delta_c$~\cite{Franciolini:2023pbf}, but there is currently no clear prescription to include those effects.
In our analysis we adopt the results of Refs.~\cite{Franciolini:2022tfm,Musco:2023dak}, obtained using the lattice code developed in Refs.~\cite{Musco:2004ak,Musco:2008hv,Musco:2012au} and display them in Fig.~\ref{fig:QCD_effects}.

\paragraph{PBH mass function (Gaussian limit).}

\begin{figure*}[t!]
    \centering
    \includegraphics[width=0.7\linewidth]{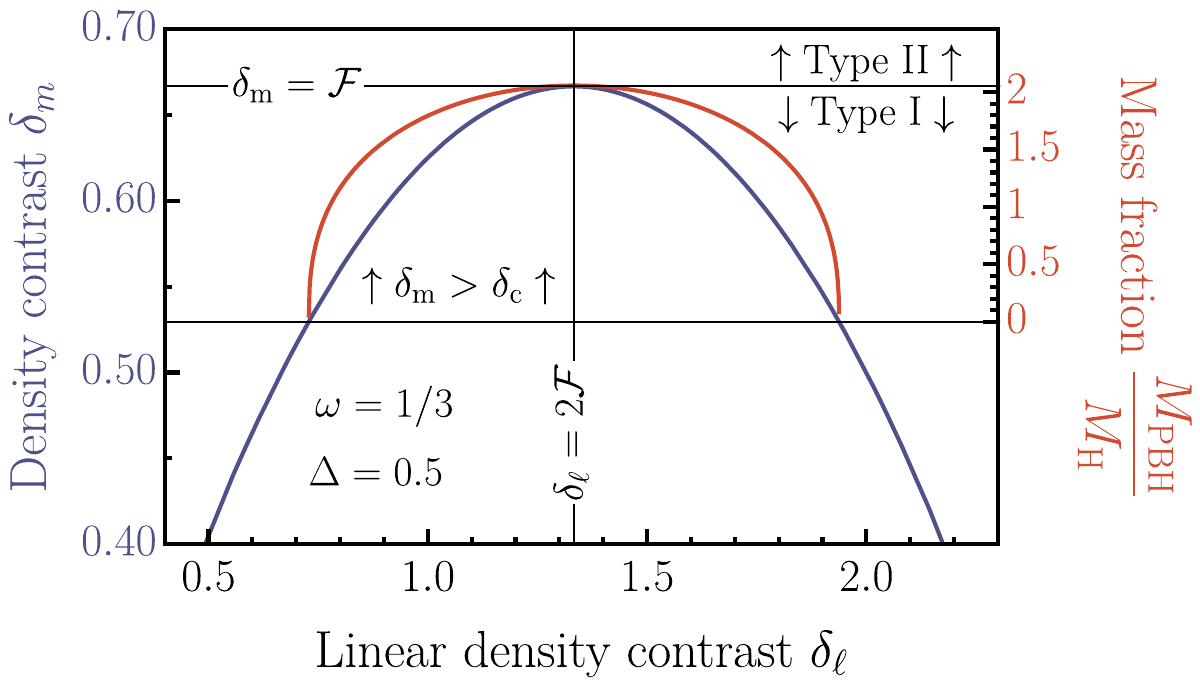}  
    \caption{Non-linear averaged density contrast $\delta_m$, defined in Eq.~\eqref{eq:delta_m_def}, as a function of the linear density contrast, introduced in Eq.~\eqref{eq:delta_ell_def}. The right axis shows the PBH mass in unit of the Hubble horizon $M_{\rm PBH}/M_H$, following Choptuik's law in Eq.~\eqref{eq:choptuik_law}. We observe that the non-linear density contrast $\delta_m$ is bounded by $\delta_c < \delta_m<\mathcal{F}$, and the PBH mass is bounded by $M_{\rm PBH} / M_H< \mathcal{K}(\mathcal{F}-\delta_c)^{\gamma_{\rm M}}\sim 2$. For a fixed $M_{\rm PBH}$, this implies the lower bound $M_H^{\rm min}$ introduced in Eq.~\eqref{eq:M_H_min} for the integral over $\ln M_H$ in Eq.~\eqref{eq:fPBH_Gauss}. \label{fig:delta_m_delta_ell}}
\end{figure*}

In order to investigate PBH formation over a wider range of scales and recover the mass distribution of the resulting PBH populations, we revisit the perturbation amplitude and introduce a smoothing function. At $t=t_H$ we may rewrite the mean mass excess in Eq.~\eqref{eq:delta_K_zeta} as~\cite{Young:2019yug, Young:2022phe}
\begin{equation}
    \bar\delta(X,t_H) = \int \frac{\delta \rho}{\rho_b}(x,t_H) W(X-x,r_m) d^3x \,, \quad W(\mathbf{x},r_m) = \frac{\Theta(r_m-|\mathbf{x}|)}{\frac{4}{3}\pi r_m^3}\,,
\end{equation}
where $X$ is the location of the peak in Cartesian coordinates and $W$ is the top-hat window function. We may express then the peak amplitude in Fourier space as
\begin{equation}
\label{eq:delta_m_fourier}
    \delta_m(k,r_m) = \tilde{W}(k, r_m)\frac{\delta\rho}{\rho_b}(k,r_m)\,,\quad \tilde{W}(k,r_m) = 3 \frac{\mathrm{sin}(k r_m)- k r_m \mathrm{cos}(k r_m)}{(k r_m)^3}\,,
\end{equation}
where the Fourier transform of Eq.~\eqref{eq:rho_contrast_zeta} at $\mathcal{O}(\zeta(r_m))$ reads
\begin{equation}
\label{eq:delta_zeta_fourier}
    \frac{\delta\rho}{\rho_b}(k,r_m) = \frac{2}{3}\mathcal{F}\left(k r_m \right)^2 \zeta(k)\,.
\end{equation}
and in this linear limit we have $\delta_m(k,r_m) = \delta_\ell(k,r_m)$.
Combining then Eqs. \eqref{eq:delta_m_fourier} and ~\eqref{eq:delta_zeta_fourier}, we obtain
\begin{equation}
\label{eq:delta_m_zeta_fourier}
    \delta_\ell(k,r_m) = \frac{2}{3}\mathcal{F} (kr_m)^2 \tilde{W}(k,r_m)\mathcal{T}(k,r_m)\zeta(k)\,,
\end{equation}
where we have also introduced the linear transfer function $\mathcal{T}$ to encode the evolution of the perturbations after horizon reentry~\cite{Blais:2002gw,Josan:2009qn,Franciolini:2021nvv} 
\begin{equation}
\mathcal{T}(k,r_m) =  3 \frac{\mathrm{sin}(c_sk r_m)- c_sk r_m\, \mathrm{cos}(c_sk r_m)}{(c_sk r_m)^3}\,, \qquad c_s=1/\sqrt{3}\,.
\label{eq:transfer_per}
\end{equation}
 We are now equipped to derive the power spectrum of $\delta_{\ell}$ as a function of the curvature power spectrum and integrate over it to obtain the variance
\begin{equation}
\label{eq:sigma_ell_def}
    \sigma_{\delta_\ell}^2(M_H) \equiv \langle \delta_{\ell}^2 \rangle = \int_0^\infty \frac{dk}{k} \mathcal{P}_{\delta_{\ell}}(k,r_m) = \frac{4}{9}\mathcal{F}^2  \int_0^\infty \frac{dk}{k} (kr_m)^4 \tilde{W}^2(k,r_m)\mathcal{T}^2(k,r_m)\mathcal{P}_{\zeta}(k)\,.
\end{equation}
We have made explicit the dependence of $\sigma_{\delta_\ell}$ on $M_H$, 
which arises through the relation between $r_m$ and $M_H$ in
Eq.~\eqref{eq:m_horizon_rm}.
The fraction of the universe that collapses to PBHs is then given by 
 \begin{equation}
 \label{eq:beta_Gauss}
     \beta(M_{\rm PBH};M_H) = \int_{\delta_{c}}^{\infty} d\delta_{\ell} \, \frac{M_{\rm PBH}}{M_H} P_f(\delta_{\ell};M_H) \, \delta_{\rm D}\left[\ln{\frac{M_{\rm PBH}}{m_{\rm PBH}(\delta_m)}}\right]\,, 
 \end{equation}
where $P_f(\delta_{\ell};M_H)$ denotes the probability distribution for the linear Hubble-averaged density contrast $\delta_\ell$ at the time when the mass within the horizon is $M_H$. The quantity $\delta_D(x)$ is the Dirac function. Substituting Eq. \eqref{eq:choptuik_law} in the quadratic Eq. \eqref{eq:delta_ell_def} and solving for $\delta_{\ell}$ we get
\begin{equation}
\delta_{\ell} = 2\mathcal{F} \left(1-\sqrt{\Lambda}\right)\,, \quad \Lambda\equiv 1-\frac{\delta_c}{\mathcal{F}} - \frac{1}{\mathcal{F}}\left(\frac{M_{\rm PBH}}{\mathcal{K}M_H}\right)^{1/{\gamma_{\rm M}}}\,,
\end{equation}
and now we can use the properties of $\delta_D(x)$ to perform the integration in Eq. \eqref{eq:beta_Gauss} and obtain
\begin{equation} \label{eq:beta_Gauss_2}
    \beta(M_{\rm PBH};M_H) = \left(\frac{M_{\rm PBH}}{\mathcal{K} M_H}\right)^{\!\!\frac{1+{\gamma_{\rm M}}}{{\gamma_{\rm M}}}}  \frac{{\cal K}}{{\gamma_{\rm M}}\sqrt{\Lambda}}  P_f(\delta_{\ell};M_H) \,.
\end{equation}
Two main statistical methods are commonly employed to estimate the probability distribution $P_{H}(\delta)$ of the fluctuations that lead to PBH formation:  
\begin{enumerate}

\item \emph{Press–Schechter formalism} (or \emph{threshold statistics})~\cite{Press:1973iz,Bond:1990iw,Niemeyer:1997mt,Dimopoulos:2019wew,Karam:2022nym}.  
In this approach, a PBH forms whenever the collapse criterion of Eq.~\eqref{eq:PBH_formation_condition} is satisfied, so the calculation reduces to the statistics of the linear density contrast $\delta_{\ell}$.  
For Gaussian perturbations one {\color{red}{assumes}}
\begin{equation} \label{eq:Prob_PS_Gauss}
    P_f(\delta_{\ell};M_H)
    = \frac{1}{\sqrt{2\pi}\,\sigma_{\delta_\ell}}
      \exp\!\left(-\frac{\delta_{\ell}^2}{2\sigma_{\delta_\ell}^2}\right)\,.
\end{equation}

\item \emph{Peak theory}~\cite{Bardeen:1985tr,Green:2004wb,Young:2014ana,Young:2019yug,DeLuca:2019qsy,MoradinezhadDizgah:2019wjf,Wu:2020ilx,Gow:2020bzo,Young:2022phe,Iovino:2024tyg}.  In contrast to Press--Schechter, peak theory does not sample $\delta_\ell$ at random points, but instead counts local maxima of the smoothed density field.
The relevant quantity is the differential number density of maxima of the compaction function, given by~\cite{Bardeen:1985tr}
\begin{equation}
    \mathcal N_{\rm pk}(\delta_{\ell};M_H)
    = \frac{1}{3^{3/2}(2\pi)^2 r_m^3}
      \left(\frac{\tilde{\sigma}_{\delta_\ell}}{\sigma_{\delta_\ell}}\right)^3
      \left(\frac{\delta_{\ell}}{\sigma_{\delta_\ell}}\right)^3
      \exp\!\left(-\frac{\delta_{\ell}^2}{2\sigma_{\delta_\ell}^2}\right)\,,
\end{equation}
with
\begin{equation}
    \tilde{\sigma}_{\delta_\ell}^2(M_H)
    = \frac{4 \mathcal{F}^2}{9}
      \int_0^\infty \frac{dk}{k}\,(k r_m)^6\,
      \tilde{W}^2(k,r_m)\,\mathcal{T}^2(k,r_m)\,\mathcal{P}_{\zeta}(k)\,.
\end{equation}
In the \emph{high-peak limit} $\delta_{\ell}/\tilde{\sigma}_{\delta_\ell} \gg 1$, we can safely assume that peaks in $\delta_\ell$ correspond to high peaks in $\zeta$. The probability of finding at least one such peak inside a Hubble volume is
\begin{equation} \label{eq:Prob_PT_Gauss}
    P_f(\delta_{\ell};M_H)
    = \frac{4\pi r_m^3}{3}\,\mathcal N_{\rm pk}(\delta_{\ell},M_H)\,.
\end{equation}
Relative to the Press--Schechter estimate in Eq.~\eqref{eq:Prob_PS_Gauss}, peak theory predicts an \emph{enhancement} of rare overdensities by a characteristic factor $\propto (\delta_{\ell}/\sigma_{\delta_\ell})^3$, reflecting the preferential formation of PBHs around highly curved density peaks rather than generic regions above threshold. 
\end{enumerate}
The PBH mass function can be expressed as
\begin{equation}
\label{eq:fPBH_Gauss}
    \frac{df_{\rm PBH}}{d\ln M_{\rm PBH}} 
    \equiv \frac{1}{\Omega_{\rm DM}} 
    \int_{M_H^{\rm min}}^\infty d \ln M_H \,
    \left( \frac{M_{\rm eq}}{M_H}\right)^{1/2}  
    \beta(M_{\rm PBH};M_H)\,,
\end{equation}
where $M_{\rm eq} \simeq 2.8 \times 10^{17} M_\odot$ is the horizon mass at matter–radiation equality.  
The lower integration limit is fixed by the fact that the function $\delta_m(\delta_\ell)$ in Eq.~\eqref{eq:delta_ell_def} reaches a maximum at $(\delta_\ell,\delta_m)=(2\mathcal{F},\mathcal{F})$, see Fig.~\ref{fig:delta_m_delta_ell}.
Using Eq.~\eqref{eq:choptuik_law} this gives the lower bound on $M_H$
\begin{equation}
\label{eq:M_H_min}
    M_{H}^{\rm min}(M_{\rm PBH}) 
    = \frac{M_{\rm PBH}}
    {\mathcal{K}\,(\mathcal{F}-\delta_c)^{\gamma_{\rm M}}}\,.
\end{equation}
Configurations with $\delta_m>\mathcal{F}$ are referred to as \emph{type II} fluctuations~\cite{Kopp:2010sh,Carr:2014pga}\footnote{We emphasize that the terminology of type~I and type~II fluctuations 
originates from Refs.~\cite{Kopp:2010sh,Carr:2014pga} and is not related to 
the nomenclature of type~I and type~II \emph{critical collapse} introduced 
in the study of critical phenomena~\cite{Gundlach:2025yje}. While the collapse 
of PBHs is known to follow type~II critical collapse, the fluctuations 
themselves can be of either type~I or type~II.} 
In contrast, \emph{type I} fluctuations, which are the focus of this work, satisfy $\delta_m<\mathcal{F}$.  
From Eq.~\eqref{eq:delta_K_zeta} the boundary $\delta_m=\mathcal{F}$ corresponds to the condition $K(\hat{r})\hat{r}^2=1$.  
In this case the conformal radius $\hat{r}$ develops a coordinate singularity, as seen in Eq.~\eqref{eq:metric_conformal}.  
This singularity is absent when expressed in terms of the physical radius $r$, where the areal radius defined in Eq.~\eqref{eq:areal_radius} reaches a minimum, $dR(r,t)/dr=0$.  
This signals the appearance of a neck-like geometry~\cite{Uehara:2024yyp}. For backgrounds with $w<-1/3$, \emph{type II} fluctuations give rise to a wormhole geometry that pinches off, producing a baby universe on the inside and a black hole as seen from the outside~\cite{Carr:2014pga}.

\paragraph{Non-Gaussian perturbations.}

Next we consider the impact of NGs on the PBH abundance. 
\emph{Local-type} NGs, e.g. $f_{\rm NL}^{\rm local}$ and $g_{\rm NL}^{\rm local}$, are the first non-linear expansion parameters around the Gaussian perturbations $\zeta$, where locality refers to the fact that the expansions does not depend on spatial derivatives of the $\zeta$~\cite{Ferrante:2022mui}. They are parameterized by the power-series expansion
\begin{equation}
\label{eq:local_NG}
\zeta_{\rm NG} = F(\zeta) = \zeta + F_{\rm NL} (\zeta ^2- \langle \zeta ^2 \rangle)  + G_{\rm NL} \zeta ^3 + ...\,,  
\end{equation}
where the parameters $F_{\rm NL}=3f_{\rm NL}^{\rm local}/5$, $G_{\rm NL}=9g_{\rm NL}^{\rm local}/25$ are sufficient to capture the  NG effects that are relevant in the current discussion. Intuitively,  the quadratic term introduces skewness to the distribution, whilst the
cubic term deforms the kurtosis. The perturbativity regime where the above expansion is typically considered valid is $|F_{\rm NL}|^2 A_{\zeta}\ll1$ and $|G_{\rm NL}|A_{\zeta}\ll1$.

Equivalently, we obtain the peak amplitude in Eq. \eqref{eq:delta_ell_def} by replacing the linear component by its NG counterpart 
\begin{equation}\label{eq:delta_ell_NG_def}
\delta_{m} = \delta_{\ell,\rm NG} -\frac{1}{4\mathcal{F}} \delta_{\ell,\rm NG}^2\,, \quad \delta_{\ell,\rm NG} = \delta_{\ell} \,\rm F'(\zeta)\,,
\end{equation}
where $\delta_{\ell}$ is again a Gaussian field.
The statistics of $\zeta$ and $\delta_{\ell}$ are described by a two-dimensional Gaussian joint probability distribution function (PDF)~\cite{Ferrante:2022mui,Young:2022phe,Gow:2022jfb,DeLuca:2022rfz}
\begin{equation}
    \textrm{P}_{\rm G}(\delta_{\ell},\zeta) 
 = \frac{1}{(2\pi)\sqrt{\det\Sigma}}
 \exp\left(
 -\frac{1}{2}Y^{\rm T}\Sigma^{-1}Y
 \right)\,,\quad
\Sigma =
\left(
\begin{array}{cc}
 \langle\delta_{\ell}^2\rangle & \langle\delta_{\ell}\zeta\rangle   \\
 \langle\delta_{\ell}\zeta\rangle & \langle\zeta^2\rangle     
\end{array}
\right)\,,
\end{equation}
where $Y =\left(
\begin{array}{c}
 \delta_{\ell}  \\
  \zeta    
\end{array}
\right)$.
By calculating the inverse of $\Sigma$ and its determinant, and by completing the square within the argument of the exponential function, we obtain~\cite{Young:2022phe}
\begin{equation}
     \label{eq:joint_PDF}
P_{\mathrm{G}}\left(\delta_{\ell}, \zeta\right)=
   \frac{1}{2 \pi \sigma_2 \sigma_0 \sqrt{1-\gamma_{m}^2}} \exp\left[- \frac{\zeta^2}{2 \sigma_0^2}-\frac{1}{2\left(1-\gamma_{m}^2\right)}\left(\frac{\delta_{\ell}}{\sigma_2}-\frac{\gamma_{m} \zeta}{\sigma_0}\right)^{\!2}\right]\,,
\end{equation}
where the correlation coefficients are defined by $ \sigma_0^2\equiv\left< \zeta^2\right>$, $ \sigma_1^2\equiv\left< \delta_{\ell}\zeta\right>$, $ \sigma_2^2\equiv\left< \delta_{\ell}^2\right>=\sigma_{\delta_{\ell}}^2$, and $\gamma_{m} \equiv \sigma_{1}^2 / (\sigma_0 \sigma_2)$. They are calculated by 
\begin{align}
    & \sigma_n^2(M_H)=\left(\frac{2 f}{3}\right)^n \int_0^{\infty} \frac{\rm d k}{k}\left(k r_m\right)^{2n} \!\tilde{W}^{n}\!\!\left(k, r_m\right) \!\tilde{W}_s^{2-n}\!\left(k, r_m\right) \!\mathcal{T}^2\left(k, r_m\right) P_\zeta(k)\,, \quad n=0,1,2\,.
\end{align}
Again, the dependence on $M_H$ arises from the relation $r_m(M_H)$ obtained from inverting Eq.~\eqref{eq:m_horizon_rm}.
The linear transfer function $\mathcal{T}$ is given in Eq. \eqref{eq:transfer_per}, and $\tilde{W}$ and $\tilde{W}_s$ are the top-hat and spherical-shell window functions, respectively, in Fourier space. The former is defined in Eq. \eqref{eq:delta_m_fourier}, while the latter is
\begin{align} \label{eq:window_spherical}
\tilde{W}_s\left(k, r_m\right)= \frac{\sin(k r_m)}{kr_m}\,.
\end{align}
The coefficient ${\gamma_{m}}$ denotes the correlation coefficient between $\delta_\ell$ and $\zeta$, and the limit ${\gamma_{m}} \to 1$ corresponds to a monochromatic power spectrum $\mathcal P_\zeta(k) = \mathcal A_\zeta k \delta_D (k-k_\star)$. It is worth mentioning that the perturbative expansion in Eq. \eqref{eq:local_NG} is shown to be unreliable for ${\gamma_{m}} \ll 1$, which correspond to broader power spectra and the abandonment of the high-peak limit~\cite{Young:2022phe,Ferrante:2022mui}. More generally, as emphasized in Ref.~\cite{Ferrante:2022mui}, whenever a closed relation exists between $\zeta$ and the underlying Gaussian field (as in ultra-slow roll or curvaton scenarios), the power-series expansion breaks down already for relatively narrow spectra with width $\Delta \gtrsim 0.2$ (see their Fig.~4), and for broader spectra it is never exact (see their Fig.~7). While we will not place special emphasis on this point in the subsequent analysis, it is worth noting here that, \emph{a posteriori}, the regime with $\Delta<0.5$, which is consistently captured within the perturbative framework, is also of heightened phenomenological interest.

\begin{figure*}[t!]
    \centering
\includegraphics[width=0.7\linewidth]{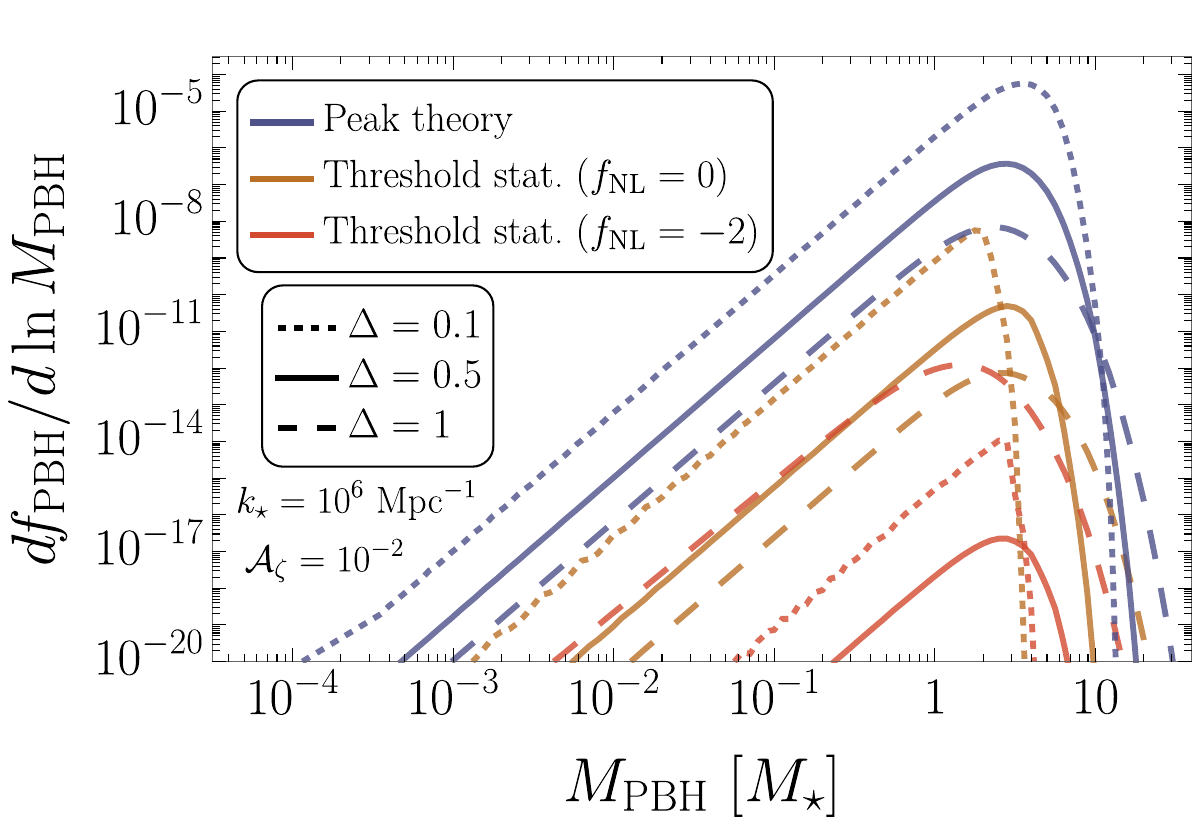}  
    \caption{PBH mass function for different benchmarks for the width $\Delta=0.1,0.5,1$ (dotted, solid, dashed curves) both for the Gaussian peak theory (blue) and the threshold statistics calculation with $f_{\rm NL}=-2,0$ (red, orange). $M_\star$ is the mass of the universe when the peak scale $k_\star$ crosses the horizon $k_\star=\mathcal{H}$. \label{fig:fPBH_log_normal}}
\end{figure*}

\begin{figure*}[t!]
    \centering
    \includegraphics[width=0.495\linewidth]{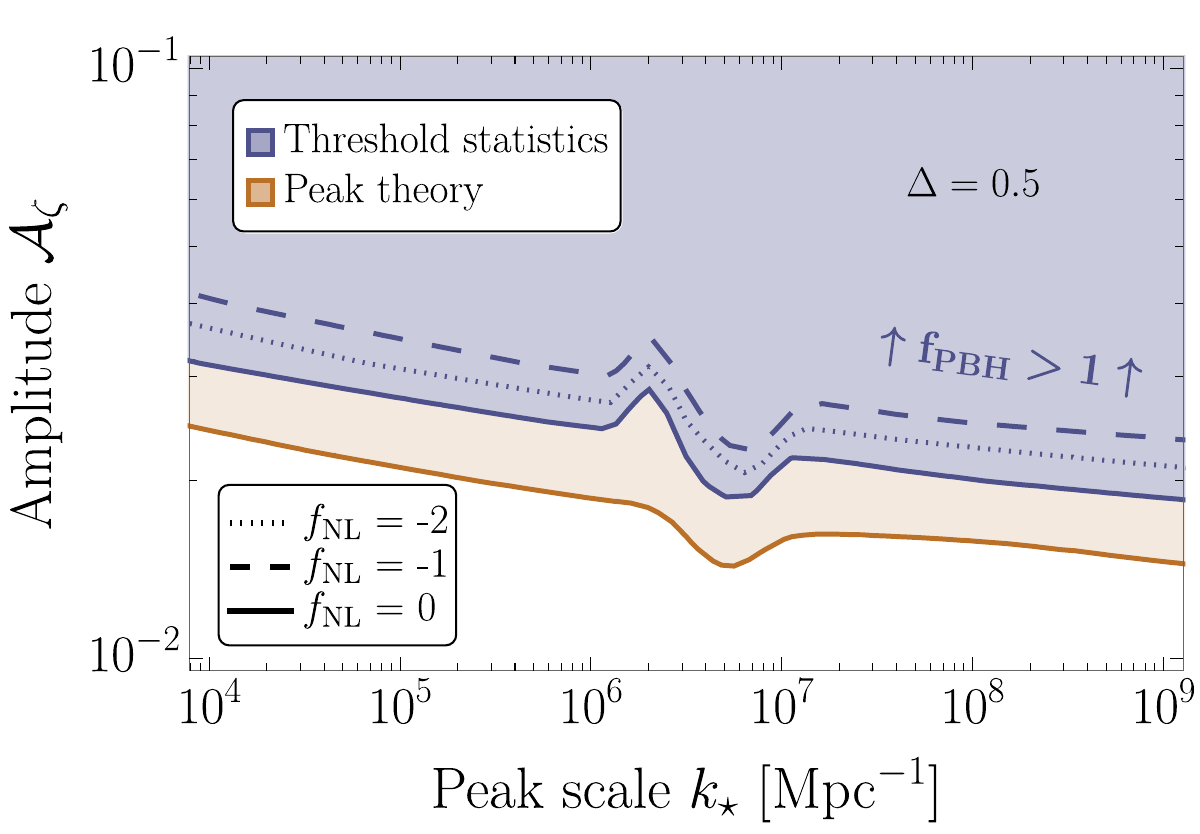}  
    \includegraphics[width=0.495\linewidth]{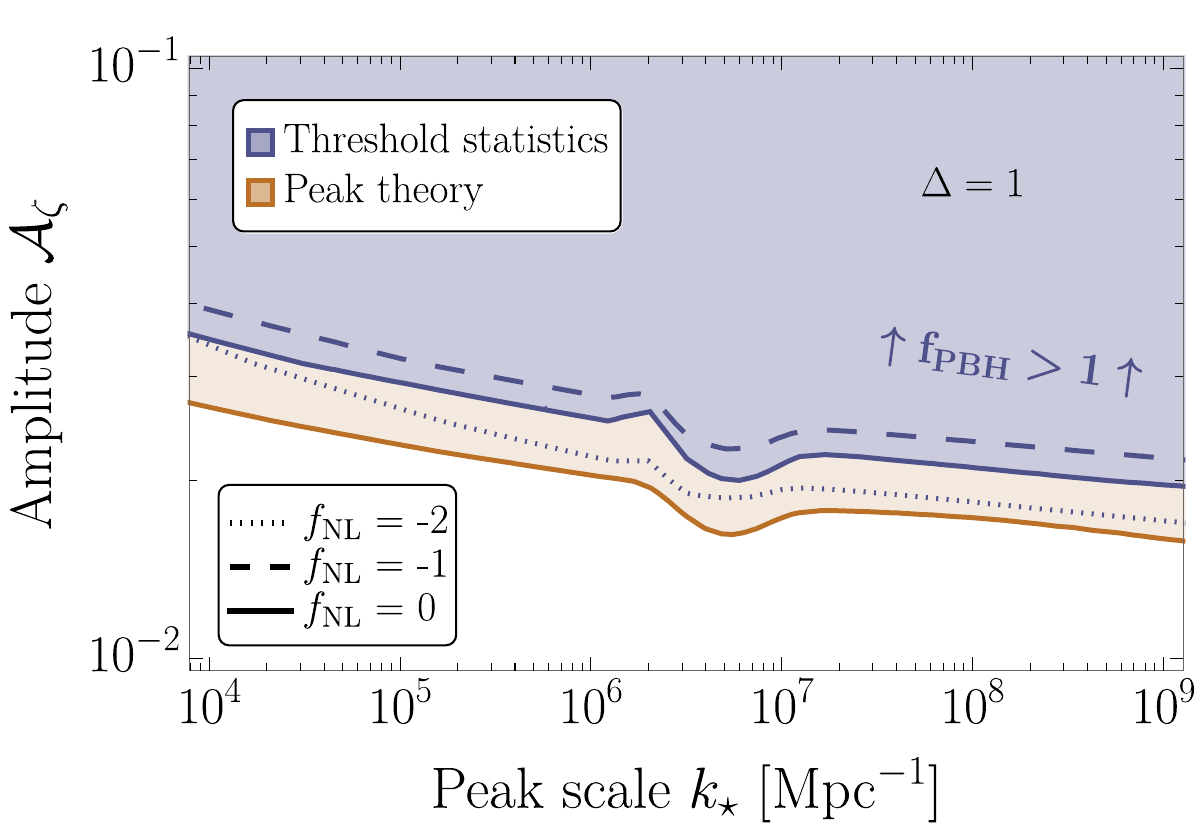}  
    \includegraphics[width=0.495\linewidth]{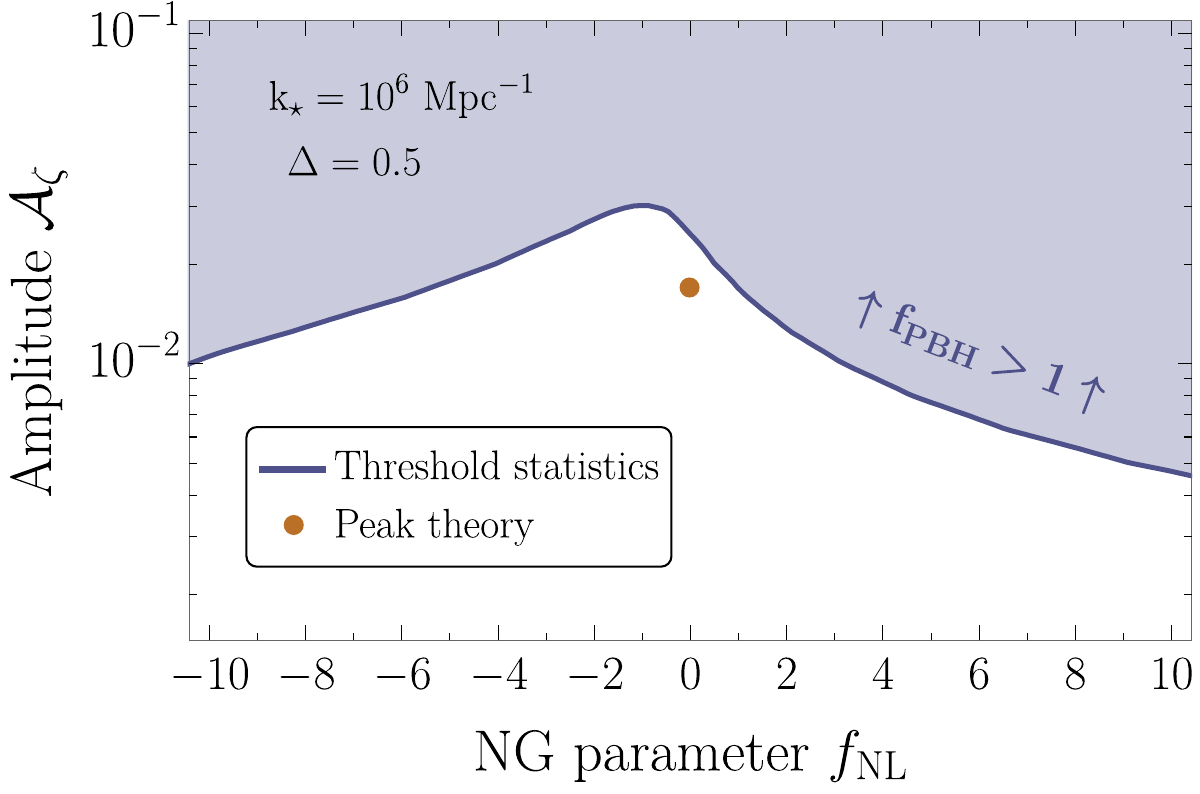}  
    \includegraphics[width=0.495\linewidth]{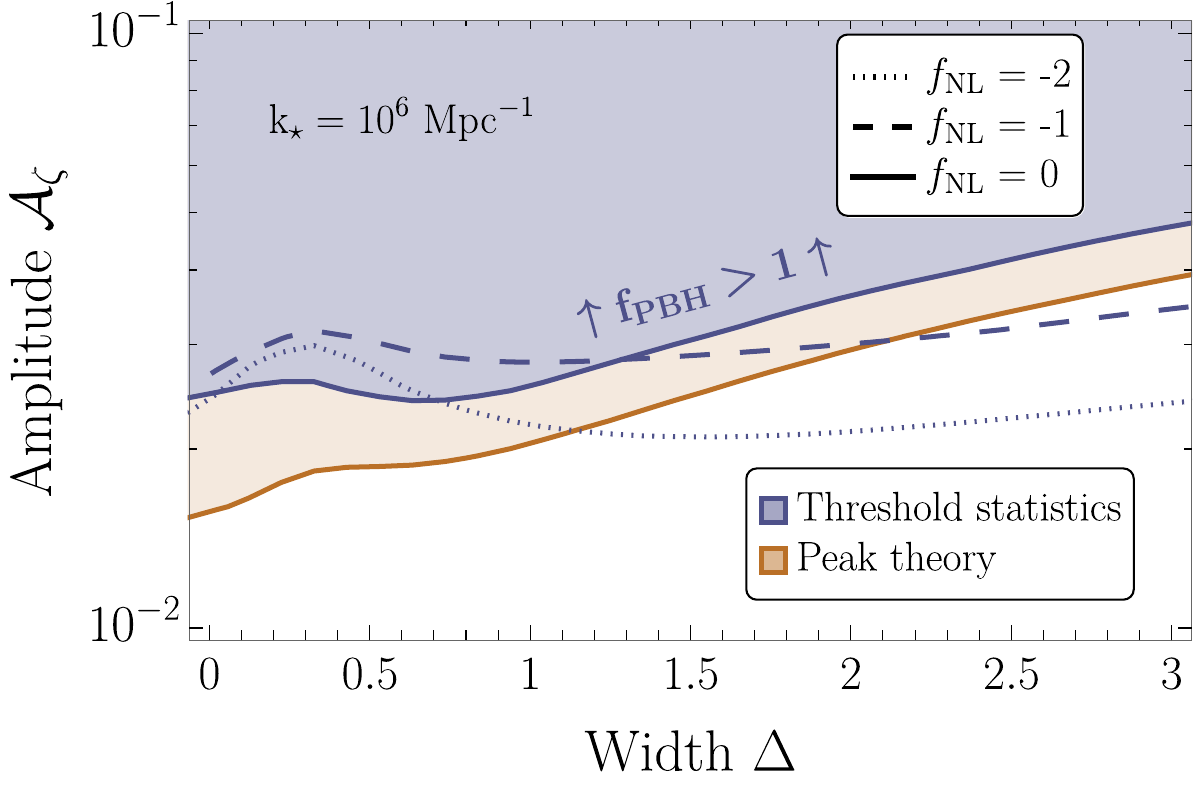}

    \caption{Dark matter overclosure bound on the amplitude $\mathcal{A}_{\zeta}$ of the log-normal curvature distribution in Eq.~\eqref{eq:log_normal}. The bound is given as a function of the comoving scale $k_\star$ for two width benchmarks $\Delta=0.5,1$ (\textbf{top-left} and \textbf{top-right}), respectively, the NG parameter $f_{\rm NL}$ (\textbf{bottom-left}), and the width $\Delta$ (\textbf{bottom-right}) using both  threshold statistics (blue) and peak theory (orange). In the case of threshold statisticsthe top figures illustrate also the effect the NG parameter for three benchmarks $f_{\rm NL}=-2,-1, 0$ (resp. dotted, dashed and solid). \label{fig:fPBH_log_normal_DM}}
\end{figure*}

\begin{figure*}[t!]
    \centering
    \includegraphics[width=0.495\linewidth]{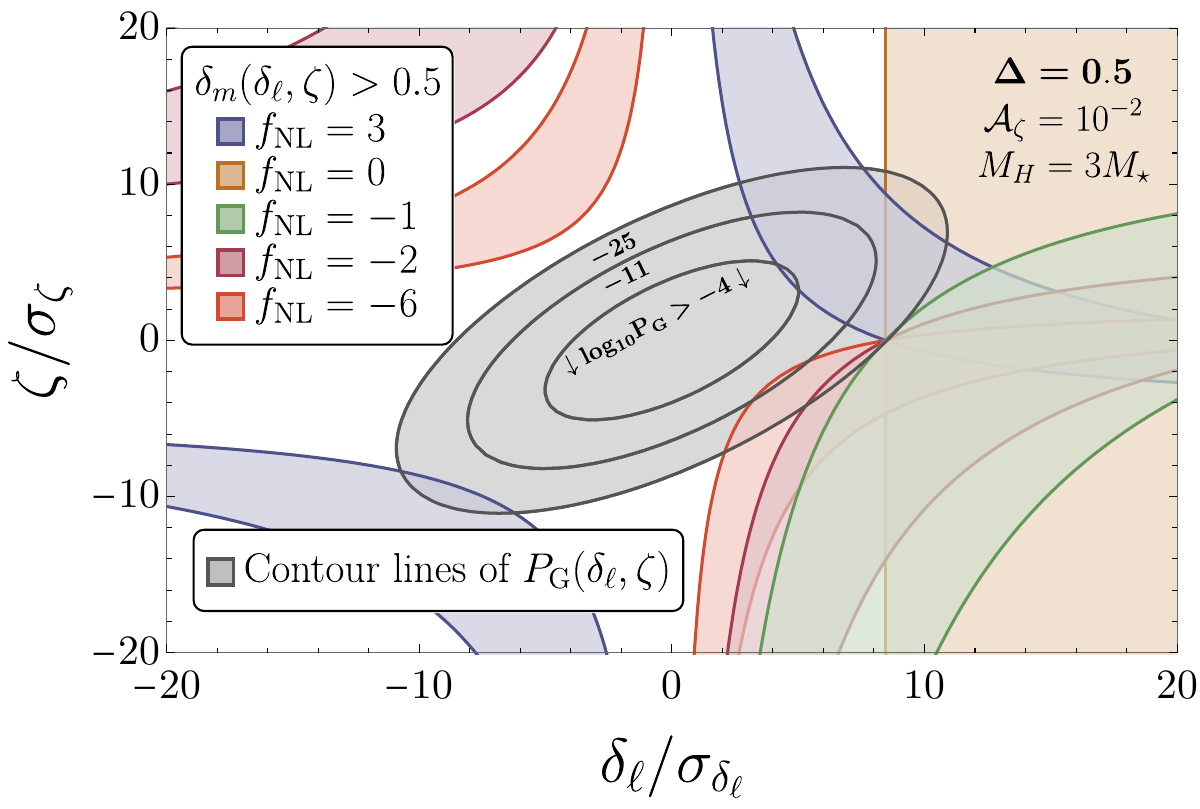} 
    \includegraphics[width=0.495\linewidth]{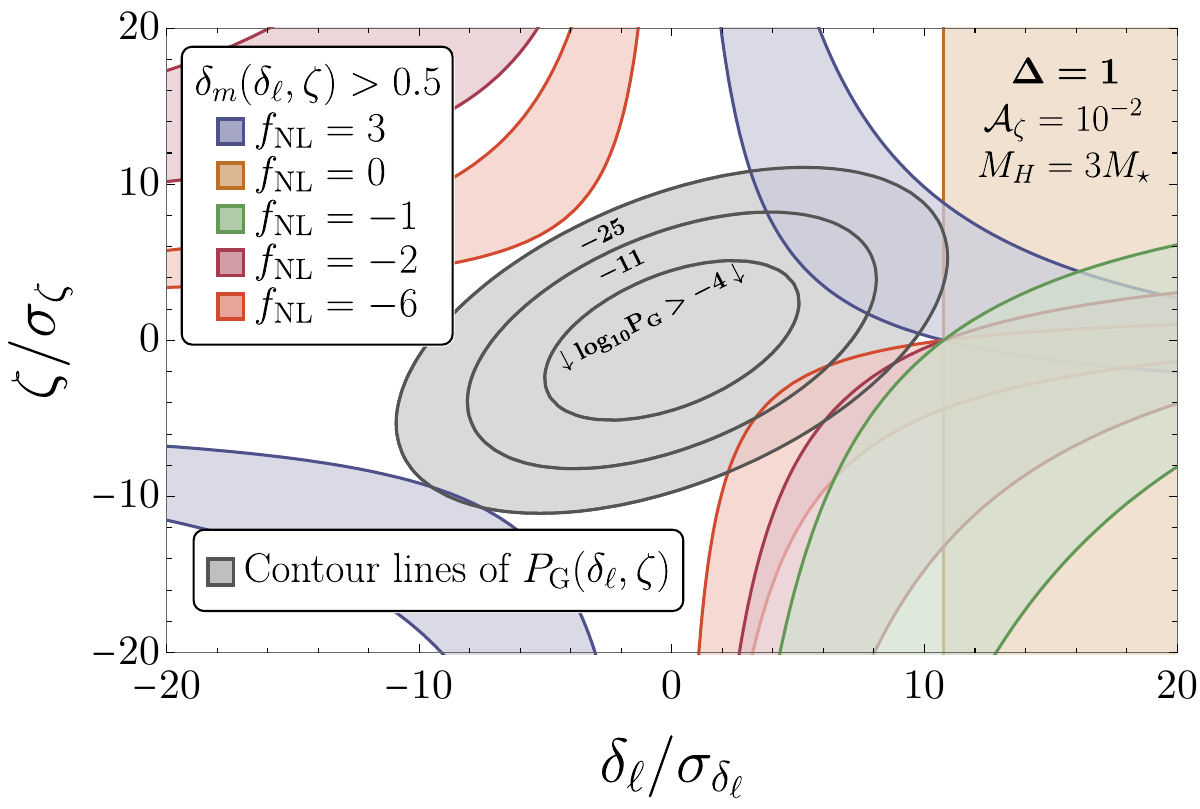} 
    \caption{Contours (black lines) of the joint probability distribution $P_G(\delta_{\ell},\zeta)$ defined in Eq.~\eqref{eq:joint_PDF}, shown for $\Delta = 0.5$ (\textbf{left}) and $\Delta = 1$ (\textbf{right}). The shaded regions (colored) correspond to the parameter space where the non-linear averaged density contrast $\delta_{m}(\delta_{\ell},\zeta)$, given in Eq.~\eqref{eq:delta_ell_NG_def}, exceeds the collapse threshold $\delta_c \simeq 0.5$. This defines the \emph{PBH region}. The PBH abundance at formation can be approximated as $f_{\rm PBH}\propto \max_{\delta_{\ell},\zeta}\!\left[P_G(\delta_{\ell},\zeta)\right]$ under the condition $\delta_{m} > \delta_c$. Increasing $|f_{\rm NL}|$ shifts the PBH region towards smaller values of $(\delta_\ell,\zeta)$. For $f_{\rm NL}>0$, the statistically favored values of $\delta_\ell$ and $\zeta$ are positively correlated, whereas for $f_{\rm NL}<0$ the threshold condition enforces an anti-correlation in the PBH region. As a result, there exists an intermediate range $f_{\rm NL}\in[-2,-1]$ in which the joint probability density at threshold is minimized, leading to a suppression of PBH formation. The width $\Delta$ controls the strength of the statistical correlation between $\delta_\ell$ and $\zeta$: smaller $\Delta$ produces stronger correlations, reflected in more elongated (oval) black contour lines. In evaluating Eq.~\eqref{eq:fPBH_NG}, we fix the horizon mass to $M_H\simeq M_{\rm PBH}$, with $M_{\rm PBH}$ chosen as the peak value $M_{\rm PBH}\sim 3M_{\star}$ of the mass distribution shown in Fig.~\ref{fig:fPBH_log_normal}.}
    \label{fig:why_fnl_m2}
\end{figure*}

In the threshold statistics approach, the PBH formation probability is defined as
\begin{equation}
\label{eq:Prob_PS_NG}
P_f(\delta_{\ell};M_H) =   \int_{-\infty}^{\infty} d\zeta~  \textrm{P}_{\rm G} (\delta_\ell,\zeta) \,F'(\zeta)^{-1}\,.
\end{equation}
where $\delta_\ell=2\mathcal{F} F'(\zeta)^{-1} \left(1-\sqrt{\Lambda}\right)$.
While, the impact of primordial NGs on the Press–Schechter prediction has been extensively investigated~\cite{Bugaev:2013vba, Nakama:2016gzw,Byrnes:2012yx,Young:2013oia,Kawasaki:2019mbl,DeLuca:2019qsy,Young:2019yug,Taoso:2021uvl,Meng:2022ixx,Escriva:2022pnz,Ferrante:2022mui}.  
By contrast, much less attention has been devoted to understanding how NGs modify the peak theory prediction~\cite{Yoo:2019pma,Riccardi:2021rlf,Young:2022phe,Ianniccari:2024bkh,Iovino:2024tyg} with the notable exception of Ref.~\cite{Kitajima:2021fpq}. That work extended peak theory to include local-type NGs, but only in the case of a monochromatic curvature power spectrum. Since in our analysis we adopt a log-normal spectrum, applying the relevant formalism is not entirely straightforward. While Ref.~\cite{Kitajima:2021fpq} notes that the monochromatic results can be extended to very narrow spectra, a consistent treatment for finite-width distributions would require adapting their derivation of the PBH mass function and re-implementing it within our Bayesian framework. We therefore restrict our use of the peak theory approach to the Gaussian limit, and leave a systematic incorporation of the results of Ref.~\cite{Kitajima:2021fpq} to future work.

Including NGs we may rewrite the PBH mass function in Eq. \eqref{eq:fPBH_Gauss} as
\begin{equation}
\label{eq:fPBH_NG}
\frac{df_{\rm PBH}}{d \ln  M_{\rm PBH} } = \frac{1}{\Omega_{\rm DM}} \int_{M_H^{\rm min}}^\infty d \ln M_H \left( \frac{M_{\rm eq}}{M_H}\right)^{1/2} \left(\frac{M_{\rm PBH}}{\mathcal{K} M_H}\right)^{\!\!\frac{1+\gamma_{\rm M}}{\gamma_{\rm M}}}  \frac{{\cal K}}{\gamma_{\rm M}\sqrt{\Lambda}} \int_{-\infty}^{\infty} d\zeta  \,\textrm{P}_{\rm G} (\delta_\ell,\zeta) F'(\zeta)^{-1}\,.
\end{equation}
In Fig. \ref{fig:fPBH_log_normal} we plot the mass function both for the Gaussian peak theory (blue) calculation as well as the one with threshold statistics both in the Gaussian limit (orange) as well as for $f_{\rm}=-2$ (red) and for different values of the width $\Delta$ (where the exact numerical values depends on the normalization of Eq. \eqref{eq:log_normal}). First, we observe that peak theory predicts generically larger abundances than threshold statistics when NGs are negligible. Even though we do not adopt the NG-extended peak theory results here, Ref.~\cite{Kitajima:2021fpq} finds that for narrowly-peaked power spectra the impact of NGs on PBH formation is in fact more pronounced in peak theory than in threshold statistics. As we will see in Sec.~\ref{sec:results}, negative NGs play a crucial role in alleviating phenomenological constraints, so a stronger suppression in peak theory would actually be beneficial. This highlights that our comparison is conservative, and that incorporating NGs into peak theory could potentially provide even more favorable predictions.

Threshold statistics also predict even smaller abundances once negative NGs are included. This can be understood from the fact that negative NGs skews the primordial curvature perturbation distribution toward smaller values, thereby reducing the probability of large fluctuations in the tail. Since PBH formation is exponentially sensitive to the abundance of rare, large-amplitude perturbations (see Eq. \ref{eq:Prob_PS_Gauss}), this suppression in the high-$\zeta$ tail directly translates into a reduced PBH abundance. Moreover, in the Gaussian case, narrower power spectra (smaller $\Delta$) yield larger PBH abundances. This arises because concentrating the power around $k_\star$ maximizes the variance at the corresponding horizon scale, thereby enhancing the probability of crossing the collapse threshold $\delta_c$ in Eq. \eqref{eq:choptuik_law}. Conversely, broader spectra (larger $\Delta$) distribute the power across many scales, reducing the variance at any given mass and thus suppressing the abundance. This tendency is inverted in the NG case for larger widths, further indicating the limitations of the perturbative approximation for this choice of parameters (see discussion below Eq.~\eqref{eq:window_spherical}).

Furthermore, we will simplify discussion of the PBH phenomenology by replacing the extended PBH distributions with monochromatic populations characterized by two quantities: the integrated PBH mass function and the mean PBH mass,
\begin{align}
\label{eq:f_PBH_final}
  \langle f_{\rm PBH} \rangle
   = \int \frac{d M_{\rm PBH}}{M_{\rm PBH}}\frac{df_{\rm PBH}}{d \ln M_{\rm PBH} }\, ,\quad 
   \langle M_{\rm PBH} \rangle 
   =  \langle f_{\rm PBH} \rangle \left(\int \frac{d M_{\rm PBH}}{M_{\rm PBH}^2} \frac{df_{\rm PBH}}{d \ln M_{\rm PBH} }\right)^{-1} \,.
\end{align}
In this context and for the remainder of the article, we will also reserve the notation $f_{\rm PBH}$ and $M_{\rm PBH}$ to denote $\langle f_{\rm PBH} \rangle$ and $\langle M_{\rm PBH} \rangle$, respectively. 

In Fig.~\ref{fig:fPBH_log_normal_DM}, we present the dark matter overclosure bound ($f_{\rm PBH} > 1$) on $\mathcal{A}_\zeta$ using threshold statistics (blue) and peak theory (orange). As noted previously, peak theory consistently predicts larger abundances than threshold statistics, leading to a more restrictive bound on $\mathcal{A}_\zeta$. Then we observe that narrower spectra (smaller $\Delta$) require smaller amplitudes to saturate the bound, reflecting the enhanced variance at the peak scale. Additionally, smaller $k_\star$ values correspond to fluctuations that enter later the horizon and thus produce heavier PBHs that contribute more to the dark matter abundance. Thus the bound forces the amplitude to be smaller, although we see that the dependence is mild as the variation of $\mathcal A_{\zeta}$ is at the percent level, while $k_\star$ values span several orders of magnitude. Finally, for sufficiently narrow spectra we find that negative NGs suppresses the PBH abundance, thereby relaxing the overclosure bound. This suppression is most pronounced near $f_{\rm NL}\sim -1$. When combined with the impact of NGs on the induced GW amplitude (discussed in Sec.~\ref{sec:SIGW}), this leads to $f_{\rm NL}\sim -2$ emerging as the preferred value in our analysis, see Table~\ref{tab:BF_NG}. One may wonder why the PBH abundance reaches a minimum around $f_{\rm NL}\sim -1$ instead of continuing to decrease for $f_{\rm NL}\ll -1$. The physical origin of this behavior can be traced to the value of the joint PDF $P_{G}(\delta_\ell,\zeta)$ within the PBH-forming region $\delta_m\gtrsim 0.5$, as illustrated in Fig.~\ref{fig:why_fnl_m2}; see also discussions in Refs.~\cite{Young:2022phe,Ferrante:2022mui,Franciolini:2023pbf,Iovino:2024tyg}.

\subsection{Scalar-induced gravitational waves}
\label{sec:SIGW}

At second order in perturbation theory interactions between scalar and tensor perturbations are present. In particular, this non-linear coupling leads to the generation of tensor modes from the scalar modes, which are the SIGW~\cite{Ananda:2006af,Baumann:2007zm,Espinosa:2018eve,Kohri:2018awv,Domenech:2021and, Dandoy:2023jot, Franciolini:2023pbf,Vagnozzi:2023lwo,Franciolini:2023wjm,Wang:2023ost,Liu:2023ymk,Firouzjahi:2023lzg,Balaji:2023ehk,Harigaya:2023pmw,Tagliazucchi:2023dai,Chen:2023uiz,Chen:2024fir,Chen:2024twp,Domenech:2024rks,Iovino:2024tyg,Ferrante:2023bgz,Frosina:2023nxu,Clesse:2024epo}.  As we have already commented on, at CMB and galaxy formation scales, the power spectrum of the scalar fluctuations is observed to be of order $\mathcal{P}_{\zeta} \sim 10^{-9}$, which implies a negligible scalar induced tensor power spectrum $\mathcal{P}_{h}\lesssim \mathcal{P}^2_{\zeta} \sim 10^{-18}$.  In contrast, enhanced curvature peaks similar to the ones producing large population of PBHs, as discussed in the previous section, can lead to sizable and potentially detectable SIGW background signal. 

To obtain an expression for the SIGW energy density, we start by writing the perturbed metric in the Newtonian gauge (neglecting vector perturbations) as
\bea 
ds^2 = a^2(\tau) \bigg[-(1+ 2\Phi)d\tau^2 + \bigg((1-2 \Psi)\delta_{ij}+\frac{1}{2}h_{ij}\bigg)dx^i dx^j\bigg] \, ,
\eea
where $\Phi$, $\Psi$ are the Newtonian potentials at first order and $h_{ij}$ are the tensor perturbations. 
For super-horizon modes $k \ll H$, the Fourier transform of the Newtonian potentials can be related to the primordial curvature perturbation $\zeta$ introduced in the previous section via
\bea \label{eq:Phi_potential}
 \Phi(\mathbf{k}, \tau) = \Psi(\mathbf{k}, \tau) = \mathcal F(w) \phi(k\tau) \zeta(\mathbf{k}) \, ,
\eea 
where we introduced the new transfer function $\phi(k\tau)$. 
 The Fourier transform of the tensor perturbation $h_{ij}$ is given by 
\bea 
h_{ij}(\tau, \mathbf{x}) =  \sum_{\lambda=+,\times}\int \frac{d^3k}{(2\pi)^{3/2}} e^{i \mathbf{k} \cdot \mathbf{x}} \epsilon_{ij}^\lambda (\mathbf{k}) h_{ij} (\tau, \mathbf{k})  \, ,
\eea 
where 
\bea 
\epsilon_{ij}^+ = \frac{1}{\sqrt{2}}(\epsilon_i (\mathbf{k})\epsilon_j(\mathbf{k}) - \bar\epsilon_i (\mathbf{k})\bar\epsilon_j(\mathbf{k}) )\,  , \qquad \epsilon_{ij}^\times = \frac{1}{\sqrt{2}}(\epsilon_i (\mathbf{k})\bar\epsilon_j(\mathbf{k}) + \bar\epsilon_i (\mathbf{k})\epsilon_j(\mathbf{k}) ) \, ,
\eea 
and $\epsilon_j$ and $\bar \epsilon_j$ form a basis constructed in such a way that  $\epsilon_j$ and $\bar \epsilon_j$ are orthonormal, traceless, and transverse to $\mathbf{k}$. We define the tensor power spectrum and $\mathcal{P}_{h,\lambda_1} (\tau, \mathbf{k}_1 )$ the \emph{dimensionless} tensor power spectrum\footnote{Notice that what we define here as the dimensionless power spectrum $\mathcal{P}_{h,\lambda_1} (\tau, \mathbf{k}_1 )$ is often denoted $\Delta_h^2(\tau, k)$ in the GW literature.}
\bea 
\label{eq:power_spectrum_GW}
\langle h_{ \lambda_1} (\tau, \mathbf{k}_1) h_{ \lambda_2} (\tau, \mathbf{k}_2)\rangle \equiv \delta^{(3)}(\mathbf{k}_1 + \mathbf{k}_2 )\delta^{\lambda_1 \lambda_2} \frac{2\pi^2}{k_1^3}  \mathcal{P}_{h,\lambda_1} (\tau, \mathbf{k}_1 )\, , 
\eea 

and the fractional energy density in GWs per logarithmic wavenumber as~\cite{Baumann:2007zm,Espinosa:2018eve}
\bea \label{eq:Omega_GW_def1}
\Omega_{\rm GW}(\tau, k) \equiv \frac{\rho_{\rm GW}(\tau, k) }{\rho_{\rm tot}(\tau)} =  \frac{\overline{\dot h_{ij}\dot h^{ij}}}{32\pi G a^2(\tau)\rho_{\rm tot}(\tau)} \, , 
\eea 
where the overbar means an average over time. We obtain at the end 
\bea 
\Omega_{\rm GW}(\tau, k) =  \frac{1}{48}\bigg(\frac{k}{a(\tau)H(\tau)}\bigg)^2 \sum_{\lambda=+, \times}  \overline{\mathcal{P}_{h,\lambda} (\tau, \mathbf{k} ) } \, ,
\eea 
The stochastic GW power spectrum at present time is redshifted as follows~\cite{Kohri:2018awv}
\begin{equation}
\label{eq:Omega_SIGW_master}
\Omega_{\rm GW}(\tau_0 ,k)h^2 = \Omega_{r,0}h^2\Delta_{\rm R}(T_c)\Omega_{\rm GW}(\tau_c ,k)\,, \qquad \Delta_{\rm R}(T_c) \equiv \bigg(\frac{g_{\star, r}(T)}{g_{\star, r}(T_{\rm eq})}\bigg)\bigg(\frac{g_{\star, s}(T_{\rm eq})}{g_{\star, s}(T_{})}\bigg)^{4/3}\,,
\end{equation}
where $g_{\star, s(r)}$ is the number of entropy (radiation) degrees of freedom, $T_{\rm eq}$ is the temperature at matter-radiation equality, $\tau_c$ represents the conformal time after the generation of the GW signal, and $\Omega_{r,0}h^2 \simeq 4.2 \times 10^{-5}$ is the fractional radiation density today~\cite{Planck:2018vyg}.  The frequency of the GW today (see Eq. \eqref{eq:f0_Hstar}) is given in terms of $k$ as  
\begin{equation}
    f_0 = \frac{k}{2\pi} \simeq 1.6\, {\rm nHz} \left( \frac{k}{10^6\, {\rm Mpc}^{-1}} \right)\,.
\end{equation}
Let us now sketch the calculation of $\Omega_{\rm GW}(\tau_c ,k)$. The first step is to apply the projector on the tensor modes $\mathcal{T}^{lm}_{ij}$ on both sides of the Einstein field equations 
\bea 
\mathcal{T}^{lm}_{ij} G_{lm}^{(2)} = 8\pi G  \mathcal{T}^{lm}_{ij} T^{(2)}_{lm} \, . 
\eea 
where the subscript $(2)$ indicates that the respective quantity is computed up to second order in the scalar $\Phi$ and tensor $h_{ij}$ perturbations.  
In the absence of anisotropic stress from the stress-energy tensor, expanding the above equation yields
\bea 
\label{eq:eq_of_motion}
h''_{ij} + 2 \mathcal{H} h'_{ij} - \nabla^2 h_{ij} = - 4   \mathcal{T}^{lm}_{ij}  \mathcal{S}^{ij}\,, \qquad \mathcal{H} \equiv \frac{a'(\tau)}{a(\tau)} \, .
\eea 
where $'$ denotes a derivative with respect to the conformal time $\tau$  and the source term $\mathcal{S}^{ij}$ contains the square of the scalar perturbations~\cite{Adshead:2021hnm}
\bea 
\label{eq:source}
\mathcal{S}_{ij} = 4 \Phi \partial_i \partial_j \Phi + \frac{2(1+3 w)}{3(1+w)}\partial_i \Phi \partial_j \Phi  - \frac{4}{3(1+w)\mathcal{H}^2} \bigg[\partial_i \Phi' \partial_j \Phi' + \mathcal{H}\partial_i \Phi \partial_j \Phi+ \mathcal{H}\partial_i \Phi \partial_j \Phi' \bigg] \, . 
\eea 
One can solve Eq. \eqref{eq:eq_of_motion} for the tensor modes in the presence of the source term using Green’s functions and compute the power spectrum according to Eq. \eqref{eq:power_spectrum_GW}. Below we discuss the various contributions to the $\Omega_{\rm GW}(\tau_c ,k)$ that arise both in the case where the curvature perturbation $\zeta$ in Eq.~\eqref{eq:Phi_potential} is a purely Gaussian field, and also in the case in which $\zeta_{\rm NG}$ contains local NGs as introduced in Eq.~\eqref{eq:local_NG}.

\paragraph{Gravitational wave spectrum (Gaussian component).}

\begin{figure}[h!]
\centering
  \begin{tikzpicture}
    \begin{feynman}
      \vertex (i1) at (-3,0);
      \vertex (o1) at (3,0);
      \vertex (v1) at (-2,0);
      \vertex (v2) at (2,0);
      
      \vertex (a) at (-1,1.2);
      \vertex (b) at (1,1.2);
      \vertex (c) at (-1,-1.2);
      \vertex (d) at (1,-1.2);

      \diagram*{
        (i1) -- [graviton, edge label=$\vec{k}$] (v1),
        (v2) -- [graviton, edge label=$\vec{k}'$] (o1),

        (v1) -- [charged scalar, edge label=$\vec{q}_1 - \vec{k}$] (a),
        (b) -- [charged scalar, edge label=$\vec{q}_1 - \vec{k}'$] (v2),

        (a) -- [fermion, edge label=$\vec{q}_1 - \vec{k}$] (b)
            -- [fermion, style={opacity=0}] (d)
            -- [fermion, edge label=$\vec{q}_1$] (c)
            -- [fermion, style={opacity=0}] (a),

        (c) -- [charged scalar, edge label=$\vec{q}_1$] (v1),
        (v2) -- [charged scalar, edge label=$\vec{q}_1$] (d),
      };
    \end{feynman}
  \end{tikzpicture}
  \caption{Diagram of the \textbf{Gaussian} contribution to the SIGW. }
  \label{fig:Gaussian_contribution}
\end{figure}
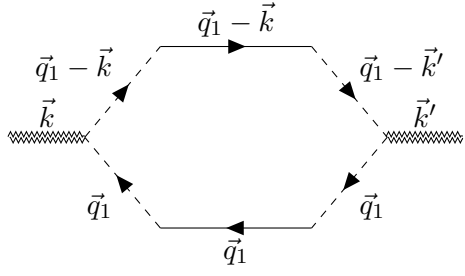

First, applying the results above, the tensor correlation reads
\bea 
\langle h_{\lambda_1} (\bm{k}_1) h_{\lambda_2} (\bm{k}_2) \rangle 
= 16 \int \frac{\mathrm{d}^3 \bm{q}_1}{(2\pi)^{3/2}} \frac{\mathrm{d}^3 \bm{q}_2}{(2\pi)^{3/2}} 
\langle \zeta_{\bm{q}_1} \zeta_{\bm{k}_1 - \bm{q}_1} \zeta_{\bm{q}_2} \zeta_{\bm{k}_2 - \bm{q}_2} \rangle 
Q_{\lambda_1}(\bm{k}_1, \bm{q}_1) Q_{\lambda_2}(\bm{k}_2, \bm{q}_2)
\nonumber
\\
\times I(|\bm{k}_1 - \bm{q}_1|, q_1, \tau_1) 
I(|\bm{k}_2 - \bm{q}_2|, q_2, \tau_2),
\eea

where we have defined the projector
\bea 
Q_\lambda(\mathbf{k}, \mathbf{q}) \equiv \epsilon_{ij}^\lambda(\mathbf{k}) q_i q_j \, ,
\eea 

and the transfer function
\bea
I(p, q, \tau) = \int_{\tau_0}^{\tau} \mathrm{d}\tau' \, G_k(\tau, \tau') 
\frac{a(\tau')}{a(\tau)} f(p, q, \tau') \, , 
\eea 
where the $f$ function is given by  

\bea
f(p, q, \tau) = 2\mathcal F (w)\bigg[
 \, \phi(p\tau)\phi(q\tau) 
+ \tau^2\frac{(1 + 3w)^2}{2(5+3w)} \phi'(p\tau)\phi'(q\tau)
+ \tau \frac{(1 + 3w)}{(5+3w)}\left( \phi(p\tau)\phi'(q\tau) + \phi'(p\tau)\phi(q\tau) \right)
\bigg].
\eea

To systematize the computation of the SIGW spectrum, one can represent the 
different contributions using diagrammatic rules, derived in 
Refs.~\cite{Adshead:2021hnm,Perna:2024ehx} and summarized in 
Table~\ref{tab:Feynman_rules}. The procedure is then straightforward: one draws 
the allowed diagrams, assigns the corresponding kernels, and integrates 
over the internal loop momenta. The external legs carry the momentum of the GW, while internal solid lines represent the curvature power spectrum and the dashed lines the transfer function.  Momentum is conserved at each vertex. The 
determination of the correct multiplicity factors can, however, be more 
involved.

The \emph{Feynman-like} diagram of the Gaussian contribution is shown on Fig.~\ref{fig:Gaussian_contribution}. In addition, each momentum unconstrained by conservation of momenta should be integrated over.

\begin{table}[ht!]
\renewcommand{\arraystretch}{1.5}
  \begin{center}
    \begin{tblr}{|Q[c,m,4cm]|Q[c,m,6cm]|Q[c,m,6cm]|} 
      \hline
      \SetCell[r=2]{c}{\textbf{Description}} 
      & \SetCell[r=2]{c}{\textbf{Diagram}} 
      & \SetCell[r=2]{c}{\textbf{Rule}} \\
      \\ 
      \hline \hline 
      \SetCell[r=2]{c}{Dashed line connected to a solid line}
      & \SetCell[r=2]{c}{%
        \begin{tikzpicture}
          \begin{feynman}
            \vertex (a) at (0,0);
            \vertex (b) at (2,0);
            \vertex (c) at (4,0);
            \diagram*{
              (a) -- [charged scalar, , edge label=$\vec{q}$] (b),
              (b) -- [fermion, , edge label=$\vec{q}$] (c),
            };
          \end{feynman}
        \end{tikzpicture}
      }
      & \SetCell[r=2]{c}{1} \\
      \\
      \hline
    \SetCell[r=2]{c}{Oriented solid line}
      & \SetCell[r=2]{c}{%
        \begin{tikzpicture}
          \begin{feynman}
            \vertex (a) at (0,0);
            \vertex (b) at (4,0);
            \diagram*{
              (a) -- [fermion, edge label=$\vec{q}$] (b),
            };
          \end{feynman}
        \end{tikzpicture}
      }
      & \SetCell[r=2]{c}{$\mathcal{P}_{\zeta}(q) \frac{q^3}{2\pi^2}$} 
      \\
      \\
      \hline
    \SetCell[r=2]{c}{Oriented dashed line}
      & \SetCell[r=2]{c}{%
        \begin{tikzpicture}
          \begin{feynman}
            \vertex (a) at (0,0);
            \vertex (b) at (4,0);
            \diagram*{
              (a) -- [charged scalar, edge label=$\vec{q}$] (b),
            };
          \end{feynman}
        \end{tikzpicture}
      }
      & \SetCell[r=2]{c}{$1$} \\
      \\
      \hline 
      \SetCell[r=2]{c}{ Vertex between 2 solid lines and 1 dashed line}
      & \SetCell[r=2]{c}{%
        \begin{tikzpicture}
          \begin{feynman}
            \vertex (a) at (0,0);
            \vertex (b) at (2,0);
            \vertex (c) at (4,1);
            \vertex (d) at (4,-1);
            \diagram*{
              (a) -- [charged scalar, edge label=$\vec{q}_1$] (b),
              (b) -- [fermion, edge label=$\vec{q}_1+ \vec{q}_2$] (c),
              (d) -- [fermion, edge label=$\vec{q}_2$] (b),
            };
          \end{feynman}
        \end{tikzpicture}
      }
      & \SetCell[r=2]{c}{$F_{\rm NL}$} \\
      \\
      \hline
      \SetCell[r=2]{c}{ Vertex between 1 graviton and 2 dashed lines}
      & \SetCell[r=2]{c}{%
        \begin{tikzpicture}
          \begin{feynman}
            \vertex (a) at (0,0);
            \vertex (b) at (2,0);
            \vertex (c) at (4,1);
            \vertex (d) at (4,-1);
            \diagram*{
              (b) -- [graviton, edge label=$\vec{k}$] (a),
              (b) -- [charged scalar, edge label=$\vec{q}$] (c),
              (d) -- [charged scalar, edge label=$\vec{q}-\vec{k}$] (b),
            };
          \end{feynman}
        \end{tikzpicture}
      }
      & \SetCell[r=2]{c}{$$4 \int_{\tau_0}^{\tau} \mathrm{d}\bar{\tau} \, \frac{a(\bar{\tau})}{a(\tau)} 
      G_{\bm{k}}(\tau, \bar{\tau}) Q_{\lambda}(\bm{k}, \bm{q}) 
      $$
      $$
      \times f(|\bm{k} - \bm{q}|, q, \bar{\tau})$$
      $$
      $$
      } \\
      \\
      \hline
    \end{tblr}

    \caption{\label{tab:Feynman_rules}
    Diagrammatic rules for the SIGW calculation~\cite{Adshead:2021hnm}. 
    }
  \end{center}
\end{table}

It is conventional to introduce two new variables $ u = \frac{|k - q|}{k} $ and $ v = \frac{q}{k} $ to simplify the above equations. Then, combining the results above and considering the limit $ x \to \infty $ as we are interested in the GW spectrum observed today, we finally obtain \footnote{We emphasize that the curvature power spectrum used in this formula is the \emph{dimensionlesss} power spectrum defined in Eq.~\eqref{eq:zeta_PS_def}, which is also called $\Delta_\zeta^2$ in the literature.}
\bea 
\Omega_{\mathrm{GW}}^{\mathrm{G}}(k) = \frac{2}{3} \int_0^\infty dv \int_{|1-v|}^{1+v} du \, \mathcal{I}(u,v,u,v,x \to \infty) \frac{\mathcal{P}_\zeta(vk)}{v^2} \frac{\mathcal{P}_\zeta(uk)}{u^2} \, ,
\eea
where we introduced the functions~\cite{Atal:2021jyo, Liu:2023ymk}
\begin{align}
\,\mathcal{I}(u_1,v_1, u_2, v_2, x \to \infty ) &= \frac{([v_1+u_1]^2-1)([v_2+u_2]^2-1)(1-[v_1-u_1]^2)(1-[v_2-u_2]^2) }{128}  J(u_1,v_1, u_2, v_2)\, ,  \notag
\\
J(u_1,v_1, u_2, v_2) &= I_A(u_1, v_1)I_A(u_2, v_2)\bigg(I_B(u_1, v_1)I_B(u_2, v_2)+\pi^2 I_C(u_1, v_1)I_C(u_2, v_2)\bigg)\, , 
\end{align}
and 
\begin{align}
&I_A(u, v) = \frac{3(u^2+v^2 -3)}{4 u^3v^3}, \qquad I_B(u, v) = -4uv + (u^2+v^2 -3)\log \bigg|\frac{3-(u+v)^2}{3-(u-v)^2}\bigg|\,, \notag
\\
&I_C(u,v) = (u^2+v^2 -3)\Theta(u+v - \sqrt{3}) \, . 
\end{align}

It is numerically convenient to change the integration variables from $ (u_i, v_i) $ to $ (t_i, s_i) $, defined by 
\bea 
s_i = u_i - v_i\,, \qquad t_i = u_i+v_i -1\,, \qquad u_i = (t_i+s_i +1)/2\, , \qquad  v_i = (t_i-s_i +1)/2 \, .
\eea 
with the relation 
\begin{equation}
\int_0^\infty \mathrm{d}v_i \int_{|1 - v_i|}^{1 + v_i} \mathrm{d}u_i\, (\ldots)
= \frac{1}{2} \int_0^\infty \mathrm{d}t_i \int_{-1}^{1} \mathrm{d}s_i\, (\ldots) \, .
\end{equation}

The GW energy fraction at time $\tau_c$ can be expressed as a function of the curvature power spectrum~\cite{Kohri:2018awv} 
\begin{align}
\label{eq:GW_spectrum}
\Omega^{\rm G}_{\rm GW} (x) &= \frac{1}{3} \int_{0}^{\infty}  dt \int_{1}^{-1} ds~ \mathcal{I}(u, v, u, v, x \to \infty) \frac{\mathcal{P}_\zeta(xu)\mathcal{P}_\zeta(xv)}{u^2 v^2} \, ,
\end{align}
where we defined $x\equiv k \tau = 2k/(1+3w)\mathcal{H}$ with $\mathcal{H}(\tau_c)=1/\tau_c$. For a log-normal power spectrum and in the limit $\tau_c \to \infty$ the integral in Eq. \eqref{eq:GW_spectrum} can be approximated analytically~\cite{Pi:2020otn}. In particular, for $\Delta \gtrsim 0.1$  we may write
{\small
\begin{align}
\frac{\Omega_\text{GW}^\text{G}}{\mathcal{A}_\zeta^2}
&\simeq\frac{4\kappa ^3}{5\sqrt\pi}\frac{e^{\tfrac{9 \Delta ^2}{4}}}{\Delta}
\left[
\left(\ln^2K+\frac{\Delta^2}{2}\right)
\text{erfc}\left(\frac{\ln \left( \sqrt{\frac{3}{2}}K\right)}{\Delta}\right)
-\frac{\Delta}{\sqrt\pi}\exp\left(-\frac{\ln^2 \left( \sqrt{\frac{3}{2}}K\right)}{\Delta^2}\right)
\ln \left( \sqrt{\frac{2}{3}}K\right)\right]\\
+&\frac{0.0659}{\Delta^2}\kappa^2 e^{\Delta^2}\exp\left(-\frac{\left(\Delta^2+\ln \kappa\sqrt{\frac{3}{4}}\right)^2}{\Delta^2}\right)
+\frac1{3}\sqrt{\frac2\pi}\kappa^{-4}\frac{e^{8\Delta^2}}{\Delta}
\exp\left(-\frac{\ln^2\kappa}{2\Delta^2}\right)
\text{erfc}\left(\frac{4\Delta^2-\ln(\kappa/4)}{\sqrt2\Delta}\right),\notag
\end{align}
}
where $\kappa\equiv k/k_\star$ and $K\equiv \kappa e^{3\Delta^2/2}$. Conversely, for very narrowly-peaked spectra with $\Delta \lesssim 0.1$, the GW power spectrum is given by
{\small
\begin{align}
\frac{\Omega_\text{GW}^\text{G}}{\mathcal{A}_\zeta^2}&\simeq3\kappa^2e^{\Delta^2}
\left[\text{erf}\left(\frac{1}{\Delta}\text{arcsh}\frac{\kappa e^{\Delta^2}}2\right)-
\text{erf}\left(\frac{1}{\Delta}\mathrm{Re}\left(\text{arcch}\frac{\kappa e^{\Delta^2}}2\right)\right)\right]\left(1-\frac14\kappa^2e^{2\Delta^2}\right)^2
\left(1-\frac32\kappa^2e^{2\Delta^2}\right)^2\\\label{Omega4}
&\quad\cdot\left[\left[\frac12\left(1-\frac32\kappa^2e^{2\Delta^2}\right)\ln\left|1-\frac4{3\kappa^2e^{2\Delta^2}}\right|-1\right]^2
+\frac{\pi^2}4\left(1-\frac32\kappa^2e^{2\Delta^2}\right)^2\Theta\left(2-\sqrt3\kappa e^{\Delta^2}\right)
\right]\,.
\end{align}
}
In the monochromatic limit $\Delta \to 0$ we recover the form~\cite{Kohri:2018awv,Domenech:2019quo}
\begin{equation}
\frac{\Omega_{\rm GW}(x_c)}{\mathcal{A}_\zeta^2} = 3\frac{y_\star^2}{v_\star^2}\left(\frac{4v_\star^2-1}{4v_\star^2}\right)^2\left[ \frac{\pi^2 y_\star^2}{4} \Theta(1+y_\star)  + \left(1-\frac{y_\star}{2}\ln{\left|1-\frac{4}{3}v_\star^2  \right|} \right)^2\right]\Theta(2v_\star-1)\,,
\end{equation}
where $y_\star \equiv 1-3/2v_\star^2$ and $v_\star\equiv k_\star/k$.

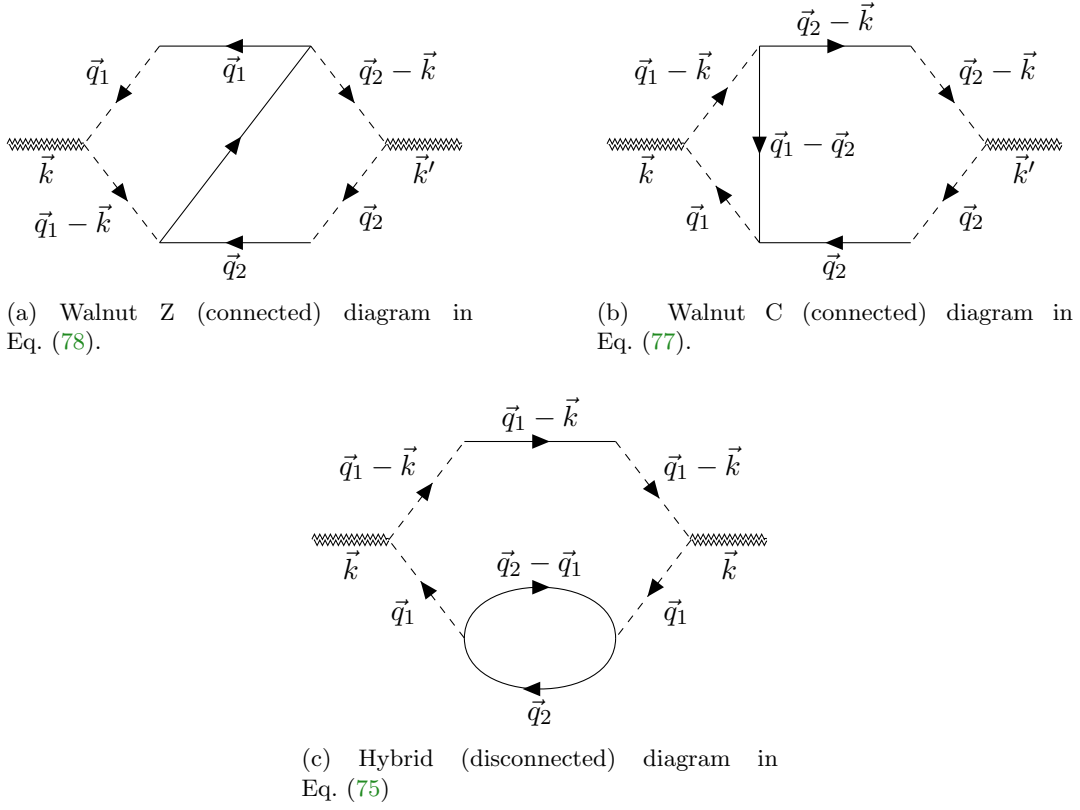
\begin{figure}
\centering
    \subfloat[Walnut Z (connected) diagram in Eq.~\eqref{eq:fnl22}.]
{\begin{tikzpicture}
  \begin{feynman}
    \vertex (i1) at (-3,0);
    \vertex (o1) at ( 3,0);
    \vertex (v1) at (-2,0);
    \vertex (v2) at ( 2,0);
    \vertex (tL) at (-1, 1.3);
    \vertex (tR) at ( 1, 1.3);
    \vertex (bL) at (-1,-1.3);
    \vertex (bR) at ( 1,-1.3);
    \vertex (mC) at ( 0, 0);  

    \diagram*{
      (i1) -- [graviton, edge label'=$\vec{k}$] (v1),
      (v2) -- [graviton, edge label'=$\vec{k}'$] (o1),

      (tL) -- [charged scalar, edge label'=$\vec{q}_1$] (v1),
      (v1) -- [charged scalar, edge label'=$\vec{q}_1-\vec{k}$] (bL),

      (tR) -- [fermion,  edge label=$\vec{q}_1$] (tL),
      (bL) -- [fermion] (tR),
      (bR) -- [fermion, edge label=$\vec{q}_2$] (bL),

      (tR) -- [charged scalar, edge label=$\vec{q}_2-\vec{k}$] (v2),
      (v2) -- [charged scalar, edge label=$\vec{q}_2$] (bR),
    };
  \end{feynman}
\end{tikzpicture}
}
\hspace{1.5cm}
    \subfloat[ Walnut C (connected) diagram in Eq.~\eqref{eq:fnl2}.]
    {
 \begin{tikzpicture}
    \begin{feynman}
      \vertex (i1) at (-3,0);
      \vertex (o1) at ( 3,0);
      \vertex (v1) at (-2,0);
      \vertex (v2) at ( 2,0);
      \vertex (a)  at (-1, 1.3);
      \vertex (b)  at ( 1, 1.3);
      \vertex (c)  at (-1,-1.3);
      \vertex (d)  at ( 1,-1.3);

      \diagram*{
        (i1) -- [graviton, edge label'=$\vec{k}$] (v1),
        (v2) -- [graviton, edge label'=$\vec{k}'$] (o1),

        (v1) -- [charged scalar, edge label=$\vec{q}_1-\vec{k}$] (a),
        (c) -- [charged scalar, edge label=$\vec{q}_1$] (v1),

        (a) -- [fermion, edge label=$\vec{q}_2-\vec{k}$] (b),
        (d) -- [fermion, edge label=$\vec{q}_2$] (c),

        (b) -- [charged scalar, edge label=$\vec{q}_2-\vec{k}$] (v2),
        (v2) -- [charged scalar, edge label=$\vec{q}_2$] (d),

        (a) -- [fermion, edge label=$\vec{q}_1-\vec{q}_2$] (c),
      };
    \end{feynman}
  \end{tikzpicture}
}

    \subfloat[Hybrid (disconnected) diagram in Eq.~\eqref{eq:fnl2hybrid}]
    {
\begin{tikzpicture}
  \begin{feynman}
    \vertex (i1) at (-3,1);
    \vertex (o1) at ( 3,1);
    \vertex (v1) at (-2,1);
    \vertex (v2) at ( 2,1);
    \vertex (a)  at (-1, 2.3);   
    \vertex (b)  at ( 1, 2.3);   
    \vertex (c)  at (-1,-0.3);   
    \vertex (d)  at ( 1,-0.3);   

    \diagram*{
      (i1) -- [graviton, edge label'=$\vec{k}$] (v1),
      (v2) -- [graviton, edge label'=$\vec{k}$] (o1),

      (v1) -- [charged scalar, edge label=$\vec{q}_1-\vec{k}$] (a),
      (c) -- [charged scalar, edge label=$\vec{q}_1$] (v1),
      (b)  -- [charged scalar, edge label=$\vec{q}_1-\vec{k}$] (v2),
      (v2)  -- [charged scalar, edge label=$\vec{q}_1$] (d),

      (a) -- [fermion, edge label=$\vec{q}_1-\vec{k}$] (b),

      (c) -- [fermion, half left,  looseness=1.15, edge label=$\vec{q}_2-\vec{q}_1$] (d), 
      (c) -- [anti fermion, half right, looseness=1.15, edge label'=$\vec{q}_2$] (d),     
    };
  \end{feynman}
\end{tikzpicture}
}






    \caption{Diagram contributing at $\mathcal{O}(F_{\rm NL}^2)$ to the SIGW.}
  \label{fig:FNL2_contribution}
\end{figure}

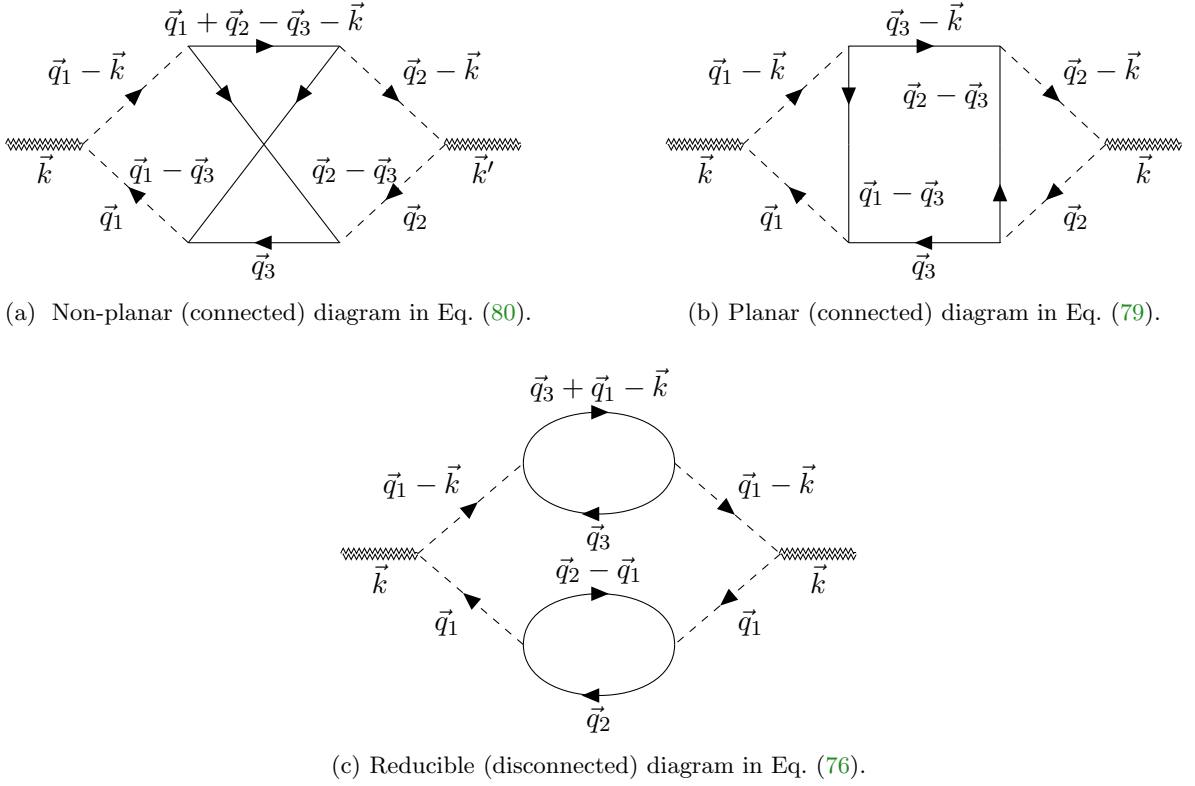
\begin{figure}
\centering
    \subfloat[ Non-planar (connected) diagram in Eq.~\eqref{eq:fnl24}.]
{\begin{tikzpicture}
  \begin{feynman}
    \vertex (i1) at (-3.4,0);
    \vertex (o1) at ( 3.4,0);
    \vertex (v1) at (-2.4,0);
    \vertex (v2) at ( 2.4,0);
    \vertex (tL) at (-1, 1.3);
    \vertex (tR) at ( 1, 1.3);
    \vertex (bL) at (-1,-1.3);
    \vertex (bR) at ( 1,-1.3);
    \vertex (mC) at ( 0, 0);  

    \diagram*{
      (i1) -- [graviton, edge label'=$\vec{k}$] (v1),
      (v2) -- [graviton, edge label'=$\vec{k}'$] (o1),

      (v1) -- [charged scalar, edge label=$\vec{q}_1-\vec{k}$] (tL),
      (bL) -- [charged scalar, edge label=$\vec{q}_1$] (v1),

      (tL) -- [fermion, edge label=$\vec{q}_1+\vec{q}_2- \vec{q}_3-\vec{k}$] (tR),
      (tR) -- [fermion] (mC) -- [plain, edge label'=$\vec{q}_1-\vec{q}_3$] (bL),
      (tL) -- [fermion] (mC) -- [plain, edge label=$\vec{q}_2-\vec{q}_3$] (bR),
      (bR) -- [fermion, edge label=$\vec{q}_3$] (bL),

      (tR) -- [charged scalar, edge label=$\vec{q}_2-\vec{k}$] (v2),
      (v2) -- [charged scalar, edge label=$\vec{q}_2$] (bR),
    };
  \end{feynman}
\end{tikzpicture}
}
\hspace{1.5cm}
    \subfloat[Planar (connected) diagram in Eq.~\eqref{eq:fnl23}.]
    {
\begin{tikzpicture}
  \begin{feynman}
    \vertex (i1) at (-3.4,0);
    \vertex (o1) at ( 3.4,0);
    \vertex (v1) at (-2.4,0);
    \vertex (v2) at ( 2.4,0);
    \vertex (mC) at ( 1, 0);
    \vertex (mC2) at ( -1, 0);

    \vertex (A) at (-1, 1.3);   
    \vertex (B) at ( 1, 1.3);   
    \vertex (C) at (-1,-1.3);   
    \vertex (D) at ( 1,-1.3);   

    \diagram*{
      (i1) -- [graviton, edge label'=$\vec{k}$] (v1),
      (v2) -- [graviton, edge label'=$\vec{k}$] (o1),

      (v1) -- [charged scalar, edge label=$\vec{q}_1-\vec{k}$] (A),
      (C) -- [charged scalar, edge label=$\vec{q}_1$] (v1),
      (B)  -- [charged scalar, edge label=$\vec{q}_2-\vec{k}$] (v2),
      (v2)  -- [charged scalar, edge label=$\vec{q}_2$] (D),

      (A) -- [fermion, edge label=$\vec{q}_3- \vec{k}$] (B),

    (A) -- [fermion] (mC2) -- [plain, edge label=$\vec{q}_1-\vec{q}_3$] (C),
    (D) -- [fermion]  (mC) -- [plain, edge label=$\vec{q}_2-\vec{q}_3$] (B), 
    (D) -- [fermion, edge label=$\vec{q}_3$] (C)
    };
  \end{feynman}
\end{tikzpicture}

}

    \subfloat[Reducible (disconnected) diagram in Eq.~\eqref{eq:fnl4reducible}.]
    {
\begin{tikzpicture}
  \begin{feynman}
    \vertex (i1) at (-3.4,0);
    \vertex (o1) at ( 3.4,0);
    \vertex (v1) at (-2.4,0);
    \vertex (v2) at ( 2.4,0);
    \vertex (a)  at (-1, 1.2);   
    \vertex (b)  at ( 1, 1.2);   
    \vertex (c)  at (-1,-1.2);   
    \vertex (d)  at ( 1,-1.2);   

    \diagram*{
      (i1) -- [graviton, edge label'=$\vec{k}$] (v1),
      (v2) -- [graviton, edge label'=$\vec{k}$] (o1),

      (v1) -- [charged scalar, edge label=$\vec{q}_1-\vec{k}$] (a),
      (c) -- [charged scalar, edge label=$\vec{q}_1$] (v1),
      (b)  -- [charged scalar, edge label=$\vec{q}_1-\vec{k}$] (v2),
      (v2)  -- [charged scalar, edge label=$\vec{q}_1$] (d),


      (c) -- [fermion, half left,  looseness=1.15, edge label=$\vec{q}_2-\vec{q}_1$] (d), 
      (c) -- [anti fermion, half right, looseness=1.15, edge label'=$\vec{q}_2$] (d),     
      (a) -- [fermion, half left,  looseness=1.15, edge label=$\vec{q}_3+\vec{q}_1- \vec{k}$] (b), 
      (a) -- [anti fermion, half right, looseness=1.15, edge label'=$\vec{q}_3$] (b),     
    };
  \end{feynman}
\end{tikzpicture}
}

    \caption{Diagrams contributing at $\mathcal{O}(F_{\rm NL}^4)$ to the SIGW. }
  \label{fig:FNL4_contribution}
\end{figure}

\paragraph{Non-Gaussian components.}
We now examine the effects of NGs in the expansion presented in Eq. \eqref{eq:local_NG}~\cite{Cai:2018dig,Unal:2018yaa, Meng:2022ixx}. 
There are two qualitatively different contributions to the NG part of the GW power spectrum: the disconnected and the connected components~\cite{Unal:2018yaa,Zeng:2024ovg, Adshead:2021hnm,LISACosmologyWorkingGroup:2025vdz}. One can calculate them with the diagrammatic rules presented in Table \ref{tab:Feynman_rules}, and we show the associated diagrams in Fig.~\ref{fig:FNL2_contribution} and \ref{fig:FNL4_contribution}.  The corrections from the disconnected NG diagrams, i.e. the \emph{hybrid} (Fig.~\ref{fig:FNL2_contribution}(c)) and the \emph{reducible}  (Fig.~\ref{fig:FNL4_contribution}(c)) diagrams, read 
\begin{align}
\label{eq:fnl2hybrid}
\Omega^{\rm H}_{\rm GW} (x) &= \frac{F_{\rm NL}^2}{3} \int_{0}^{\infty} dt_1 dt_2 \int_{-1}^1 ds_1 ds_2  \,\mathcal{I}(u_1, v_1, u_1, v_1) \frac{\mathcal{P}_\zeta(xv_1 v_2)}{v_2^2}\frac{\mathcal{P}_\zeta(xu_1)}{u_1^2} \frac{\mathcal{P}_\zeta(xv_1 u_2)}{(v_1 u_2)^2}\, , 
\\
\label{eq:fnl4reducible}
\Omega^{\rm R}_{\rm GW} (x) &= \frac{F_{\rm NL}^4}{12} \int_{0}^{\infty} dt_1 dt_2 dt_3 \int_{-1}^1 ds_1 ds_2  ds_3\,\mathcal{I}(u_1, v_1, u_1, v_1) 
\notag
\\
&\times\frac{\mathcal{P}_\zeta(xv_1 v_2)}{v_2^2}\frac{\mathcal{P}_\zeta(xv_1 u_2)}{v_1^2u_2^2} \frac{\mathcal{P}_\zeta(xu_1 v_3)}{v_3^2}\frac{\mathcal{P}_\zeta(xu_1 u_3)}{ u_1^2u_3^2} \, . 
\end{align}
Finally, the connected NG components are given by 
\begin{align}
\label{eq:fnl2}
\Omega^{\rm WC}_{\rm GW} (x) &= \frac{F_{\rm NL}^2}{3\pi} \int_{0}^{\infty} dt_1 dt_2 \int_{-1}^1 ds_1 ds_2 \int_0^{2\pi} d\varphi_{12} \cos 2 \varphi_{12} \,\mathcal{I}(u_1, v_1, u_2, v_2) 
\notag
\\
& \times u_1v_1\frac{\mathcal{P}_\zeta(x v_2)}{v_2^2} \frac{\mathcal{P}_\zeta(x u_2)}{u_2^2} \frac{\mathcal{P}_\zeta(x w_{12})}{w_{12}^3}\, ,
\\
\label{eq:fnl22}
\Omega^{\rm WZ}_{\rm GW} (x)  &= 
\frac{F_{\rm NL}^2}{3\pi} \int_{0}^{\infty} dt_1 dt_2 \int_{-1}^1 ds_1 ds_2 \int_0^{2\pi} d\varphi_{12} \cos 2 \varphi_{12} \,\mathcal{I}(u_1, v_1, u_2, v_2) 
\notag
\\
& \times  u_2v_1\frac{\mathcal{P}_\zeta(x v_2)}{v_2^2} \frac{\mathcal{P}_\zeta(x u_1)}{u_1^2} \frac{\mathcal{P}_\zeta(x w_{12})}{w_{12}^3}\, ,
\\
\label{eq:fnl23}
\Omega^{\rm P}_{\rm GW} (x)  &= 
\frac{F_{\rm NL}^4}{12\pi^2} \int_{0}^{\infty} dt_1 dt_2 dt_3 \int_{-1}^1 ds_1 ds_2 d_3\int_0^{2\pi} d\varphi_{12}d\varphi_{23} \cos 2 \varphi_{12} \,\mathcal{I}(u_1, v_1, u_2, v_2) 
\notag
\\
& \times  u_1u_2v_1v_2\frac{\mathcal{P}_\zeta(x u_3)}{u_3^2} \frac{\mathcal{P}_\zeta(x v_3)}{v_3^2} \frac{\mathcal{P}_\zeta(x w_{13})}{w_{13}^3} \frac{\mathcal{P}_\zeta(x w_{23})}{w_{23}^3}\, ,
\\
\label{eq:fnl24}
\Omega^{\rm NP}_{\rm GW} (x)  &= 
\frac{F_{\rm NL}^4}{24\pi^2} \int_{0}^{\infty} dt_1 dt_2 dt_3 \int_{-1}^1 ds_1 ds_2 d_3\int_0^{2\pi} d\varphi_{12}d\varphi_{23} \cos 2 \varphi_{12} \,\mathcal{I}(u_1, v_1, u_2, v_2) 
\notag
\\
& \times  u_1u_2v_1v_2u_3\frac{\mathcal{P}_\zeta(x v_3)}{v_3^2} \frac{\mathcal{P}_\zeta(x w_{13})}{w_{13}^3} \frac{\mathcal{P}_\zeta(x w_{23})}{w_{23}^3} \frac{\mathcal{P}_\zeta(x w_{123})}{w_{123}^3} \, ,
\end{align}
where the different superscripts denote \emph{walnut C} (Fig.~\ref{fig:FNL2_contribution}(b)), \emph{walnut Z} (Fig.~\ref{fig:FNL2_contribution}(a)), \emph{planar} ( Fig.~\ref{fig:FNL4_contribution}(b)) and finally \emph{non-planar} (Fig.~\ref{fig:FNL4_contribution}(a)), respectively, and we also used
\begin{align}
&w_{ij} = \sqrt{v_i^2 +v_j^2 - 2y_{12}} \, , \qquad \qquad w_{123} = \sqrt{v_1^2 + v_2^2 + v_3^2 +2y_{12} - 2 y_{13} - 2y_{23}} \\
&y_{ij} = \frac{\cos \varphi_{ij}}{4}\sqrt{t_i(t_i+2)(1-s_i^2)t_j(t_j+2)(1-s_j^2)} + \frac{(1- s_i(t_i+1))(1- s_j(t_j+1))}{4} \, , \\
&\varphi_{ij} = \varphi_i - \varphi_j\, . 
\end{align}
Unlike the case of PBH formation, we notice that the effects of NGs on the SIGW production is independent of the sign of $F_{\rm NL}$. 
Following the discussion presented in Ref.~\cite{Unal:2018yaa}, the planar and the reducible diagrams give contributions being numerically close to each other, while the non-planar is much smaller~\cite{Garcia-Bellido:2017aan}. All those contributions are incorporated in our numerical analysis. 
For the simplicity of the discussion and the Bayesian analysis that will follow, we will neglect the possible contribution from the cubic correction $G_{\rm NL}$. In practice, it has been shown that $G_{\rm NL}$ is a subdominant correction with respect to $F_{\rm NL}$ if $F_{\rm NL} \gtrsim \sqrt{10 G_{\rm NL}}$~\cite{Perna:2024ehx}, which we will implicitly assume in what follows. We however emphasize that the largest contribution to the tensor power spectrum from the cubic correction $G_{\rm NL}$ can be reexpressed as a correction to the Gaussian contribution.  $G_{\rm NL}$ at this order is thus partly degenerate with $\mathcal{A}_\zeta$.

We neglected corrections of order 
$F_{\rm NL}\,\mathcal{P}_{\mathcal{R}_{\rm G}}^3$, 
which arise at third order in cosmological perturbation 
theory~\cite{Malik:2008im,Arroja:2009sh,Hwang:2017oxa,Tomikawa:2019tvi,Chang:2023aba,Picard:2025bwq}. We also draw attention to the fact that, as in the case of the PBH formation, not all deviations from Gaussianty can be parametrized with an expansion around a Gaussian field. This can possibly have consequences for the SIGW \cite{Iovino:2024sgs, Zeng:2025cer}.
We finally note that throughout this paper we adopt a perturbative approach to the study of SIGW, 
treating small inhomogeneous perturbations on top of a homogeneous and isotropic 
FLRW background. This splitting is gauge dependent, which in turn induces a 
potential gauge dependence in the computation of tensor modes sourced by scalar 
perturbations~\cite{Domenech:2017ems,Hwang:2017oxa,Yuan:2019fwv,Gong:2019mui,Inomata:2019yww,Tomikawa:2019tvi,DeLuca:2019ufz,Domenech:2020xin,Kugarajh:2025pjl}.It was however shown that for gravitational waves emitted during a radiation dominated universe, the $\Omega_{\rm GW}$ at late time is gauge-independent.

\section{Binary merger events}
\label{sec:GW_from_PBH_mergers}

\subsection{Merger rates}
\label{sec:merger_rates}

The starting point for evaluating the probability of GW merger events from PBH binaries is to determine their merger rates.
Following their formation and in the case of purely Gaussian curvature perturbations, PBHs are distributed randomly and sparsely in space, adhering to Poisson statistics. NGs in the curvature perturbations will induce corrections to the behavior, which have been shown to be mild~\cite{Crescimbeni:2025ywm}. Consequently, we will ignore such corrections to the Poisson distribution in this paper.

To a good approximation, individual PBHs follow then the Hubble flow, as they are separated by distances much larger than the Hubble distance. As the Universe expands, the Hubble distance eventually exceeds the average separation between PBHs during the radiation-dominated era. Binary systems can now form by decoupling from the expansion of the universe, which occurs when the gravitational attraction of two nearby PBHs overtakes the Hubble flow of the cosmic expansion. For ther formation of a binary, a necessary condition is the presence of an external torque by a third PBH, which prevents an immediate collision, and causes the system to transition into an orbital configuration~\cite{Nakamura:1997sm, Ioka:1998nz}. This mechanism is referred to as \emph{early binary} formation~\cite{Sasaki:2016jop, Sasaki:2018dmp, Raidal:2017mfl, Raidal:2018bbj, Vaskonen:2019jpv}. Notice that binaries can also form at later times during matter domination era~\cite{Mouri:2002mc,Bird:2016dcv,Clesse:2016vqa}, via the so-called \emph{late binary} formation. However, this formation process is largely subleading with respect to the early time binary formation~\cite{Escriva:2022duf}. Consequently, our focus will remain on the early binary formation mechanism. 

\paragraph{Two-body interaction.}
 After the decoupling of the binary during radiation-dominated era, the two PBHs in the binary will orbit around each other and lose energy mostly via gravitational radiation until they finally merge. Approximating the PBH mass distribution as being monochromatic, the merger rate at time $t$ for binaries formed in the early universe when accretion effects are neglected~\cite{DeLuca:2020qqa} is given by \cite{Peters:1964zz,Ali-Haimoud:2017rtz,Chen:2018czv,Raidal:2018bbj}
\bea
\label{eq:merger_rate}
\mathcal R_2 (f_{\rm PBH}, M_{\rm PBH}, t) =\frac{3\times 10^6}{\text{Gpc}^3 \text{yr}} f_{\rm PBH}^{53/37} \bigg(\frac{t}{t_0}\bigg)^{-34/37} \bigg(\frac{M_{\rm PBH}}{M_{\odot}}\bigg)^{-32/37} S( f_{\rm PBH}, t) \, ,
\eea 
where $t_0$ denotes the present time. The suppression factor $S$  accounts for the disruption of binaries after their formation from encounter with other PBH or matter overdensities and can be factorized into early and late time contributions
\bea \label{eq:supp_factor}
S (f_{\rm PBH}, t) \equiv S_{\rm early}(f_{\rm PBH}) \times S_{\rm late} (f_{\rm PBH}, t)\,.
\eea
Let us first discuss the component $S_{\rm early}$, which encodes the disruption of binaries due to close encounters with nearby PBHs during the radiation-dominated era. It can be approximated by~\cite{Hutsi:2020sol}
\bea 
\label{eq:S1}
S_{\rm early} \simeq 1.42 \left(\frac{1}{\bar{N}_{\rm m}+C}+ \frac{\sigma_M^2}{f_{\rm PBH}^2}\right)^{-21/74} e^{-\bar{N}_{\rm m}}\,, \qquad 
\eea 
where
\bea 
C \equiv \frac{f_{\rm PBH}^2}{\sigma_M^2} \left[ \left(\frac{\Gamma(29/37)}{\sqrt{\pi}} U \bigg(\frac{21}{74}, \frac{1}{2}, \frac{5 f_{\rm PBH}^2}{6 \sigma_M^2} \bigg) \right)^{-74/21} -1\right]^{-1}\, , \quad \sigma_M\equiv \frac{\Omega_{\rm M}}{\Omega_{\rm DM}} \sqrt{\langle \delta_M^2\rangle}\,,
\eea 
$U$ denoting the confluent hypergeometric function and $\Gamma$ the gamma functions, and $\sigma_M \simeq 0.005$~\cite{Ali-Haimoud:2017rtz} is the rescaled variance of matter density perturbation,
and $\bar{N}_{\rm m}$ is the expected number of PBHs within the sphere of minimal volume where we do not expect any disruption of the binary system (from a third PBH or an overdensity). The latter quantity is estimated as~\cite{Raidal:2018bbj}
\bea 
\bar N_{\rm m} =\frac{2 f_{\rm PBH}}{f_{\rm PBH} + \sigma_M} \, ,, 
\eea 
where the prefactor of 2 comes from the the fact that we consider mergers between PBH of the same mass.
Moreover, when $f_{\rm PBH} < \sigma_M$, then the tidal torque between PBHs is negligible and the primary contributions to the disruption is due to inhomogeneities in the matter density~\cite{Eroshenko:2016hmn, Hayasaki:2009ug, Raidal:2018bbj}. In the opposite regime $f_{\rm PBH} > \sigma_M$ the large density of PBH make disruption from a third encounter likely.

\begin{figure*}[t!]
  \centering 
    
    \includegraphics[width=0.50\linewidth]{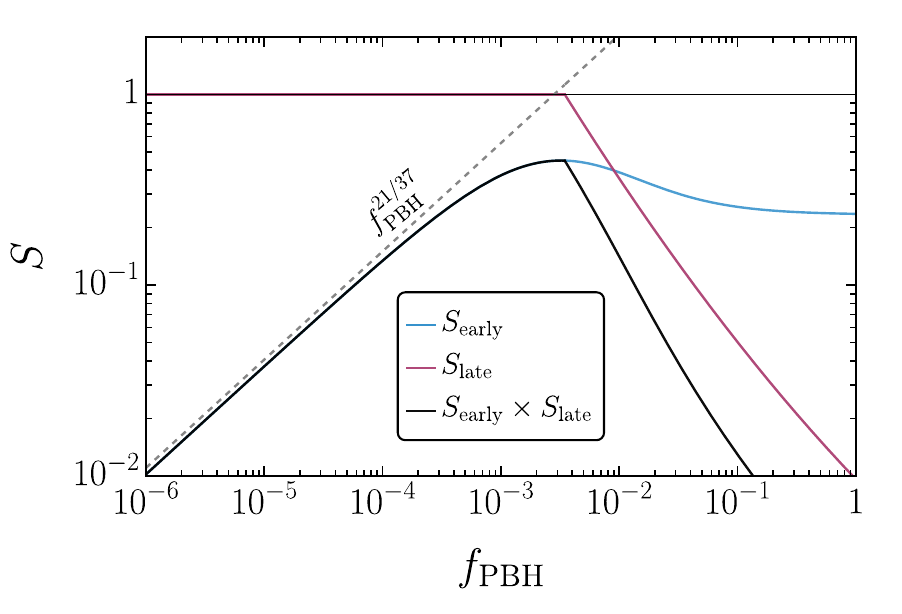} 
    \includegraphics[width=0.49\linewidth]{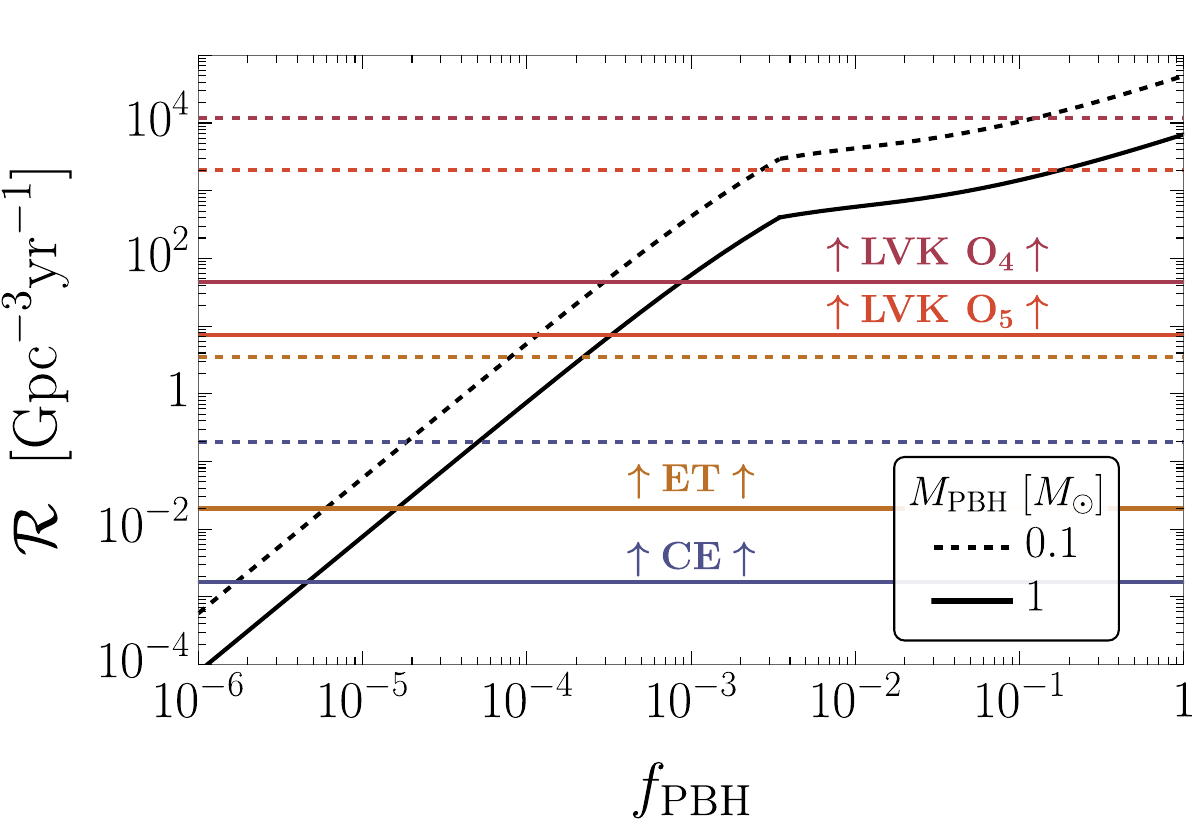} 
        
  \caption{\textbf{Left Panel}:  Suppression factors of the merger rate (see Eq. \eqref{eq:supp_factor}). Each factor $S_{\rm early}$ (red), $S_{\rm late}$ (orange) and the product $S_{\rm early} \times S_{\rm late}$ (black) are shown separately as well as the scaling of the low $f_{\rm PBH}$ tail (dotted gray). \textbf{Right Panel}: Merger rate at time $t= t_0$ for three different PBH populations with $M_{\rm PBH} =  0.1, 1~M_{\odot}$ (dashed, solid). The horizontal colored lines designate the \emph{minimal} merger rate for the population to be detectable by LVK O4 (purple), O5 (red), ET (orange) and CE (blue). The arrows show the part of the parameter space which is detectable. These thresholds are split in two different lines for each observatory, for respectively $M_{\rm PBH} =  0.1, 1~M_{\odot}$ (dashed, solid). }
  \label{fig:Supp}
\end{figure*}

The second factor in Eq.~\eqref{eq:supp_factor} accounts for the disruption from late-time effects, e.g. when the PBH binary is absorbed by a PBH cluster. This effect has been studied numerically in Refs.~\cite{Raidal:2018bbj, Jedamzik:2020ypm}. Retaining only the   fraction of binaries that avoids dense enough clusters in simulations, the following expression is obtained
\bea 
\label{eq:S2}
S_{\rm late}(f,t) \simeq \text{min}\big[1,  0.01 f(t)^{-0.65} e^{0.03 \log^2 f(t)}\big] \, , \qquad f(t) \equiv f_{\rm PBH} (t/t_0)^{0.44} \, .   
\eea 
The suppression factors as functions of $f_{\rm PBH}$ are presented on the left panel of Fig. \ref{fig:Supp}. For the region of $f_{\rm PBH} \lesssim 10^{-3}$,  we observe that the suppression factor scales like $f_{\rm PBH}^{21/37}$, which implies that the merger rate scales exactly like $\mathcal R\propto f_{\rm PBH}^{2}$. 

\paragraph{Three-body interaction.}
Assuming a monochromatic PBH mass distribution, the merger rate of binaries formed through early-Universe three-body interactions is~\cite{Vaskonen:2019jpv, Raidal:2024bmm}
\begin{equation}
    \label{eq:R3}
   \mathcal R_3 (f_{\rm PBH}, M_{\rm PBH}, t)
   \simeq \frac{2.4 \times 10^3}{\rm Gpc^{3}\,yr } 
    \left[ \frac{t}{t_0}\right]^{\frac{\gamma_a}{7} - 1}
    f_{\rm PBH}^{\frac{144 \gamma_a}{259}+\frac{47}{37}}
    \left[ \frac{\langle M_{\rm PBH}\rangle}{M_{\odot}}\right]^{\frac{5 \gamma_a -32}{37}}    \mathcal{K}_a \,
    \frac{e^{-3.2 (\gamma_a - 1)}\gamma_a}{28/9-\gamma_a},
\end{equation}
where $\gamma_a \in [1,2]$ parametrizes the angular momentum distribution $P(j) = \gamma_a j^{\gamma_a-1}$ after ejection of the lightest PBH. While $\gamma_a = 2$ would correspond to thermal equilibrium, simulations favor $\gamma_a = 1$~\cite{Raidal:2018bbj,2019Natur.576..406S}. The factor $\mathcal{K}_a$ accounts for the hardening of the early binary in encounters with other PBHs, with numerical simulations suggesting $\mathcal{K}_a=4$~\cite{Raidal:2018bbj}. Due to the suppression factors in Eqs.~\eqref{eq:S1} and~\eqref{eq:S2} which suppress the 2-body channel contribution to the merging rate in Eq.~\eqref{eq:merger_rate}, the three-body channel in Eq.~\eqref{eq:R3} can dominate for $f_{\rm PBH} \gtrsim 0.1$~\cite{Franciolini:2022ewd,Andres-Carcasona:2024wqk}. Since it is already excluded by microlensing constraints EROS and OGLE, in this work, we do not include the 3-body channel in the calculation of the merging rate.
The merger rate could in principle be modified if PBHs are already clustered at formation. While initial clustering enhances the local PBH density and thus the merger rate~\cite{Ballesteros:2018swv,DeLuca:2021hde}, it also increases the likelihood of binary disruption, suppressing mergers~\cite{Raidal:2018bbj,Vaskonen:2019jpv,Jedamzik:2020ypm}. However, no known mechanism generates clustering strong enough to significantly affect the rate~\cite{Crescimbeni:2025ywm}, and we therefore neglect it in this work and denote $\mathcal R$ to mean $\mathcal R_2$. In the right panel of Fig. \ref{fig:Supp} we plot this quantity against $f_{\rm PBH}$ for different PBH mass benchmarks.

\subsection{Detectability of transient signals}
\label{sec:GW_detectability}

The evolution of a binary system undergoes three distinct phases during which transient GW signals are emitted:
\begin{enumerate}
    \item \textbf{Inspiral Phase:} During this phase, the binary components orbit each other while gradually losing energy through GWs emission, causing the orbital radius to shrink steadily.   
    \item \textbf{Merger Phase:} This marks the point where the two black holes come into contact and begin to overlap, generating a burst of intense GW emission.    
    \item \textbf{Ringdown Phase:} After the merger, the resulting black hole settles into a stable state by radiating GWs at characteristic frequencies associated with its oscillations.
\end{enumerate}
The strain amplitude in the frequency domain is determined by the empirical expression~\cite{Ajith:2007kx, Ajith:2009bn}
\begin{align}
\tilde h(f, z) = \sqrt{\frac{5\eta}{24}} \, \frac{[G (1+z)(2M_{\rm PBH})]^{5/6}}{\pi^{2/3}d_L(z)} 
\times   \,
\begin{cases}
f^{-7/6}~ , \quad f < f_{\rm merg} 
\\
f^{-2/3}f_{\rm merg}^{-1/2}~ , \quad  f_{\rm merg} < f < f_{\rm ring}
\\
f_{\rm ring}^{-2/3}f_{\rm merg}^{-1/2} \bigg(\frac{\sigma^2}{\sigma^2+4(f-f_{\rm ring})}\bigg)~ , \quad  f_{\rm ring} < f < f_{\rm cut} 
\end{cases}\, ,
\end{align} 
where $G$ is the Newton constant, $d_L$ is the luminosity distance to the source, which is given as function of the redshift $z$ by
\bea \label{eq:lumi_distance}
d_L(z) = d_c(z)(1+z) = (1+z)\int^z_0 \frac{dz'}{H[z']} \, .
\eea 
Finally, the transition frequencies that separate the inspiral, merger and ringdown regimes, $f_{\rm merg}$, $f_{\rm ring}$, and $f_{\rm cut}$, respectively, are fitted to template waveforms as
\bea 
f_i =
\frac{a_i \eta^2 + b_i \eta  + c_i}{
2\pi G M_{\rm PBH}(1 + z)} \, , 
\eea 
where the numerical values of $ \sigma, a_i, b_i, c_i$ given in Table 1 of Ref.~\cite{Ajith:2007kx} and $\eta \equiv 1/4$.

\subsection{Reach of future experiments}

\begin{figure*}[t!]
  \centering 
    \includegraphics[width=0.51\linewidth]{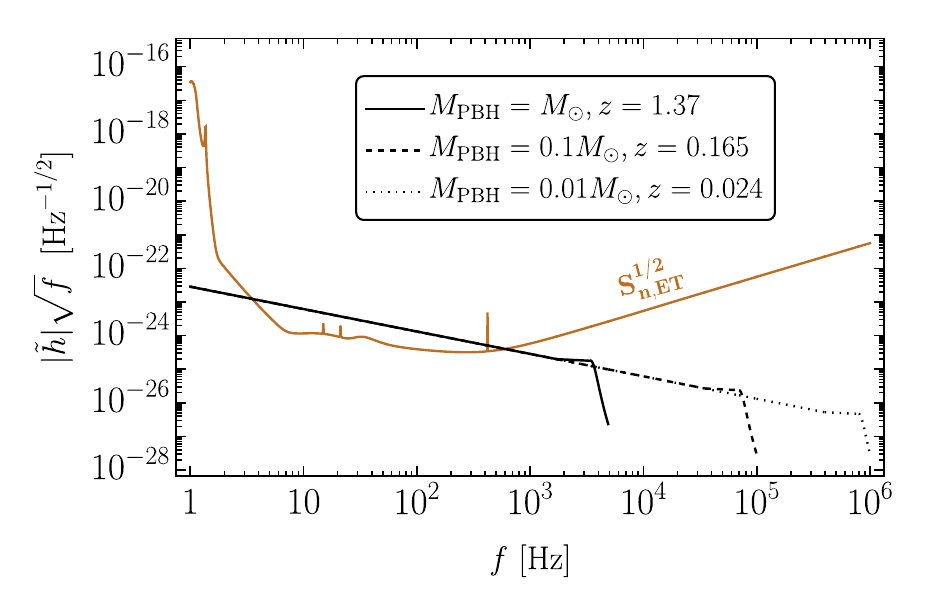} 
\includegraphics[width=0.48\linewidth]{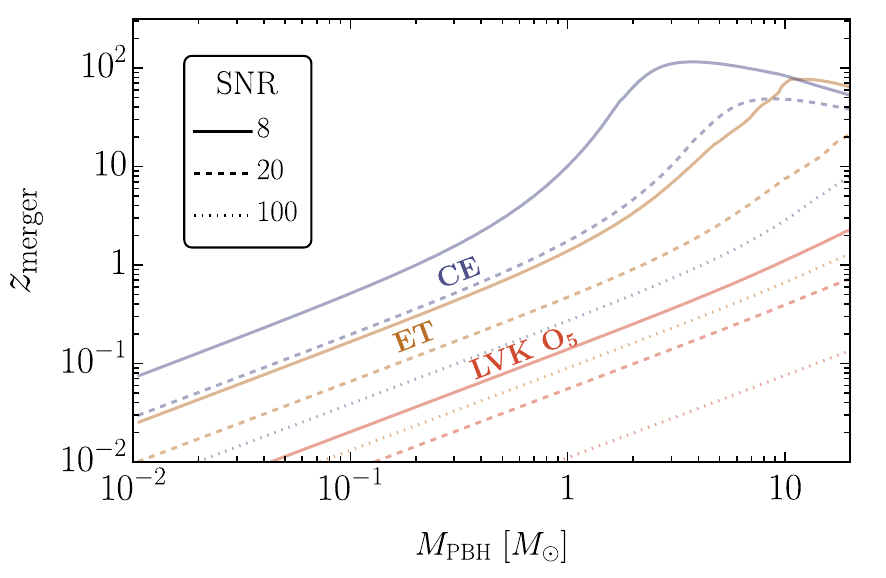}

  \caption{\textbf{Left panel}:  Strain amplitude in Fourier space as a function of frequency for different PBH mass benchmarks. We also display the expected noise power spectrum $S_{n}$ of the ET future observatory. On this plot the redshift of the merger $z$ is fixed so that the signal induces a signal-to-noise ratio $\text{SNR} = 8$ in the ET detector. 
  \textbf{Right panel}: Contour plot of SNR in the $M_{\rm PBH}-z$ plane for the ET (orange), CE (blue) and LVK O5 (red) experiments for three different SNR benchmarks. }
  \label{fig:DetecandSNR}
\end{figure*}

The detectability of a merger event in a terrestrial or space-based detector is quantified via the \emph{signal-to-noise ratio} (SNR) defined as
\bea
\text{SNR}(z) \equiv 2\sqrt{\int df\frac{|\tilde h(f , z)|^2}{S_n(f)}} \, , 
\eea 
where $S_n$ is the noise power spectrum of the GW detector. We consider the following future experiments: future runs of LVK~\cite{LIGOScientific:2020ibl,LIGOScientific:2021job,LIGOScientific:2025slb}, ET~\cite{Hild:2010id,Sathyaprakash:2012jk,Maggiore:2019uih}, and CE~\cite{Evans:2021gyd, Srivastava:2022slt}. ET and CE are the flagship projects representing the third generation ground-based laser interferometers, and in what follows we collectively refer to them as future kHz GW observatories.\footnote{Other proposals operating with prospects of detecting SIGW or transients emitted by PBH binaries include: Laser Interferometer Space Antenna (LISA)~\cite{Caprini:2019egz}, DECi-hertz Interferometer Gravitational Wave Observatory (DECIGO)~\cite{Seto:2001qf, Yagi:2011wg, Isoyama:2018rjb}, Big Bang Observer (BBO)~\cite{Corbin:2005ny, Crowder:2005nr}, Square Kilo-
metre Array (SKA)~\cite{Janssen:2014dka}, Atom Interferometer Observatory and Network (AION)~\cite{Badurina:2019hst}, Atomic Experiment for Dark Matter and Gravity Exploration in Space (AEDGE)~\cite{Bertoldi:2019tck}, and experiments based on resonant absorption in binary systems~\cite{Foster:2025nzf,Blas:2025lzc}.} The respective spectra $S_n$ used in the analysis can be found in the above references.

On the left panel of Fig. \ref{fig:DetecandSNR}, we display $\sqrt{f} |\tilde h|$ for three PBH merger events with a mass respectively of $M_{\rm PBH} = 0.01,0.1, 1 M_{\odot}$ and the distance in Eq. \eqref{eq:lumi_distance} is chosen so that $\text{SNR} \simeq 8$ in the ET experiment, to illustrate the window of mass to which the observatory will be sensitive~\cite{Franciolini:2023opt}. Additionally, on the right panel of Fig.~\ref{fig:DetecandSNR} we show the dependence of SNR on the PBH mass and the merging redshift by drawing contours for different benchmark values of SNR for the LVK O5 (red) and ET (orange) and CE (blue) experiments. 

We notice that the SNR presented in Fig.~\ref{fig:DetecandSNR} is given for an optimal orientation of the detector with respect to the merger event. To account for the random angular orientation of the observatory with respect to the sky location of the binary, and average over the various antenna patterns and signal polarizations, the detector detection probability $p_{\rm det}(r_{\rm det}) < 1$ is employed. This detection probability is a function of the ratio $r_{\rm det} \equiv\text{SNR}_c/\text{SNR}(z)$ where $\text{SNR}_c$ is the detection threshold value, with a standard benchmark of $\text{SNR}_c=8$~\cite{LIGOScientific:2016dsl}, and has only support in the range $r \in [0, 1]$, or equivalently $\text{SNR}>\text{SNR}_c$.  The behavior of the function $p_{\rm det}(r)$, has been studied in Ref. \cite{Pujolas:2021yaw}, and then more in depth in \cite{Gerosa:2019dbe,Hutsi:2020sol, DeLuca:2021wjr}. 

Moreover, the formation of large structures, such as galaxies and clusters, leads to a local increase of the density of PBHs and consequently also to the density of binaries~\cite{Pujolas:2021yaw}. In an overdense region like our galaxy,\footnote{We model the Milky Way halo by a Navarro-Frenk-White profile~\cite{Navarro:1995iw, Navarro:1996gj}
\[\rho_{\rm DM}(r) = \frac{\rho_0}{\frac{r}{r_0} \bigg(1+\frac{r}{r_0}\bigg)^2} \, ,\]
where $r_0 = 15.6 \text{ kpc} $ and the reference energy density $\rho_0$ is fixed such that the local dark matter density at $r_{\odot} =8 \text{kpc}$ is given by 
 $\rho_{\rm DM}(r=r_\odot )=7.9 \times 10^{-3} M_{\odot}/\text{ pc}^3$.} the average density contrast at distance $d$ of the binary from the solar system is
 \bea 
 \label{eq:over_coll}
 \delta_{\odot}(d) = \text{Max} \bigg[1,~ \frac{1}{2} \int^{1}_{-1} d \cos \theta\, \frac{\rho_{\rm DM}(r(d,\theta))}{\bar\rho_{\rm DM}} \bigg] \, , 
 \eea 
where  $\bar\rho_{\rm DM}$  
is the average DM density. The \emph{number of detected merger} can then be obtained by putting together the merger rate (see Eq. \eqref{eq:merger_rate}), the enhancement from the galactic overdensity (see Eq. \eqref{eq:over_coll}) and the average over detector orientation
\bea
N = T \int \frac{dz}{1+z} \frac{dV_c}{dz} \,\mathcal R(t(z)) \, \delta_{\odot} (d(z))  \, p_{\rm det}(\text{SNR}_c/\text{SNR(z)})\,,
\eea
 where $T$ is the observation time and 
the factor $V_c(z)$ is the volume of the comoving Hubble horizon at redshift $z$, 
\bea 
V_c \equiv \frac{4\pi}{3} d_c(z)^3 =\frac{4\pi}{3} \bigg(\int^z_0 dz'\frac{1}{H[z']}\bigg)^3 \, . 
\eea 

With those tools in hand, we can now obtain the exclusion regions in the PBH parameter space that we can deduce from present and future observatories.

\paragraph{Exclusion region for $M_{\rm PBH} \lesssim M_{\odot}$.}

\begin{figure*}[t!]
  \centering 
        \includegraphics[width=0.8\linewidth]{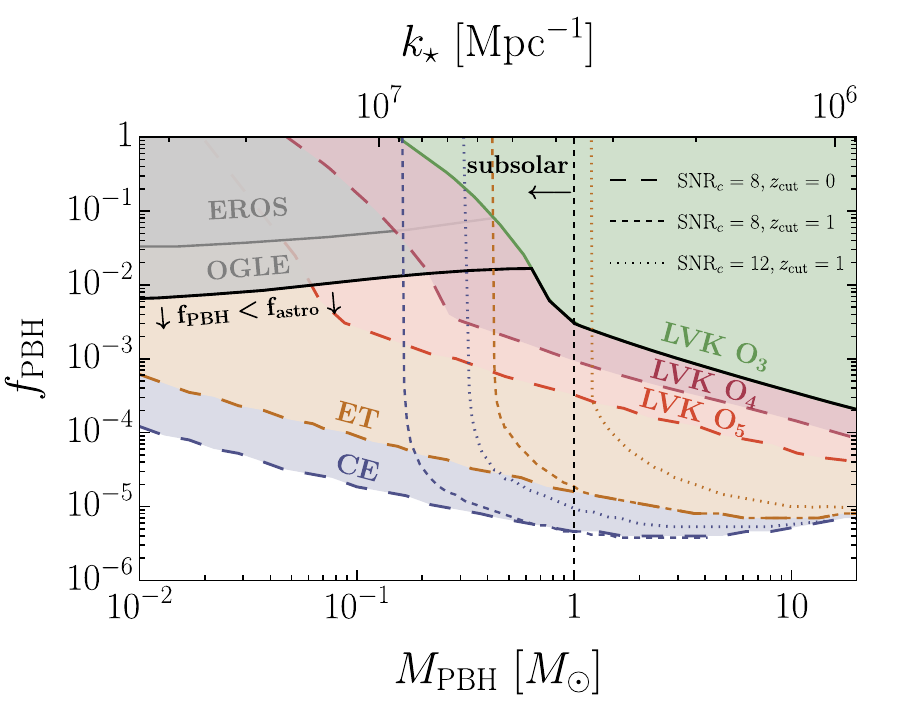} 
        
  \caption{ The reach of future kHz GW observatories derived in this work: LVK O5 (red), ET (orange) and CE (blue). The boundary corresponds to an expected observation of 3 events $N=3$ in the given observation time, which translates into an  exclusion limit at $95\%$ confidence level if the detectors does not report any merger event. We also report the microlensing bounds coming from OGLE and EROS as well as the bounds from the non-observation of sub-solar binaries during LVK O3 (assuming $z_{\rm cut} = 0$). The bounds from the recently released LVK O4 is also presented (purple). We consider three detection criteria: an optimistic with $\text{SNR}_c = 8$ and $z_{\rm cut} = 0$ (long dashed), an intermediate with $\text{SNR}_c = 8$ and $z_{\rm cut} = 0$, where one needs $\text{SNR}_c = 12$ and $z_{\rm cut} = 0$ (dashed) and the conservative $\text{SNR}_c = 8$ and $z_{\rm cut} = 1$ (dotted), which ensures the primordial origin of the black holes. Importantly, the vertical dashed line separates the region where the computation is more reliable, namely for $M_{\rm PBH} \lesssim M_{\odot}$  (see text for details). }
  \label{fig:future_reach}
\end{figure*}

In Fig.~\ref{fig:future_reach} we report the boundary curves corresponding to an expectation value of $N=3$ PBH binary merger events within an observation period of $T=4\,\text{yr}$. Long dashed lines correspond to $N=3$ by imposing a detection threshold of $\text{SNR}_c = 8$. The blue, orange, green, purple, and red curves correspond to CE, ET, LVK O3, LVK O4, and LVK O5, respectively \footnote{The LVK O3 curve is interpreted as a current bound on the parameter space. To compute it we use the sensitivities of the O3 run for LIGO Livingston (LL) and LIGO Hanford (LH), Virgo and Kagra and then calculate the total 
number of mergers by summing up the numbers corresponding to each survey. The observation periods in month are $T^{\rm O3}_{\rm LL} = 11,~T^{\rm O3}_{\rm LH} = 11 ,~T^{\rm O3}_{\rm Virgo} = 11,~T^{\rm O3}_{\rm Kagra} = 1$. For the run O4 of LVK, we take sensitivities and the durations
$T^{\rm O4}_{\rm LL} = 18, T^{\rm O4}_{\rm LH} = 18 , T^{\rm O4}_{\rm Virgo} = 18, T^{\rm O4}_{\rm Kagra} = 10$, respectively.
For the run O5 of LVK, we take a cumulative observation time $T^{O5}_{\rm Ligo} = 4$ yr, using the design sensitivity of O5.}. 

Importantly, above these curves, the expected number of detected events $N$ increases rapidly. Following Ref.~\cite{Pujolas:2021yaw}, \emph{in case no PBH binary merger signals are observed within an observation period of $T=4\,\text{yr}$}, then under the assumption of Poissonian statistics these boundary curves can be reinterpreted as 95\% confidence-level exclusion limits on the corresponding PBH abundance. We note that strictly speaking the results on Fig.~\ref{fig:future_reach} are derived in the limit of a \emph{monochromatic} spectrum and would receive corrections for extended mass function and from NGs~\cite{Carr:2017jsz,Andres-Carcasona:2024wqk}. However, as these corrections are expected to be small in our case~\cite{Carr:2017jsz}, we will neglect them.  Additionally, the gray shaded regions indicate the bounds from microlensing of stars in the Galactic bulge by OGLE~\cite{Mroz:2024mse, Mroz:2024wag} and in the Magellanic Clouds by EROS~\cite{EROS-2:2006ryy}. Similar searches can be performed using the \emph{stochastic} GW background from the superposition of mergers \cite{Pujolas:2021yaw,Romero-Rodriguez:2024ldc,Romero-Rodriguez:2025mfr,Boybeyi:2024mhp}, leading typically to stringent bounds are larger masses.

\paragraph{Exclusion region for $M_{\rm PBH} \gtrsim M_{\odot}$.} 
The reasoning outlined above applies \emph{per se} only to sub-solar black holes, for which no well-established astrophysical channels are known to produce a significant non-primordial population. For $M_{\rm PBH} \gtrsim M_{\odot}$, however, binary merger events from astrophysical black holes constitute a background that must be disentangled by determining, on an event-by-event basis, whether the origin is astrophysical or primordial. At present, no systematic method exists to subtract this astrophysical background from the data. Making the bold (and likely unrealistic) assumption that the \emph{entire} astrophysical contribution could be removed, one may extend the same line of reasoning as above and continue the contour lines into the region $M_{\rm PBH} \gtrsim M_{\odot}$. Given the above, we adopt the conservative stance that any bounds in this mass range should be regarded as carrying substantially larger uncertainties.

\paragraph{Exclusion region for alternative sub-solar black hole formation channels.}

This paper is motivated by the observation that the detection of a sub-solar black hole merger event would strongly point toward the existence of PBHs. We emphasize, however, that although an announcement of such an event would be groundbreaking, its classification as a PBH merger would require careful scrutiny. In fact, there are new physics scenarios that could present loopholes to our goal of connecting the signals observed in LVK and PTAs through a common primordial origin.

\begin{enumerate}
    \item[i)] \emph{Stellar transmutation}: From stellar evolution, white dwarfs typically acquire masses $\mathcal O( 0.5-0.7 M_{{\odot}})$~\cite{Kepler:2006ns,Kilic:2006as}. On the other hand, even if it is highly unlikely for neutron stars to develop sub-solar masses via standard astrophysical channels, they could still remain stable for masses down to $\mathcal O( 0.1 M_{{\odot}})$~\cite{Lattimer:2000nx,Lattimer:2012nd,Suwa:2018uni,Metzger:2024ujc}. Accretion of certain types of dark matter can convert astrophysical objects, such as neutron stars~\cite{Goldman:1989nd,Bramante:2013hn,Kouvaris:2013kra,Bramante:2015dfa,Takhistov:2017bpt,Kouvaris:2018wnh,Takhistov:2020vxs,Dasgupta:2020mqg,Garani:2021gvc,Singh:2022wvw,Singh:2022wvw,Bhattacharya:2023stq} or white dwarfs  matter~\cite{Steigerwald:2022pjo,Chakraborty:2024eyx}, into mildly sub-solar black holes.  Alternatively, a core of dark matter could grow at the center of a white dwarf or a neutron star and collapse into a sub-solar mass black hole~\cite{Dasgupta:2020mqg}. It was shown however that the merging rates of binaries formed of transmuted objects peaks at typically low redshift~\cite{Dasgupta:2020mqg}. To disentangle the two populations, we introduce a lower cut-off on the admitted signals $z_{\rm cut} = 1$ (i.e. discarding all the observed mergers for $z < 1$), where the expected abundance of stellar objects that could undergo transmutation becomes negligible (e.g. see Ref.~\cite{Dasgupta:2020mqg}).
    \item[ii)] \emph{Exotic compact objects}: Q-balls~\cite{Coleman:1985ki}, boson stars~\cite{Colpi:1986ye,Liebling:2012fv}, and fermion soliton stars~\cite{Lee:1986tr,DelGrosso:2023trq} could in principle exist in the sub-solar regime. To distinguish between those objects and black holes, the study of the tidal deformabilities during the merger has been shown to be a reliable method for upcoming experiments~\cite{DeLuca:2023mio, Crescimbeni:2024cwh, Crescimbeni:2024qrq, Golomb:2024mmt}. However, larger SNRs are necessary to disentangle the nature of the compact object via the observation of the tidal deformabilities. In particular, for events such as the candidate SSM200308 from the LVK O3 run, Ref.~\cite{Crescimbeni:2024cwh} demonstrated that a signal with $\text{SNR} \sim 12$ would allow the PBH origin to be excluded at the $5\sigma$ level. Motivated by this result, we heuristically assume that events with $\text{SNR}\gtrsim 12$ can be reliably classified as either PBHs or exotic compact objects. Accordingly, we use two benchmark thresholds: $\text{SNR}_c = 8$ (non-discriminant analysis) and $\text{SNR}_c = 12$ (discriminant analysis).
    \item[iii)] \emph{PBHs from sub-horizon mechanisms}: Large primordial curvature perturbations are not the only mechanism that can produce PBHs. In fact, post-inflationary era dynamics can generate overdensities that collapse to black holes on sub-horizon scales (for a recent review see Ref.~\cite{Flores:2024eyy}).  Although merger events alone may not be sufficient to distinguish between these formation scenarios, a promising avenue for future research is to investigate whether their differing phenomenological signatures --- such as spin distributions --- could serve as distinguishing features.
\end{enumerate}

After discussing the alternative formation mechanisms for sub-solar black holes, we return to Fig.~\ref{fig:future_reach}. In addition to the \emph{optimistic} case with $\text{SNR}_c = 8$ and $z_{\rm cut} = 0$ (long dashed), we introduce two further detection criteria: an \emph{intermediate} case with $\text{SNR}_c = 8$ and $z_{\rm cut} = 1$ (dashed), and a \emph{conservative} case with $\text{SNR}_c = 12$ and $z_{\rm cut} = 1$ (dotted). For the latter we see that only Cosmic Explorer is expected to probe masses (marginally) above $0.1\, M_{\odot}$.

\section{Pulsar timing arrays}
\label{sec:PTAs}

The NANOGrav collaboration \cite{NANOGrav:2020bcs,NANOGrav:2023gor} combined with the other PTA collaborations, EPTA~\cite{Chen:2021rqp,Antoniadis:2023rey}, PPTA~\cite{,Goncharov:2021oub,Reardon:2023gzh}, CPTA~\cite{Xu:2023wog} and IPTA~\cite{Antoniadis:2022pcn} have released further evidence for the Hellings-Downs angular correlation in the common-spectrum process. This points towards the existence of a GW background in the nHz range permeating the universe. In this section, we explore different hypotheses for the explanation of the observed signal based on astrophysical SMBHs and SIGW associated with sub-solar PBH production.

\subsection{Pulsar timing response to gravitational waves}

 A GW propagating in direction $\hat{\Omega}$ induces a frequency shift on pulses emitted by pulsar $I$, which is observed along the direction $-\hat{p}$. The relative frequency change at the Earth can be expressed as \cite{Allen:1997ad,Anholm:2008wy,Maggiore:2018sht} 
\begin{equation}
\label{eq:nu_I_over_nu_ref}
\frac{\delta \nu_I}{\nu} = - \Lambda^{ab} \left[ h_{ab}(t_e, \vec{x}_e) - h_{ab}(t_I, \vec{x}_I) \right]\,, \qquad \text{with} \quad \Lambda^{ab} \equiv \frac{1}{2} \frac{\hat{p}^a \hat{p}^b}{1 + \hat{\Omega} \cdot \hat{p}}\,.
\end{equation}
Here, $(t_e, \vec{x}_e)$ and $(t_I = t_e - D, \vec{x}_I)$ are the coordinates of the Earth and the pulsar, respectively, with $D$ the Earth–pulsar separation.
The effect on the observed arrival time of pulses is quantified by the integral of the frequency shift
\begin{equation}
\label{eq:residual_R_ref}
R_I(t) \equiv - \int_0^t \frac{\delta \nu_I}{\nu} dt\,.
\end{equation}
To characterize a SGWB, we analyze the cross-correlation of timing residuals between pairs of pulsars $I$ and $J$, i.e.
\begin{equation}
\label{eq:h_ab_h_ab_tau_ref}
S_{R,IJ}(f) \equiv \int_{-\infty}^{\infty} d\tau\, e^{2\pi i f \tau} \langle R_I(t) R_J(t+\tau) \rangle\,.
\end{equation}
Combining Eqs.~\eqref{eq:nu_I_over_nu_ref} and \eqref{eq:residual_R_ref}, this spectrum becomes
\begin{equation}
\label{eq:S_R_IJ_S_h_ref}
S_{R,IJ}(f) = \frac{\Gamma_{IJ}}{12\pi^2} \frac{h_c^2(f)}{f^3} = \frac{\Gamma_{IJ}}{8\pi^4} \frac{H_0^2}{f^5} \Omega_{\rm GW}(f)\,,
\end{equation}
where $\Gamma_{IJ}$ represents the Hellings-Downs angular correlation function \cite{Hellings:1983fr}, which depends on the angle $\theta_{IJ}$ between the directions to pulsars $I$ and $J$ from the Earth
\begin{equation}
\label{eq:ORF_ref}
\Gamma_{IJ} = \frac{\delta_{IJ}}{2} + \frac{1}{2} + 3(1 - \cos\theta_{IJ}) \left[ \ln\left( \frac{1 - \cos\theta_{IJ}}{2} \right) - \frac{1}{6} \right]\,.
\end{equation}
The characteristic strain $h_c(f)$ relates to the strain power spectrum $S_h(f)$ via
\begin{equation}
h_c^2(f) = f S_h(f), \qquad S_h(f) = \int_{-\infty}^{\infty} d\tau\, e^{2\pi i f \tau} \langle h_{ij}(t) h^{ij}(t+\tau) \rangle\,.
\end{equation}
The fraction $\Omega_{\rm GW}(f)$ of the universe energy attributed to the SGWB can be computed using the definition in Eq.~\eqref{eq:Omega_GW_def1} and the fact that 
\begin{equation}
\label{eq:Omega_GW_def} \left<\dot{h}_{ij}\dot{h}^{ij}\right> = 8\pi^2 \int_0^\infty  df\,f^2S_h(f) \, . 
\end{equation}

\subsection{Stochastic background from curvature peaks}
\label{sec:GW_curvature}

\begin{table}[ht!]
    \centering
    \renewcommand{\arraystretch}{2}
    \begin{tabular}{|ccc|c|c|}
    \hline
    \textbf{Parameter} $\theta$ & \textbf{Description} & \multicolumn{3}{c|}{\textbf{Prior} $\mathcal{P}(\theta)$ }\\
    \hline
    \multicolumn{5}{|c|}{\textbf{SMBH with free amplitude}} \\
    \hline
    $A_{\rm SMBH}$ & GW amplitude & \multicolumn{3}{c|}{\text{LogUniform} $[10^{-18},10^{-14}]$} \\
    \hline
    $\gamma_{\rm SMBH}$ & Spectral tilt & \multicolumn{3}{c|}{\text{Fixed to} $13/3$} \\
    \hline
    \multicolumn{5}{|c|}{\textbf{SMBH with \texttt{GWOnly-Ext}}} \\
    \hline
    $\log_{10}(A_{\rm SMBH})$ & GW amplitude & \multicolumn{3}{c|}{\text{Gaussian}(-15.6, 0.28)}\\
    \hline
    $\gamma_{\rm SMBH}$ & Spectral tilt & \multicolumn{3}{c|}{\text{Gaussian}(4.7, 0.12)} \\
    \hline
    \multicolumn{5}{|c|}{\textbf{SIGW}} \\
\hline
$A_{\zeta}$ & Scalar amplitude & LogUniform $[10^{-3}, 10^{-0.5}]$ & \multirow{4}{*}{\shortstack{$\mathcal{P}_{\rm PBH}^{\rm astro}(\theta)$\\ see~Eq.~\eqref{eq:PBH_prior_1}}}
 & \multirow{4}{*}{\shortstack{$\mathcal{P}_{\rm PBH,X}^{\rm sub-solar}(\theta)$\\ see~Eq.~\eqref{eq:PBH_prior_2}}}\\ 
\cline{1-3}
$k_\star~[\rm Mpc^{-1}]$ & Peak frequency & LogUniform $[10^6, 10^9]$ & & \\
\cline{1-3}
$\Delta$ & Width & Uniform $[0.01, 3]$ & & \\
\cline{1-3}
$f_{\rm NL}$ & NG parameter & Uniform $[-10, 10]$ & & \\
\hline
    \end{tabular}
    \caption{Priors for the model parameters used in the Bayesian analyses performed in this work.}
    \label{tab:prior_tab}
\end{table}

\begin{figure*}[t!]
  \centering 
    \includegraphics[width=0.48\linewidth]{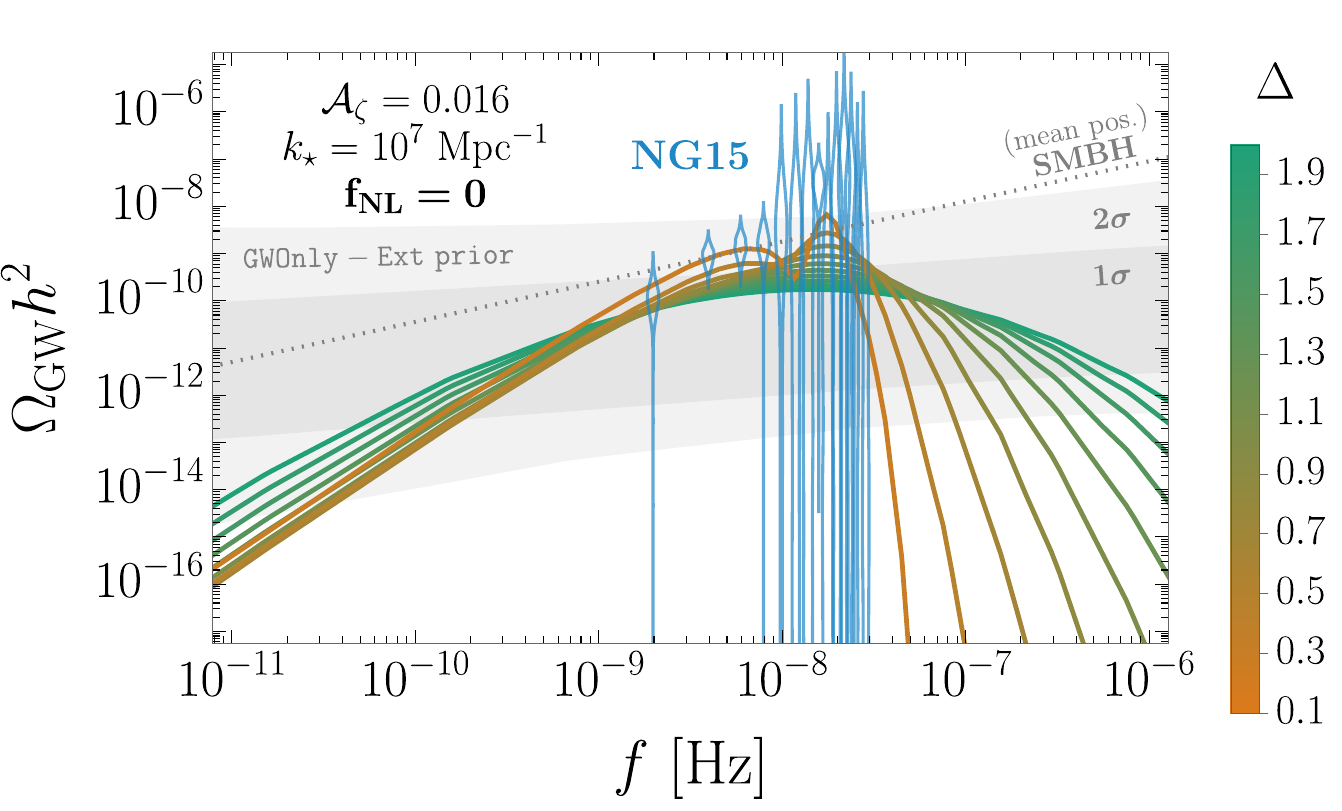}
    \includegraphics[width=0.48\linewidth]{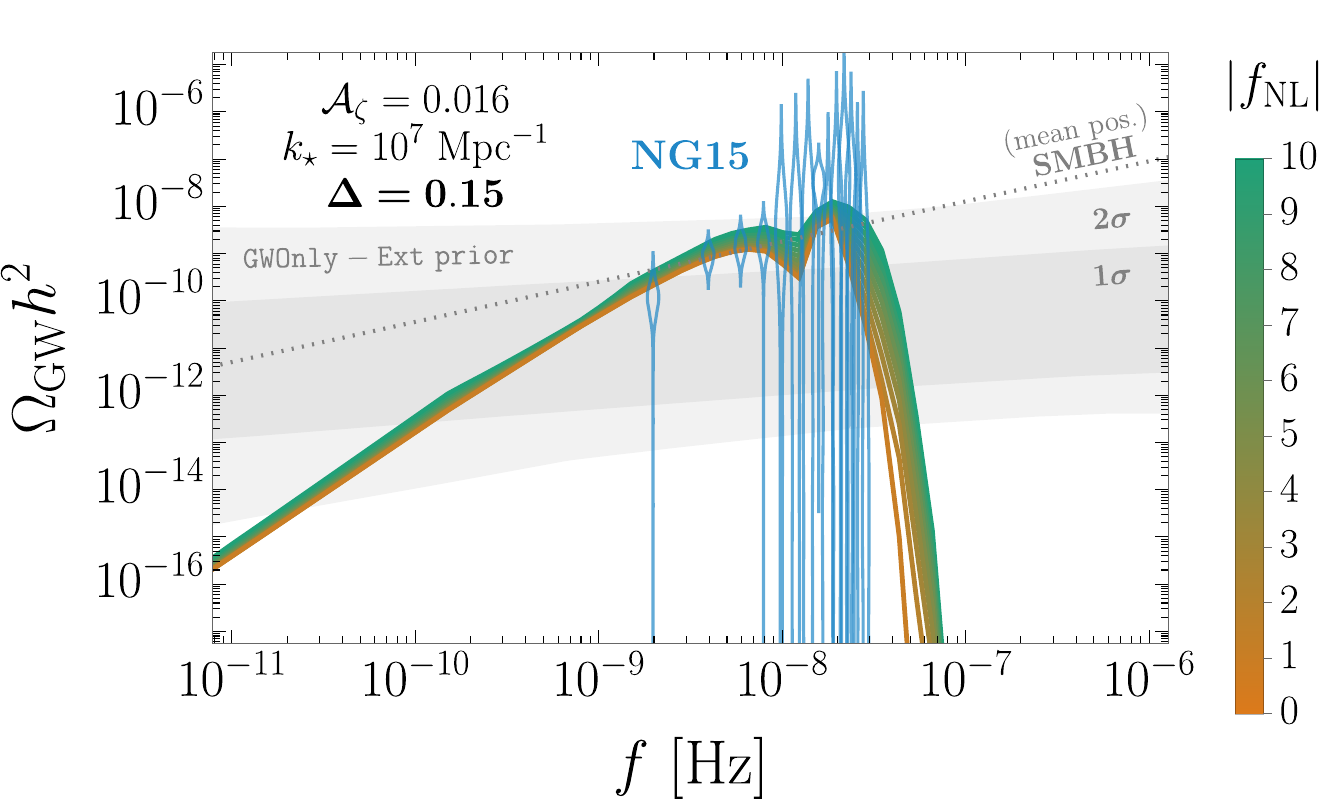}
    \includegraphics[width=0.48\linewidth]{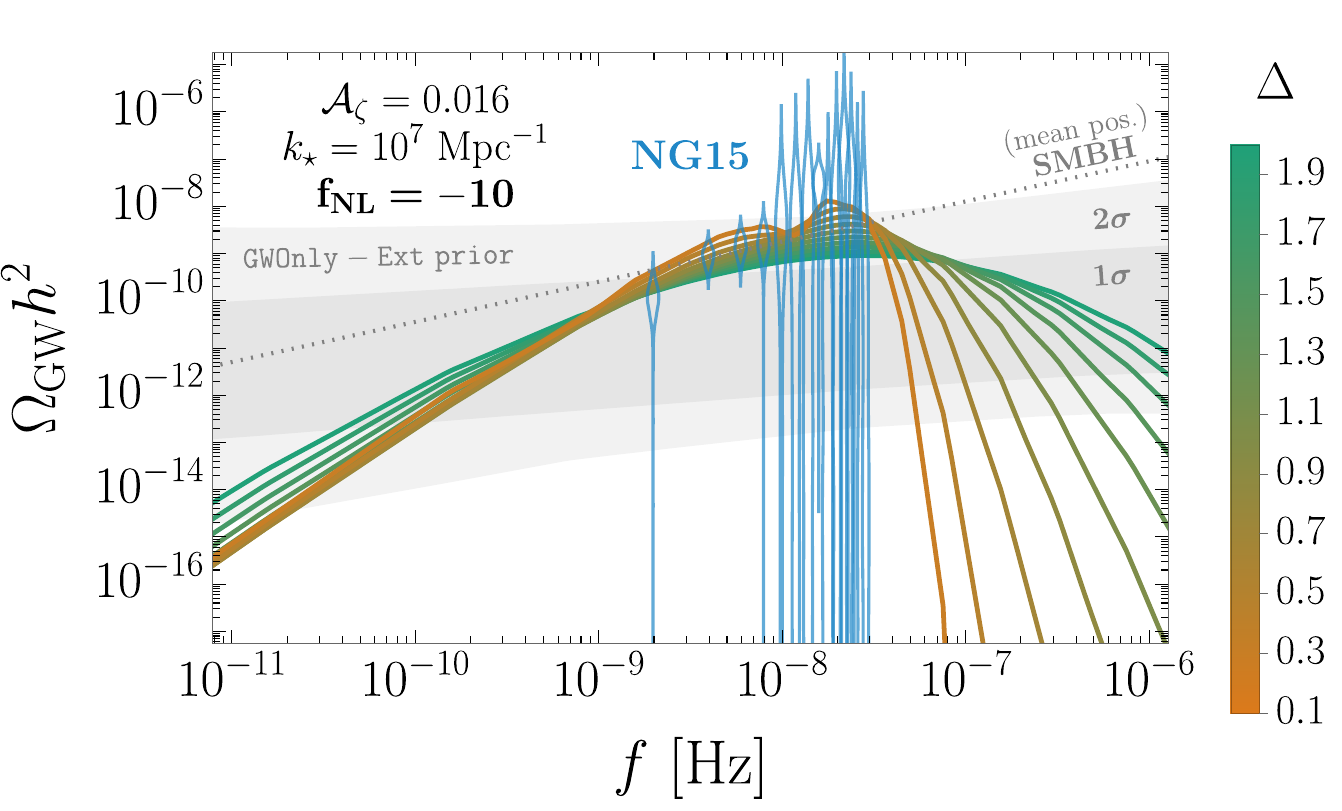}
    \includegraphics[width=0.48\linewidth]{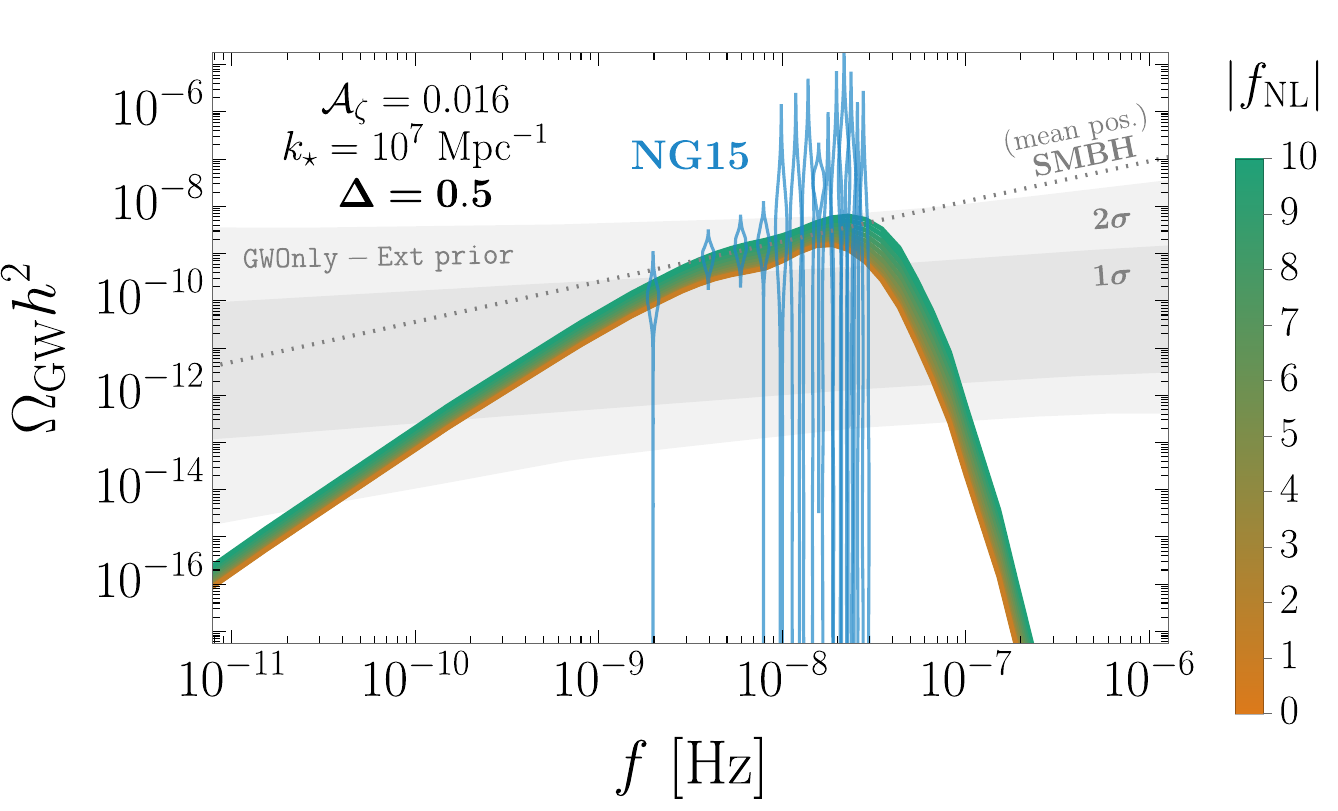}
        
  \caption{SIGW power spectra from a log-normal curvature distribution overlaid together with the NG15 posterior violin plots. The scalar amplitude and peak scale are fixed to the NG15 mean posterior values, $\log_{10}\mathcal{A}_{\zeta} = -1.8 \pm 0.075$ and $\log_{10}k_\star\,[\mathrm{Mpc}^{-1}] = 7.0 \pm 0.12$, obtained from an analysis that includes the astrophysical PBH prior (see Table~\ref{tab:meanPoseterior_SIGW}). We then vary the width $\Delta$ and the NG parameter $f_{\rm NL}$ around the mean posterior values $\Delta = 0.23^{+0.073}_{-0.15}$ and $f_{\rm NL} = -3.0^{+2.3}_{-2.9}$, to show their impact on the GW spectrum. We show the former fixing two benchmarks for the later, namely for $f_{\rm PBH} = 0,-10$ (resp. \textbf{left-top} and \textbf{left-bottom}) and vice verse for $\Delta = 0.15, 0.5$ (resp. \textbf{right-top} and \textbf{right-bottom}). For comparison, we also plot the mean posterior of the SMBH interpretation (dotted), using the \texttt{GWOnly-Ext} prior in Eq.~\eqref{eq:SMBH_prior2}, as well as the corresponding 68\% and 95\% credible regions (gray shaded bands).}
  \label{fig:SIGW_PTA_violins}
\end{figure*}

In Sec. \ref{sec:SIGW}, we presented the theoretical signal emitted by the scalar fluctuations, where the SIGW spectrum $\Omega_{\rm SIGW}$ is given by  Eq.~\eqref{eq:Omega_SIGW_master}. In this section, we study the parameter space of those fluctuations that would explain the PTA signal. For illustration purposes, we first display on Fig.~\ref{fig:SIGW_PTA_violins},  the GW signal against the \emph{violins plot} for different values of the parameters $\Delta, f_{\rm NL}, k_\star$. Moreover, we fix the scalar amplitude and peak scale to the NG15 mean posterior values, $\log_{10}\mathcal{A}_{\zeta} = -1.8 \pm 0.075$ and $\log_{10}k_\star\,[\mathrm{Mpc}^{-1}] = 7.0 \pm 0.12$, obtained from an analysis that includes the astrophysical PBH prior (see Appendix~\ref{app:meanPoseterior}). Visually, we can observe that decreasing the width $\Delta$ improves the fit to the NG15 signal. The value of $\Delta$ also determines the extent of the signal toward higher frequencies: narrower spectra (smaller $\Delta$) fall off more rapidly once outside the PTA regime, while broader spectra maintain a tail at higher frequencies. By contrast, the effect of non-Gaussianities parametrized by $f_{\rm NL}$ is less apparent in this plot, since they do not significantly alter the spectral shape. Their main impact is to lower the overall amplitude $\Omega_{\rm GW}$ as $f_{\rm NL}$ increases.

\paragraph{PBH priors.} One of the novelties of this work is that we perform the Bayesian analysis including phenomenologically motivated priors on the PBH abundance. We focus on the following three prior choices:

\begin{enumerate}
    \item[i)] \emph{No PBH prior.} As a first benchmark we study the PTA signal without imposing priors from the overproduction of PBH. We notice however that this assumption is not compatible with current astrophysical bounds, since certain parts of the parameter space are already ruled out, e.g. the region above the thick black line denoted $f_{\rm PBH} < f_{\rm astro}$ in Fig.~\ref{fig:future_reach}.
    \item[ii)] \emph{Astrophysical PBH prior.} In this case we include the astrophysical constraints as priors in our analysis.  We do this by multiplying the likelihood (see Sec.~\ref{sec:likelihood}) by the prior function
 \begin{equation}
\label{eq:PBH_prior_1}
     \mathcal{P}_{\rm PBH}^{\rm astro}(\theta)=\frac{1}{Z_{\rm astro}}\Theta( f_{\rm astro}(\theta)-f_{\rm PBH}(\theta))\,,
 \end{equation}
 where  $\theta$ are the model parameters, $f_{\rm PBH}$ given in Eq.~\eqref{eq:f_PBH_final} and $f_{\rm astro}$ are the astrophysical constraints on PBHs, dominated by OGLE and LVK O3 shown in gray and green in Fig.~\ref{fig:future_reach}.\footnote{Notice that the latest bounds from LVK O4~\cite{LIGOScientific:2025slb}, are not included in the astrophysical PBH prior. Since, the sensitivity gain of O4 over O3 in the sub-solar mass window is modest, so we do not expect our conclusions to be significantly altered by including the new exclusion region.}
 We also include in $f_{\rm astro}$, constraints from CMB at larger PBH masses, as well as EROS and HSC at lower masses~\cite{Carr:2020gox}. The normalization factor $Z_{\rm astro}$ in Eq.~\eqref{eq:PBH_prior_1} is defined such that the integration of Eq.~\eqref{eq:PBH_prior_1} over all the model parameters returns unity. The model parameters whose integration in Eq.~\eqref{eq:Z_factor_1} is non-trivial are $\theta=(\mathcal{A}_\zeta,k_\star,\Delta, f_{\rm NL})$. Hence, we get
 \begin{equation}
 \label{eq:Z_factor_1}
     Z_{\rm astro} \equiv \int d\theta ~\mathcal{P}(\theta)~\Theta( f_{\rm astro}(\theta)-f_{\rm PBH}(\theta)),
 \end{equation}
 where $\mathcal{P}(\theta)$ is the normalized ($\int d\theta\,\mathcal{P}(\theta)=1$) prior distribution of the model parameters. The choices for the priors of these parameters are given in Table~\ref{tab:prior_tab}. 
    \item[iii)] \emph{Sub-solar PBH prior in LVK-CE-ET.} Here entertain the scenario that a kHz GW observatory has detected a sub-solar PBH, and we set it as prior in the Bayesian analysis of the PTA signal. We multiply the likelihood by the prior function
 \begin{equation}
  \label{eq:PBH_prior_2}
     \mathcal{P}_{\rm PBH,X}^{\rm sub-solar}(\theta)=\frac{1}{Z_{\rm X}}\Theta(1M_{\odot}-M_{\rm PBH})\Theta(f_{\rm PBH}- f_{\rm X})\Theta( f_{\rm astro}-f_{\rm PBH})\,,
 \end{equation}
 where $f_{\rm X}$ is the experimental sensitivity threshold for $X = $~(LVK $\rm O_4$, LVK $\rm O_5$, {\rm ET}, {\rm CE}). We consider $f_{\rm X}$ to be given by the sensitivity curves in Fig.~\ref{fig:future_reach} for the optimistic criterion with $\text{SNR}_c=8$ and $z_{\rm cut}=0$. In Eq.~\eqref{eq:PBH_prior_2}, we have additionally restricted the PBH mass to be sub-solar, and imposed its abundance to be compatible with astrophysical constraints $f_{\rm astro}$. The normalization factor $Z_{\rm X}$ is calculated similarly to Eq.~\eqref{eq:Z_factor_1},  and in this case becomes
  \begin{equation}
 \label{eq:Z_factor_2}
     Z_{\rm X} \equiv \int d\theta ~\mathcal{P}(\theta)~\Theta(1M_{\odot}-M_{\rm PBH})\Theta(f_{\rm PBH}- f_{\rm X})\Theta( f_{\rm astro}-f_{\rm PBH})\,.
 \end{equation}
The resulting factors can be found in Table \ref{tab:Z_factor}. As expected, the narrower the prior range in Eq.~\eqref{eq:PBH_prior_2} the smaller is the normalization parameter.

\end{enumerate}

 \begin{table}[ht]
\centering
\renewcommand{\arraystretch}{1.05}
\begin{tabular}{|
    >{\centering\arraybackslash}p{2cm}|
    >{\centering\arraybackslash}p{2.2cm}|
    >{\centering\arraybackslash}p{2cm}|
    >{\centering\arraybackslash}p{2cm}|
    >{\centering\arraybackslash}p{2cm}|
    >{\centering\arraybackslash}p{2cm}|
}
\hline
\multicolumn{2}{|c|}{\multirow{3}{*}{\centering\textbf{Prior on PBHs}} }
& \multicolumn{4}{c|}{\textbf{Normalization factor } $\textbf{Log}_{\bf 10}~\bm{Z}~(\rm \pm  0.01)$} \\
\cline{3-6}
\multicolumn{2}{|c|}{} &
\multicolumn{3}{c|}{\textrm{Threshold statistics}} &
\multicolumn{1}{c|}{\multirow{2}{*}{\centering \makecell[c]{\textrm{Peak}\\ \textrm{theory}}}} \\
\cline{3-5}
\multicolumn{2}{|c|}{} &
$f_{\rm NL}=0$ &
$f_{\rm NL}=-2$ &
$|f_{\rm NL}|<10$ & \\
\hline\hline

\multicolumn{2}{|c|}{\multirow{1}{*}{\centering Astrophysical}} &-0.31 & -0.31 & -0.50 & -0.35  \\
\hline \hline

\multirow{4}{*}{\centering Sub-solar} & \centering CE  & -1.79 & -1.79 &-1.96 &  -1.91 \\ \cline{2-6}
& \centering ET      & -1.98 & -1.98 & -2.14 &  -2.11 \\ \cline{2-6}
& \centering LVK O5  & -2.57 & -2.57 &-2.67 &  -2.70\\ \cline{2-6}
& \centering LVK O4  &  -2.97 & -2.98 &-3.06 & -3.09 \\
\hline
\end{tabular}
\caption{\label{tab:Znorm}
Normalization factors $Z$ for the different PBH priors, as defined in Eqs.~\eqref{eq:Z_factor_1} and \eqref{eq:Z_factor_2}. 
} \label{tab:Z_factor}
\end{table}

\subsection{Stochastic background from supermassive black holes}
\label{sec:GW_SMBH}

\paragraph{Single SMBH binary.}  Observational evidence supports the presence of SMBHs at the centers of most massive galaxies \cite{Kormendy:2013dxa}. As galaxies evolve through mergers, it is plausible for multiple SMBHs to coexist within a single system. When two of these black holes approach sufficiently, they may form a gravitationally bound binary. The power emitted in GWs by such a binary, under the assumption of circular orbits is given by \cite{Peters:1963ux}
\begin{equation}
\label{eq:quadrupole_Egw_0}
\frac{dE_{\rm GW}}{dt_r} = \frac{32}{5}\frac{c^5}{G}\left( \frac{\pi G \mathcal{M} f_r}{c^3}\right)^{10/3}\,,
\end{equation}
where the chirp mass is defined as $\mathcal{M} \equiv (m_1m_2)^{3/5}/(m_1+m_2)^{1/5}$, and $t_r$ and $f_r$ denote the time and GW frequency in the rest frame of the binary. The latter relate to the observed GW frequency $f$ by $f_r = f(1+z)$ and to the orbital frequency via $f_r = 2f_{\rm orb}$. By equating the radiated power with the rate of orbital energy loss $E_{\rm orb} = -Gm_1m_2/(2R)$ and applying Kepler's third law $w_{\rm orb}^2 = G(m_1+m_2)/R^3$, one obtains the \emph{residence time} spent per logarithmic frequency interval
\begin{equation}
\label{eq:residence_time}
\frac{dt_r}{d\ln f_r} = \frac{5}{96\pi^{8/3}} \left( \frac{c^3}{G\mathcal{M}} \right)^{5/3} f_r^{-8/3}\,.
\end{equation}
Combining this result with Eq.~\eqref{eq:quadrupole_Egw_0}, the emitted GW energy per logarithmic frequency interval in the source frame becomes \cite{Thorne:1987af,Maggiore:2007ulw}
\begin{equation}
\label{eq:quadrupole_Egw}
\frac{dE_{\rm gw}}{d\ln{f_r}}  = \frac{dt_r}{d\ln{f_r}} \cdot \frac{dE_{\rm gw}}{dt_r} = \frac{\pi^{2/3}}{3G}(G \mathcal{M})^{5/3} f_r^{2/3}\,.
\end{equation}
\paragraph{Population of SMBH binaries.}
The total GW energy density per logarithmic frequency interval due to a cosmological population of SMBH binaries, described by a comoving number density $d^2n/dzd\mathcal{M}$, is given by \cite{Sesana:2008mz}
\begin{equation}
\label{eq:rho_gw_lnf}
\frac{d\rho_{\rm GW}}{d\ln f} = \frac{\pi}{4}\frac{c^2}{G}f^2h_c^2(f) = \int_0^{\infty} dz \int_0^{\infty} d\mathcal{M} \frac{d^2n}{dzd\mathcal{M}} \frac{1}{(1+z)} \frac{dt_r}{d\ln f_r} \frac{dE_{\rm gw}}{dt_r}\bigg|_{f_r = f(1+z)}\,.
\end{equation}
Substituting Eq.~\eqref{eq:quadrupole_Egw} into this expression leads to
\begin{equation}
\label{eq:h_c_dn}
h_c^2(f) =  \frac{4}{3\pi^{1/3}c^2} \int_0^{\infty}dz \int_0^{\infty}d\mathcal{M}\left[\frac{(G\mathcal{M})^5}{(1+z)f^4} \right]^{1/3}\frac{d^2n}{dzd\mathcal{M}}\,.
\end{equation}
This results in a characteristic strain spectrum that follows a power-law form~\cite{Phinney:2001di}
\begin{equation}
\label{eq:hc_SMBH-B}
h_c(f) = A_{\rm SMBH} \left( \frac{f}{f_{\rm yr}} \right)^\alpha, \qquad \alpha = -2/3, \quad f_{\rm yr} = 1~\rm yr^{-1} \simeq 31.7~\rm nHz\,.
\end{equation}
The amplitude can be approximated by
\begin{equation}
\label{eq:ASMBH_dn}
A_{\rm SMBH} =  5 \times 10^{-15} \left( \frac{\mathcal{M}_0}{10^9~\rm M_{\mathSun}} \right)^{5/3} \left( \frac{n_0}{2\times 10^{-3}~\rm Mpc^{-3}} \right) I^{1/2}\,,
\end{equation}
with the dimensionless integral
\begin{equation}
I = \int_0^{\infty} \frac{dz}{(1+z)^{1/3}} \int_0^{\infty} d\mathcal{M} \left(\frac{\mathcal{M}}{\mathcal{M}_0} \right)^{5/3} n_0^{-1} \frac{d^2n}{dzd\mathcal{M}}\,.
\end{equation}
Expressing this in terms of the fractional GW energy density (see Eq.~\eqref{eq:Omega_GW_def}), we obtain
\begin{equation}
\label{eq:ASMBH_Omega_GW}
\Omega_{\rm GW}(f) = \Omega_{\rm SMBH} \left( \frac{f}{f_{\rm yr}} \right)^{n_t}, \qquad n_t = 2(1+\alpha) = 2/3\,.
\end{equation}
Similarly, the timing residual power spectrum takes the form
\begin{equation}
\label{eq:ASMBH_S}
S_{R,IJ}(f) = \Gamma_{IJ} \frac{A_{\rm SMBH}}{12\pi^2} f_{\rm yr}^{-3} \left( \frac{f}{f_{\rm yr}} \right)^{-\gamma_{\rm SMBH}}, \qquad \gamma_{\rm SMBH} = 3 - 2\alpha = 13/3\,.
\end{equation}
\paragraph{Discretization and SMBH Priors.} PTA datasets are sampled at discrete frequency intervals $\Delta f = 1/T$, where $T$ is the total observation time. At higher frequencies, the number of binaries emitting per frequency bin (as dictated by Eq.~\eqref{eq:residence_time}) may fall below unity. In such cases, the assumption of a continuous population becomes inaccurate. To address this, the continuous distribution $d^2n/dzd\mathcal{M}$ is replaced by a discrete count $N_{ij}$ of binaries in bins of redshift and chirp mass, transforming integrals into sums \cite{Sesana:2008mz,McWilliams:2012an,Rosado:2015epa,Kelley:2017lek,Middleton:2020asl}
\begin{equation}
\label{eq:discretization}
\int dz\, d\mathcal{M} \cdots \frac{d^2n}{dz\, d\mathcal{M}} \quad \rightarrow \quad \sum_{ij} \cdots N_{ij}\,.
\end{equation}
Furthermore, two scenarios for modeling the SMBH contribution to the SGWB are explored in our analysis:

\begin{enumerate}
    \item[i)] \emph{Free amplitude:} We adopt a power-law spectrum with fixed slope $\gamma_{\rm SMBH}=13/3$, and treat the amplitude $A_{\rm SMBH}$ as a free parameter
\begin{equation}
\label{eq:SMBH_prior1}
  \gamma_{\rm SMBH}=13/3\,,\qquad \log_{10}(A_{\rm SMBH}) \in [-18, -14]\,.
\end{equation}

\item[ii)] \emph{Astrophysical prior:} Following Ref. \cite{NANOGrav:2023hvm}, we draw $A_{\rm SMBH}$ and $\gamma_{\rm SMBH}$ from
a Gaussian distribution with mean  and covariance matrix $\boldsymbol{\mu}_{\rm SMBH} $ and $ \boldsymbol{C}_{\rm SMBH} $ derived from simulated SMBH populations in the \texttt{GWOnly-Ext} catalog
\begin{equation}
\label{eq:SMBH_prior2}
\texttt{GWOnly-Ext:}\qquad \boldsymbol{\mu}_{\rm SMBH} = \begin{pmatrix} -15.6 \\ 4.7 \end{pmatrix}\,, \quad \boldsymbol{C}_{\rm SMBH} = 10^{-1} \times \begin{pmatrix} 2.8 & -0.026 \\ -0.026 & 1.2 \end{pmatrix}\,.
\end{equation}
These priors, accessible via \texttt{PTArcade} \cite{Mitridate:2023oar}, stem from simulations assuming circular, GW-driven binaries with astrophysical assumptions on galaxy properties. Deviations from the $\gamma_{\rm SMBH}=13/3$ slope arise from discretization effects \cite{Rosado:2015epa}. For simplicity, in our analyses we have neglected the non-diagonal terms of the covariance matrix $\boldsymbol{C}_{\rm SMBH}$.
\end{enumerate}

We summarize the above priors in Table~\ref{tab:prior_tab}.

\paragraph{Environmental effects.} Our analysis focuses solely on binaries evolving through GW emission in circular orbits. However, factors such as orbital eccentricity or interactions with the environment (e.g., stellar scattering, gas drag, triple interactions) may reduce the GW signal at low frequencies \cite{Sesana:2013wja,Kelley:2016gse,Burke-Spolaor:2018bvk,Taylor:2021yjx,Ellis:2023dgf,Raidal:2024tui}. These effects can mimic a steeper spectral tilt and overlap with signals expected from other sources like curvature peaks \cite{NANOGrav:2023hvm}. 

Future improvements in PTA sensitivity and the identification of individual binaries \cite{Taylor:2020zpk,NANOGrav:2023tcn} or anisotropies~\cite{Raidal:2024tui,Konstandin:2024fyo,Domcke:2025esw,Konstandin:2025ifn} will help constrain such astrophysical contributions and refine our interpretation of the SGWB.

\subsection{Noise modeling and likelihood construction}

\paragraph{Likelihood function.}
\label{sec:likelihood}
 The arrival times of pulsar signals can be predicted with a precision reaching approximately $100~\rm ns$ \cite{Hobbs:2006cd,Edwards:2006zg,Hobbs:2009yn}, thanks to timing models implemented in dedicated software such as {\tt TEMPO2} \cite{2012ascl.soft10015H} and {\tt PINT} \cite{Luo:2020ksx}, both of which are incorporated into the {\tt ENTERPRISE} analysis framework \cite{enterprise}. The timing residuals resulting from these predictions can be decomposed into a deterministic contribution $\vec{R}_{\rm det}$, optimized by the timing model, and a stochastic component capturing the signal from a SGWB as well as various noise sources
\begin{equation}
\label{eq:R_th}
\vec{R}_{\rm th} = \vec{R}_{\rm det}+ \vec{R}_{\rm GW} + \sum_n \vec{R}_{\rm noise}^{(n)}\,,
\end{equation}
Here, the index $n$ runs over distinct types of noise --- such as red noise, white noise, dispersion measure variations, and jitter noise --- each contributing to the residuals as a stochastic process \cite{NANOGrav:2015qfw,NANOGrav:2015aud,NANOGrav:2023ctt}. These noise terms are modeled as Gaussian processes characterized by covariance matrices $[S_n]$ defined in Fourier space.
The probability of observing a given dataset $\vec{R}_{\rm obs}$, assuming a model parameterized by $\theta$, is then given by a Gaussian likelihood function \cite{NANOGrav:2015qfw,NANOGrav:2015aud,NANOGrav:2023ctt}
\begin{equation}
\label{eq:likelihood}
\mathcal{P}(\vec{R}_{\rm obs}|\theta) = \frac{\exp\left[-\frac{1}{2}\vec{R}_{\rm obs}^{T} C^{-1} \vec{R}_{\rm obs} \right]}{\sqrt{(2\pi)^n \det C}}\,, \qquad \text{with} \quad C = \Delta f \cdot \vec{\mathcal{F}} \left( [S_{R,IJ}] + \sum_n [S_n] \right) \vec{\mathcal{F}}^{-1}\,.
\end{equation}
Here, $[S_{R,IJ}]$ represents the frequency-domain covariance matrix of the SGWB contribution (as given in Eq.~\eqref{eq:S_R_IJ_S_h_ref}), evaluated at sampling frequencies $f_k = k \Delta f$ with $\Delta f = 1/T$ and $T$ the total observation duration. The transformation matrix $\vec{\mathcal{F}}$ converts between Fourier and time domains using sine and cosine components. For simplicity, the deterministic part $\vec{R}_{\rm det}$ has been omitted in the expression above.
For detailed derivations of the likelihood structure, we refer the reader to Refs.~\cite{NANOGrav:2015qfw,NANOGrav:2015aud,NANOGrav:2023ctt}, and for a broader discussion on statistical methods, to Ref.~\cite{Taylor:2021yjx}. In practice, the likelihood in Eq.~\eqref{eq:likelihood} is evaluated numerically using the ${\tt ENTERPRISE}$ software suite \cite{enterprise}, along with its extended modules ${\tt enterprise\_extensions}$ \cite{enterprise_ext}.

\begin{figure*}[t!]
  \centering 
    
    \includegraphics[width=1\linewidth]{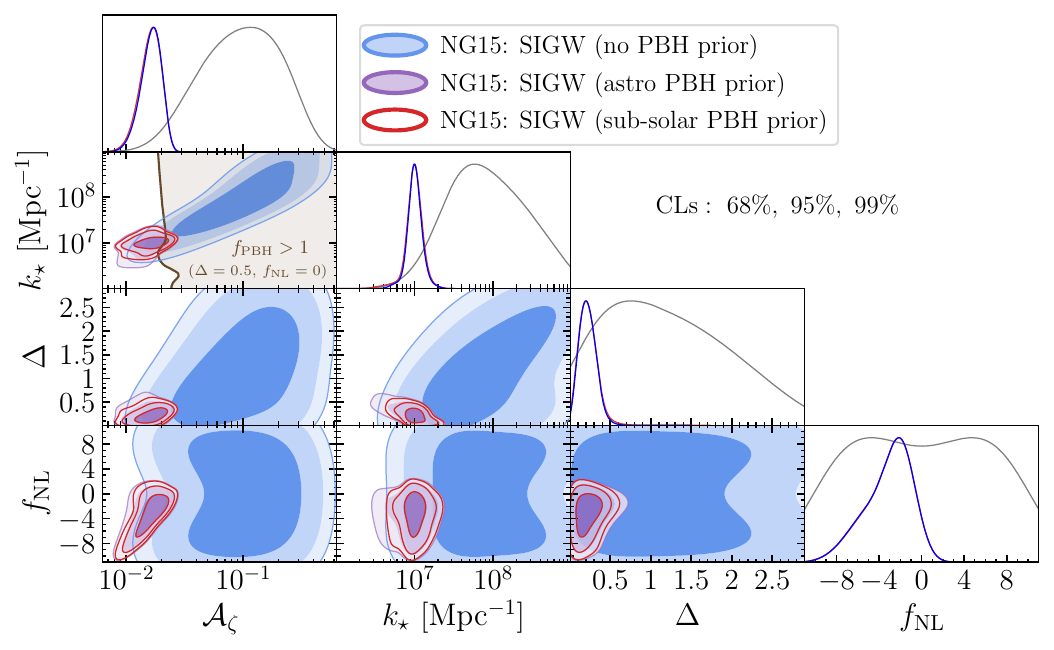}
  \caption{  Posteriors of the model parameters for the SIGW interpretation. The credible regions are given for the case of no constraints on the PBH abundance (orange), assuming priors accommodating the current astrophysical bounds in Eq.~\eqref{eq:PBH_prior_1} (blue) and assuming additionally a sub-solar event detection in CE experiments as described in Eq.~\eqref{eq:PBH_prior_1} (red).}  
  \label{fig:SIGW_posteriors}
\end{figure*}

\paragraph{Posterior sampling.} 
The likelihood function $\mathcal{P}(\vec{R}_{\rm obs}|\theta)$ in Eq.~\eqref{eq:likelihood} measures how compatible the PTA timing-residual data $\vec{R}_{\rm obs}$ are with a given point $\theta$ in parameter space.  We introduce the posterior distribution $\mathcal{P}(\theta|\vec{R}_{\rm obs})$ as the probability density for model parameters $\theta$ after taking the observed data $\vec{R}_{\rm obs}$ into account.  
The latter can be calculated using Bayes theorem 
\begin{equation}
\label{eq:posterior_ref}
\mathcal{P}(\theta|\vec{R}_{\rm obs})
= \frac{\mathcal{P}(\vec{R}_{\rm obs}|\theta)\,\mathcal{P}(\theta)}
{\mathcal{P}(\vec{R}_{\rm obs})}\,.
\end{equation}
The prior $\mathcal{P}(\theta)$, which in our case is defined by Table~\ref{tab:prior_tab}, encodes pre-data expectations, and $\mathcal{P}(\vec{R}_{\rm obs})$ ensures normalization. 
Sampling is performed via Markov Chain Monte Carlo (MCMC), specifically using the {\tt PTMCMC} sampler in {\tt enterprise\_extensions} \cite{justin_ellis_2017_1037579}.Noise parameters are sampled jointly with the signal parameters and subsequently marginalized over by analysing the reduced chains.

\begin{figure*}[t!]
  \centering 
    
    \includegraphics[width=1\linewidth]{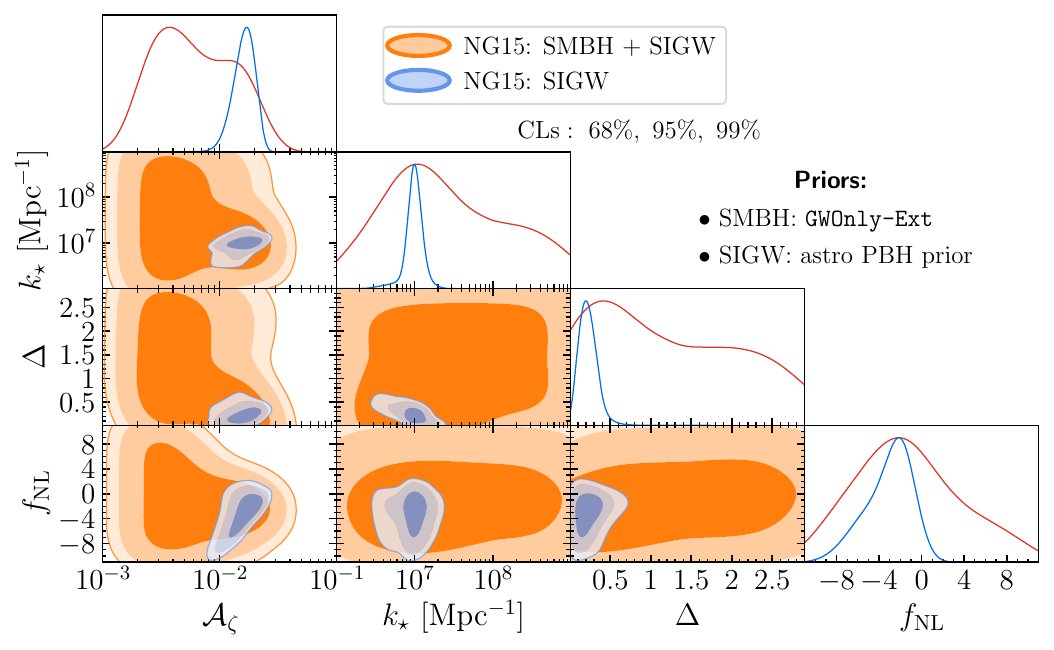}
    \includegraphics[width=1\linewidth]{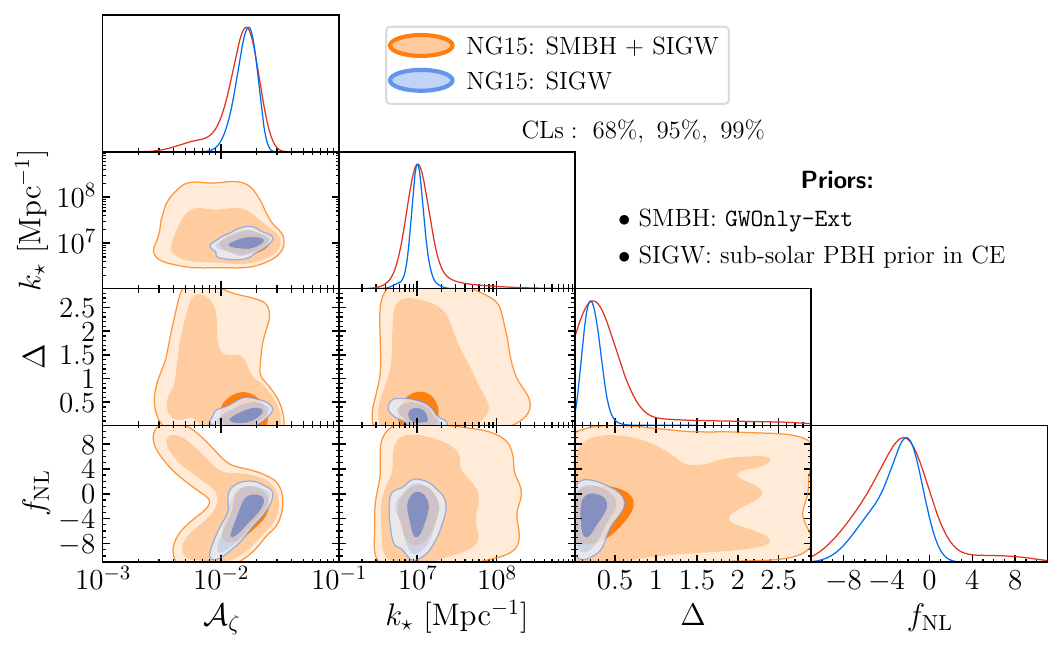}
  \caption{Posteriors of the model parameters for the SIGW+SMBH interpretation. The SMBH contribution is modeled with the \texttt{GWOnly-Ext} prior in Eq. \eqref{eq:SMBH_prior2}. The credible regions (orange) are given in the \textbf{top panel} assuming priors accommodating the current astrophysical bounds in Eq.~\eqref{eq:PBH_prior_1}), and in the \textbf{bottom panel} assuming additionally a sub-solar event detection in CE experiment as described in Eq.~\eqref{eq:PBH_prior_2}. The respective credible regions for the SIGW interpretation (blue) are juxtaposed in both panels for comparison.}
\label{fig:SIGW+SMBH_posteriors}
\end{figure*}

\paragraph{Bayesian model comparison.} 
All Bayesian analysis steps were performed using the {\tt PTArcade} interface \cite{Mitridate:2023oar}, operating in \emph{enterprise} mode to include the full likelihood and HD correlations. We analyzed the first 14 of the NANOGrav 15 (NG15) frequency bins \cite{NANOGrav:2020spf}. From this point onward, we will use “PTA signal” and “NG15” interchangeably. Posterior visualizations were produced using {\tt GetDist} \cite{Lewis:2019xzd}, with $10^6$ MCMC steps and a 25\% burn-in fraction. The posterior distributions of the different hypotheses are presented in Figs.~\ref{fig:SIGW_posteriors}, and \ref{fig:SIGW+SMBH_posteriors} and discussed in the next Section.

To assess which theoretical model better explains the PTA data, one computes the Bayes factor
\begin{equation}
\label{eq:posterior_Bayes}
\mathcal{B}_{Y,X} = \frac{\mathcal{P}(\vec{R}_{\rm obs}|Y)}{\mathcal{P}(\vec{R}_{\rm obs}|X)}\,.
\end{equation}
This quantity can be estimated within {\tt enterprise\_extensions} by introducing a model indicator parameter $m \in \theta$, allowing the posterior ratio to represent the Bayes factor
\begin{equation}
\mathcal{B}_{Y,X} = \frac{\mathcal{P}(m_1|\vec{R}_{\rm obs})}{\mathcal{P}(m_2|\vec{R}_{\rm obs})}\,.
\end{equation}
Using Jeffreys’ scale \cite{1939thpr.book.....J,kass1995bayes}, one interprets $|\log_{10}\mathcal{B}_{Y,X}| < 0.5$ as inconclusive, $>0.5$ as substantial, $>1$ as strong, $>1.5$ as very strong, and $>2$ as decisive evidence.
The Bayes factors from our analyses are reported in Tables~\ref{tab:BF} and~\ref{tab:BF_NG}. 

\begin{figure*}[t!]
    \centering
     \includegraphics[width=0.48\linewidth]{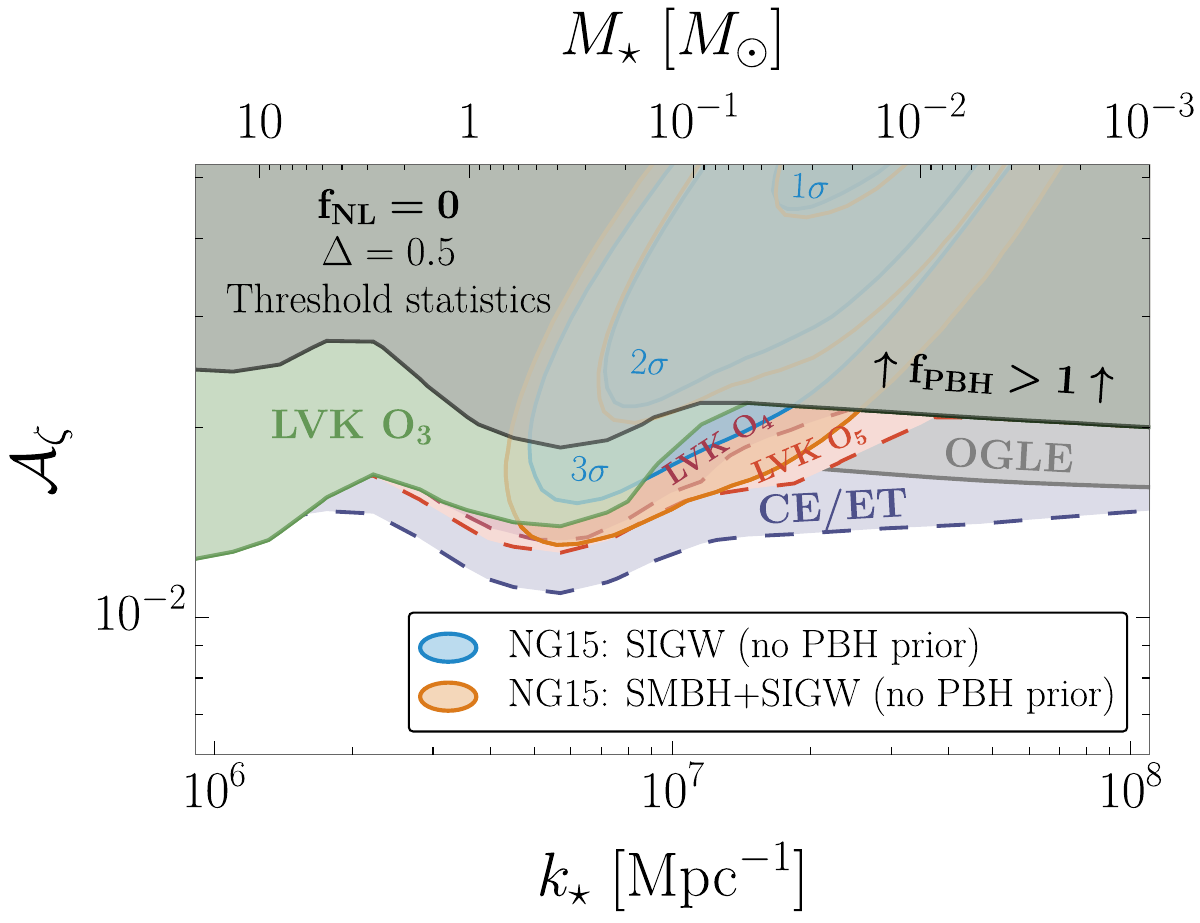}  
    \includegraphics[width=0.48\linewidth]{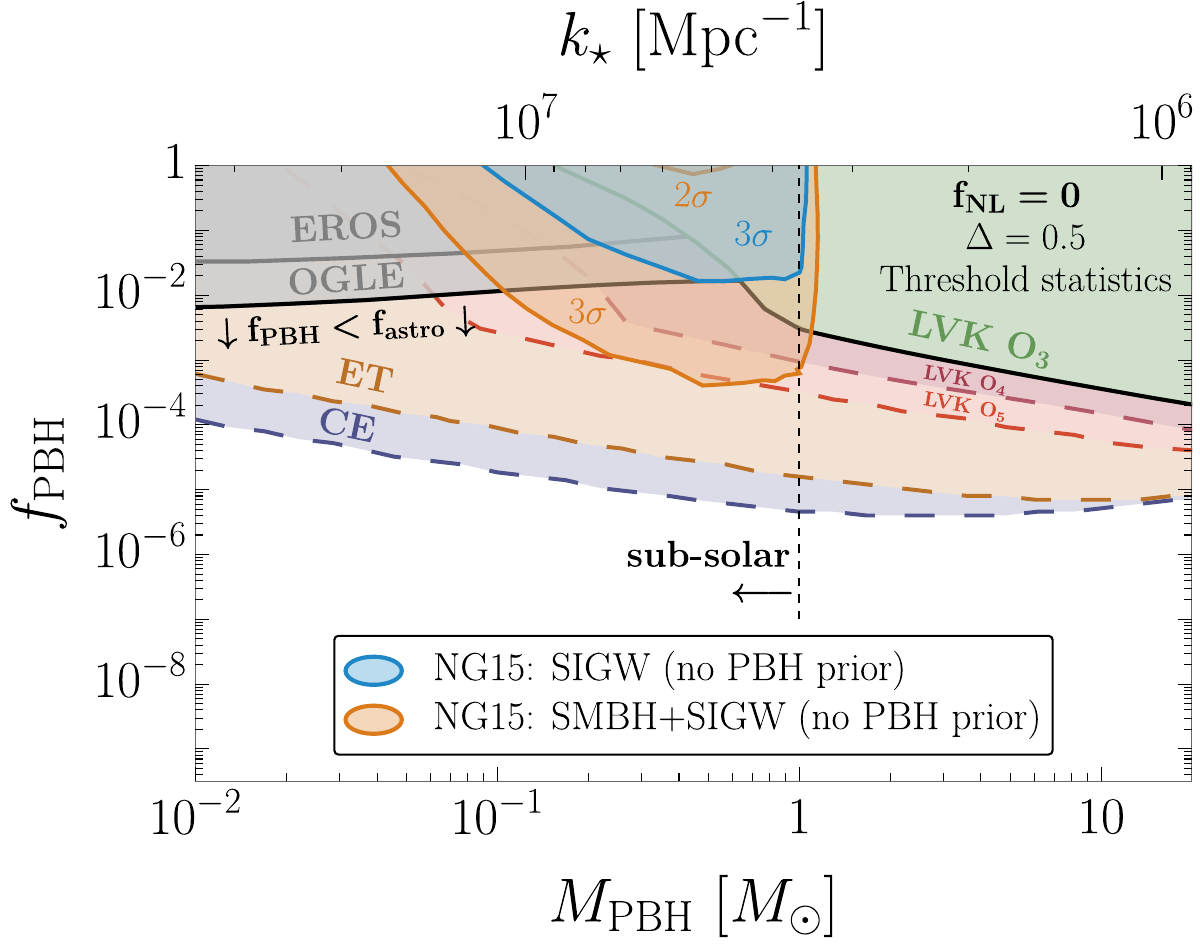}  
    \includegraphics[width=0.48\linewidth]{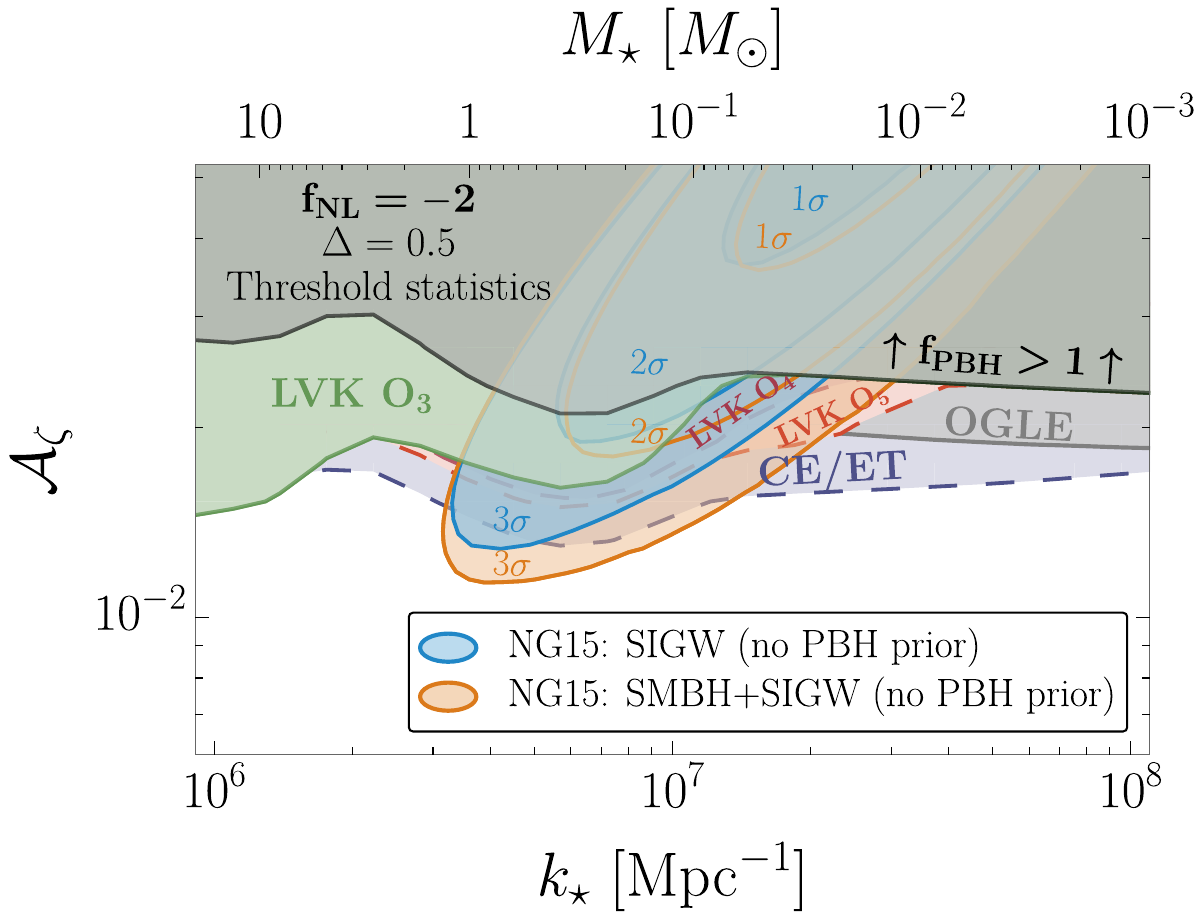} 
    \includegraphics[width=0.48\linewidth]{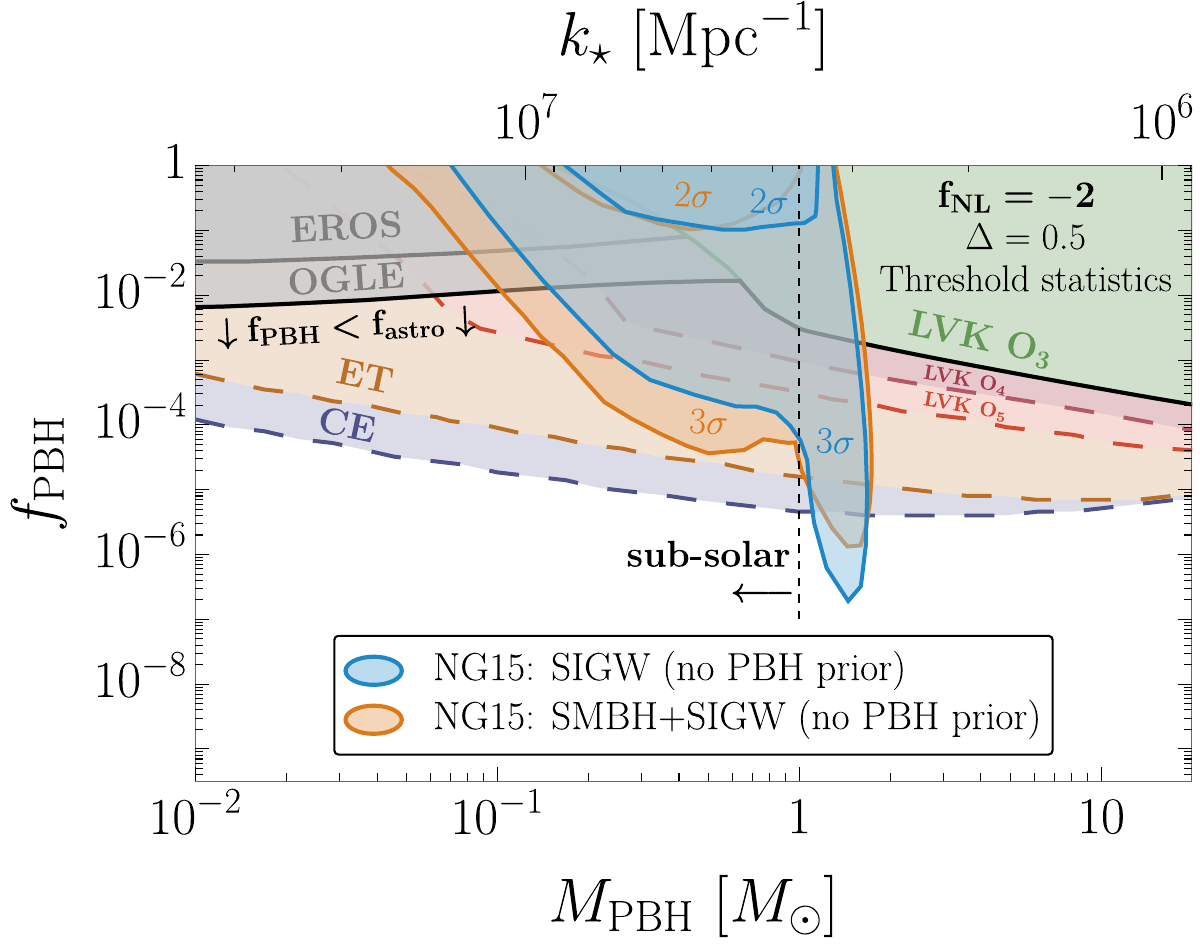} 
     \caption{Posteriors in two-dimensional planes of the log-normal curvature distribution parameters $\mathcal A_{\zeta}$ -- $k_\star$ (\textbf{left panel}), and of the PBH parameter space $f_{\rm PBH}$ -- $M_{\rm PBH}$ (\textbf{right panel}) for the SIGW interpretation (blue) and for the SIGW+SMBH interpretation (orange), with the NG parameter being fixed to $f_{\rm NL} = 0$ (\textbf{top panel}) and $f_{\rm NL} = -2$ (\textbf{bottom panel}). The SMBH contribution is modeled with the \texttt{GWOnly-Ext} prior (see Eq. \eqref{eq:SMBH_prior2}). PBH constraints are not incorporated into the Bayesian analysis but are overlaid for comparison. Contours correspond to $68\%$, $95\%$, and $99\%$ credible regions. These plots reflect the posteriors within each hypothesis individually and do not convey the relative model preference (e.g. SIGW vs SMBH), which is quantified by the Bayes factors in Table~\ref{tab:BF}. 
\label{fig:2D_Posteriors_no_priors}}
\end{figure*}

\begin{figure*}[t!]
    \centering
    \includegraphics[width=0.48\linewidth]{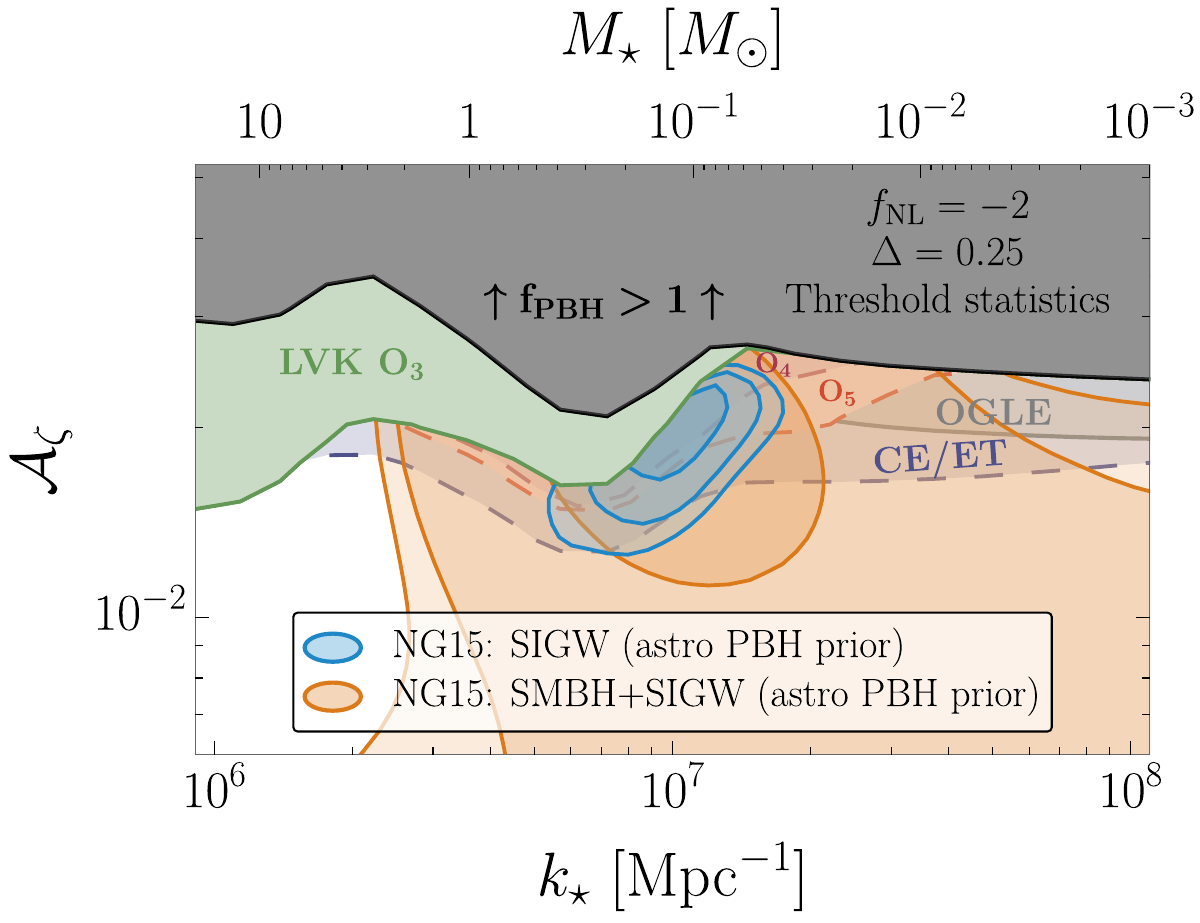}  
    \includegraphics[width=0.48\linewidth]{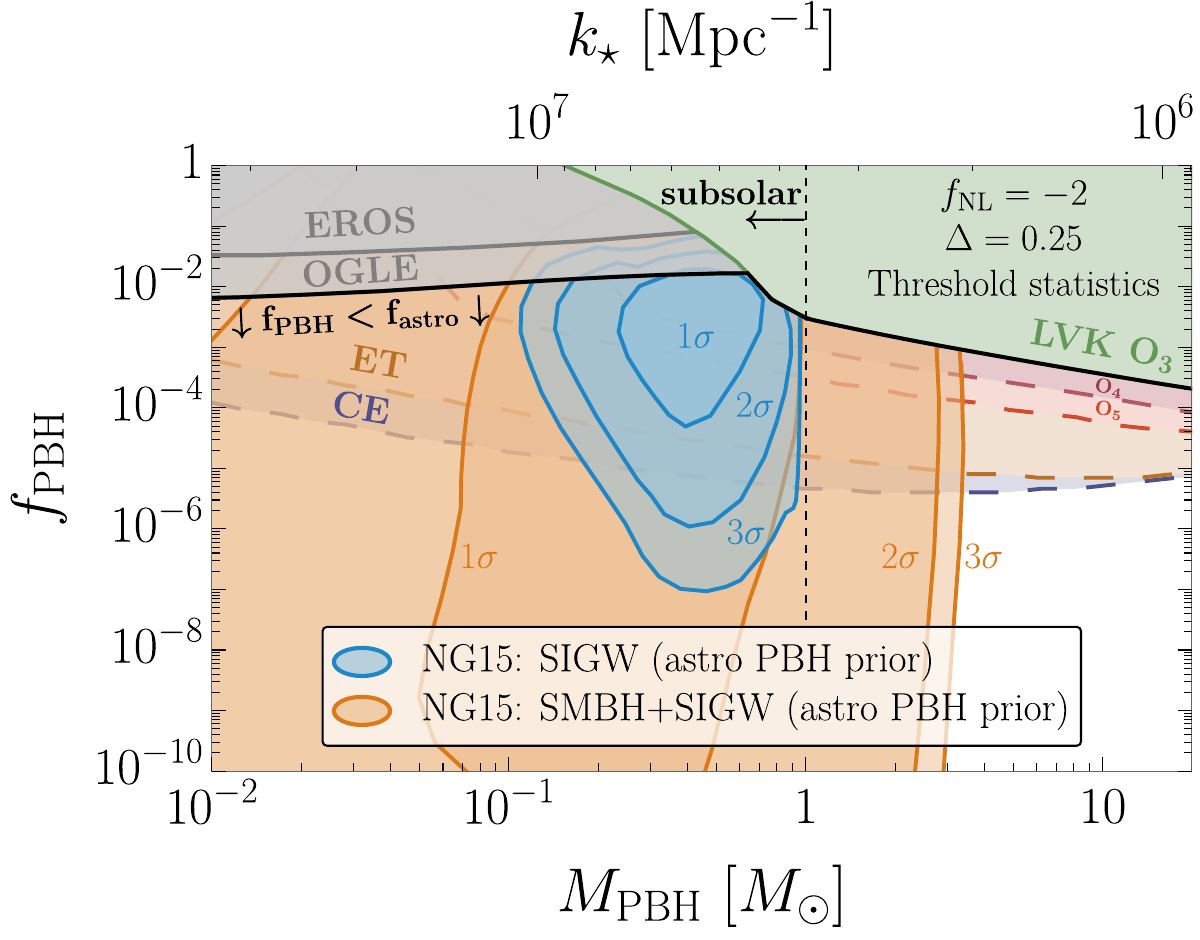}  
    \includegraphics[width=0.48\linewidth]{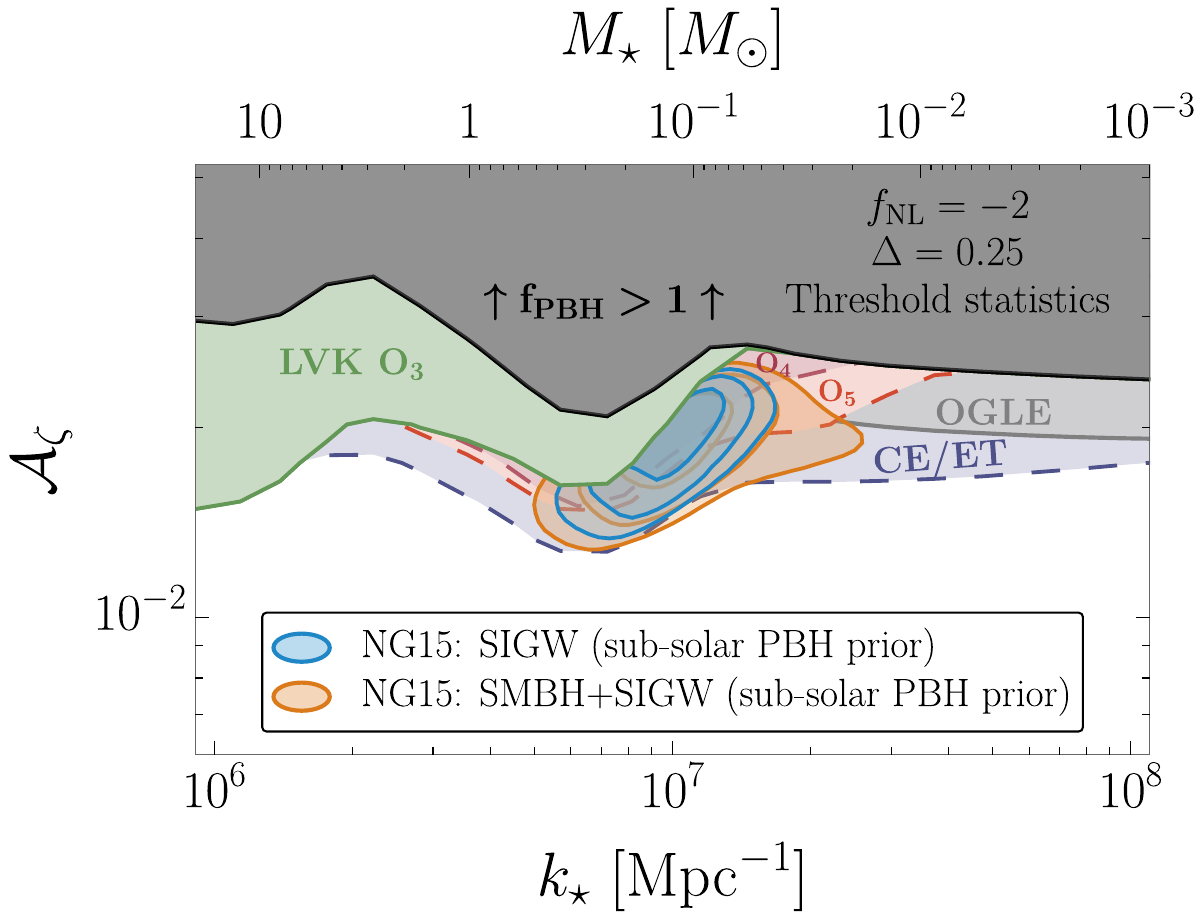}  
    \includegraphics[width=0.48\linewidth]{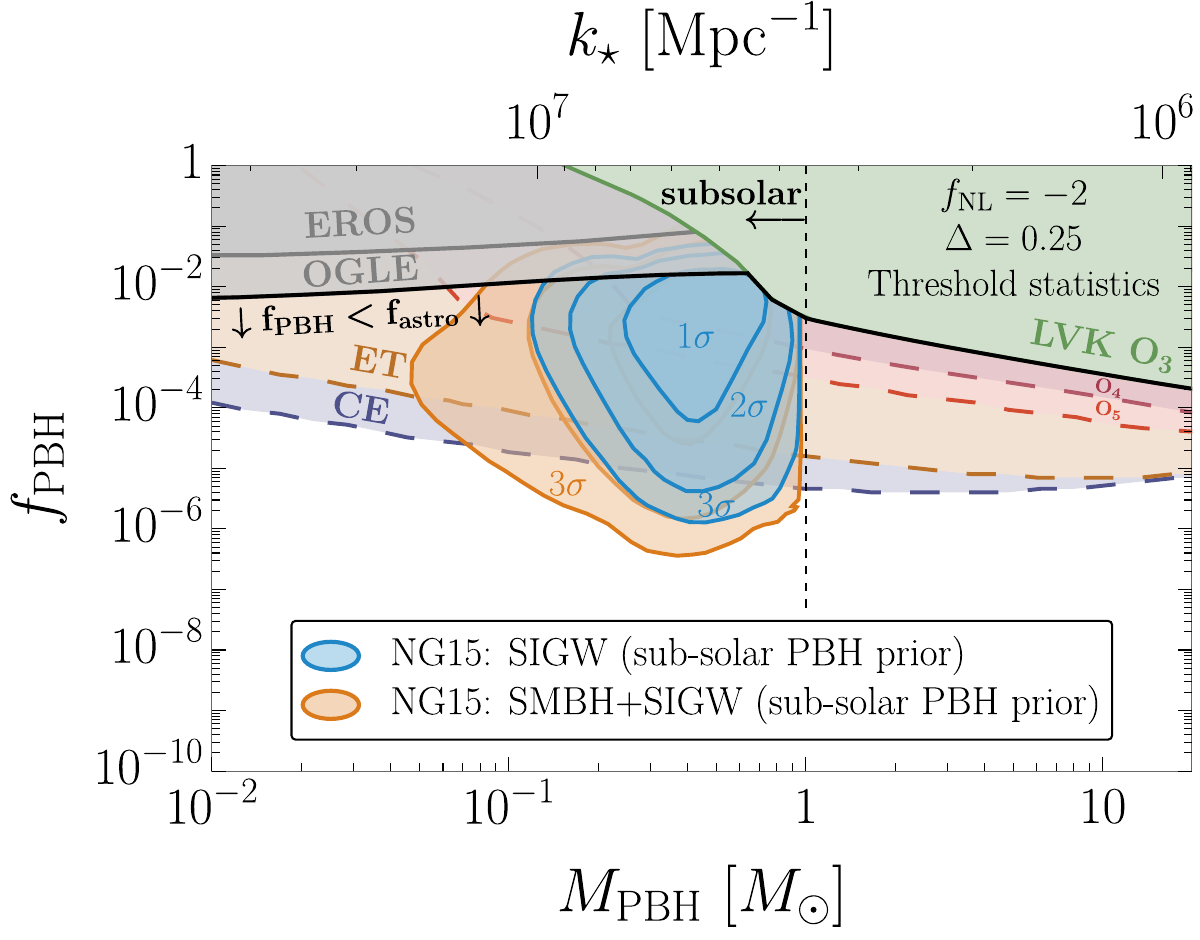}  
     \caption{\label{fig:SIGW+SMBH_2D_plots} Same as Fig.~\ref{fig:2D_Posteriors_no_priors}, but now including either the astrophysical PBH prior $f_{\rm PBH}<f_{\rm astro}$ (\textbf{top panel}) or the sub-solar PBH prior in CE (\textbf{bottom panel}),  in the Bayesian analysis, as described in Eqs.~\eqref{eq:PBH_prior_1} and \eqref{eq:PBH_prior_2}.}
\label{fig:2D_Posteriors_PBH_priors}
\end{figure*}

\section{Binary merger events confront pulsar timing arrays}
\label{sec:results}

\begin{figure*}[t!]
  \centering 
    \includegraphics[width=0.48\linewidth]{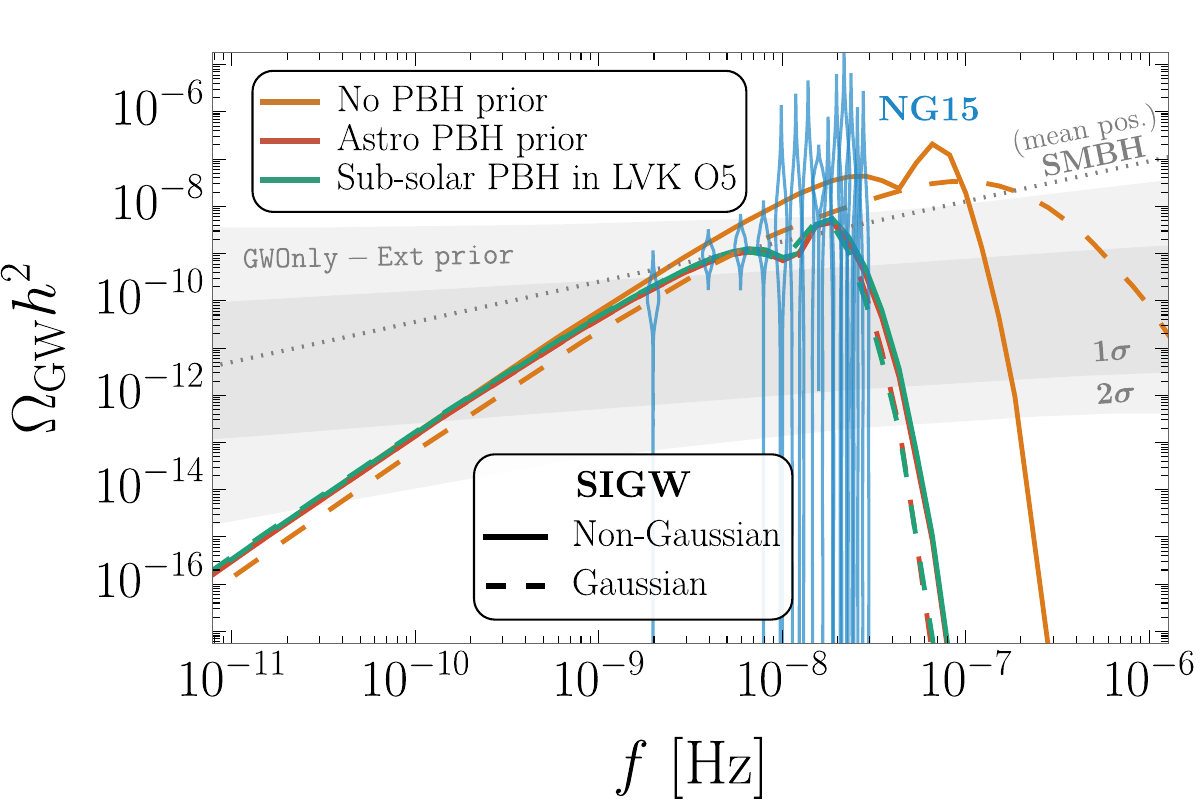}
    \includegraphics[width=0.48\linewidth]{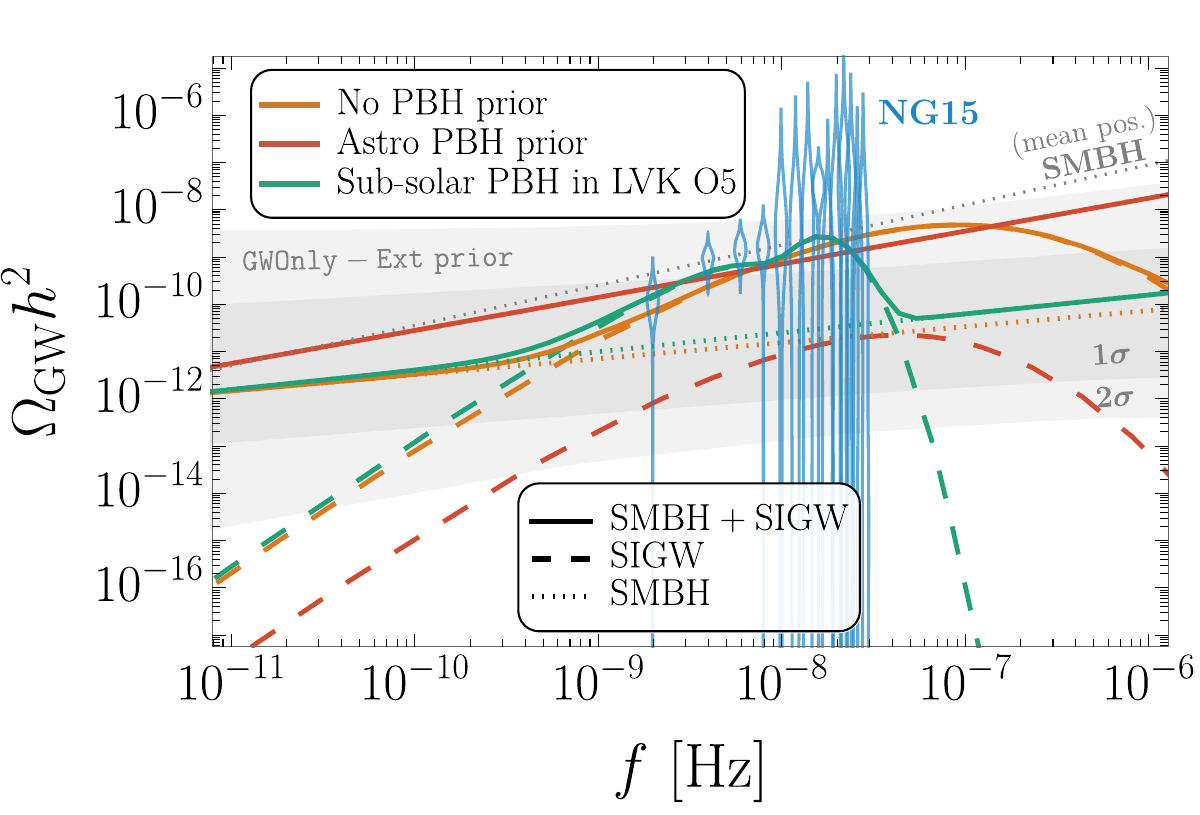}
  \caption{Stochastic GW power spectrum from SIGW (\textbf{left panel}), SIGW+SMBH (\textbf{right panel}) interpretation of the NG15 dataset.  We consider three types of prior choices for the calculation of the PBHs abundance: no priors (orange), the astrophysical bound $f_{\rm PBH}<f_{\rm astro}$ (red), and an hypothesized discovery of sub-solar PBHs in LVKO5 (green). The model parameters $\mathcal{A}_{\zeta}$, $k_\star$, $\Delta$, $f_{\rm NL}$, $A_{\rm SMBH}$ and $\gamma_{\rm SMBH}$ are set to the mean posterior values for each choice (see Appendix~\ref{app:meanPoseterior}). The SMBH contribution to the SIGW+SMBH case (gray shaded bands) is modeled with the \texttt{GWOnly-Ext} prior (see Eq. \eqref{eq:SMBH_prior2}) and we illustrate also the respective fit without any SIGW component (gray dotted).}
  \label{fig:PTA_violins}
\end{figure*}

\begin{table}[ht]
\centering
\renewcommand{\arraystretch}{1.05}
\begin{tabular}{|
    >{\centering\arraybackslash}p{1.1cm}|
    >{\centering\arraybackslash}p{1.4cm}|
    >{\centering\arraybackslash}p{2.2cm}|
    >{\centering\arraybackslash}p{2.7cm}|
    >{\centering\arraybackslash}p{1.5cm}|
    >{\centering\arraybackslash}p{1.7cm}|
    >{\centering\arraybackslash}p{1.8cm}|
    >{\centering\arraybackslash}p{1.3cm}|
}
\hline
\multicolumn{2}{|c|}{\multirow{3}{*}{\centering\makecell[c]{\textbf{Prior on} \\ \textbf{PBHs}}}} &
\multicolumn{1}{c|}{\multirow{3}{*}{\centering\makecell[c]{\textbf{Model X}\\ = SMBH\\only}}} &
\multicolumn{1}{c|}{\multirow{3}{*}{\centering \textbf{Model Y}}} &
\multicolumn{4}{c|}{\centering ${\bf\textbf{Log}_{\bf 10}~\bm{\mathcal{B}}_{Y,X}}~(\pm 0.10)$ } \\
\cline{5-8}
\multicolumn{2}{|c|}{} &
\multicolumn{1}{c|}{} &
\multicolumn{1}{c|}{} &
\multicolumn{3}{c|}{Threshold statistics} &
\multicolumn{1}{c|}{\multirow{2}{*}{\centering\makecell[c]{Peak \\ theory}}} \\
\cline{5-7}
\multicolumn{2}{|c|}{} &
\multicolumn{1}{c|}{} &
\multicolumn{1}{c|}{} &
$f_{\rm NL}=0$ &
$f_{\rm NL}=-2$ &
$|f_{\rm NL}|<10$ &
\multicolumn{1}{c|}{} \\
\hline\hline

\multicolumn{2}{|c|}{\multirow{4}{*}{\centering \makecell[c]{No prior}}} &
\multirow{2}{=}{\centering free amplitude} &
SIGW            & {\cellcolor{ForestGreen!40} $ 1.67 $} & {\cellcolor{ForestGreen!40} $ 1.74 $} & {\cellcolor{ForestGreen!40} $ 1.74 $} & {\cellcolor{ForestGreen!40} $ 1.67 $} \\
\cline{4-8}
\multicolumn{2}{|c|}{} & & 
SIGW+SMBH       & {\cellcolor{ForestGreen!40} $ 1.57 $} & {\cellcolor{ForestGreen!40} $ 1.64 $} & {\cellcolor{ForestGreen!40} $ 1.64 $} & {\cellcolor{ForestGreen!40} $ 1.57 $} \\
\cline{3-8}
\multicolumn{2}{|c|}{} &
\multirow{2}{=}{\centering $\tt{GWOnly\text{-}Ext}$} &
SIGW            & {\cellcolor{ForestGreen!40} $ 2.17 $} & {\cellcolor{ForestGreen!40} $ 2.28 $} & {\cellcolor{ForestGreen!40} $ 2.33 $} & {\cellcolor{ForestGreen!40} $ 2.17 $} \\
\cline{4-8}
\multicolumn{2}{|c|}{} & & 
SIGW+SMBH       & {\cellcolor{ForestGreen!40} $ 2.05 $} & {\cellcolor{ForestGreen!40} $ 2.16 $} & {\cellcolor{ForestGreen!40} $ 2.22 $} & {\cellcolor{ForestGreen!40} $ 2.05 $} \\
\hline\hline

\multicolumn{2}{|c|}{\multirow{4}{*}{\centering Astrophysical}} &
\multirow{2}{=}{\centering free amplitude} &
SIGW            & {\cellcolor{Maroon!50} $ -1.46 $} & {\cellcolor{Maroon!20} $ -0.70 $} & {\cellcolor{Maroon!50} $ -1.33 $} & {\cellcolor{Maroon!50} $ -3.34 $} \\
\cline{4-8}
\multicolumn{2}{|c|}{} & & 
SIGW+SMBH       & $ -0.08 $ & $ -0.01 $ & $ -0.11 $ & $ -0.10 $ \\
\cline{3-8}
\multicolumn{2}{|c|}{} &
\multirow{2}{=}{\centering $\tt{GWOnly\text{-}Ext}$} &
SIGW            & {\cellcolor{Maroon!20} $ -0.90 $} & $ -0.14 $ & {\cellcolor{Maroon!20} $ -0.75 $} & {\cellcolor{Maroon!50} $ -2.78 $} \\
\cline{4-8}
\multicolumn{2}{|c|}{} & & 
SIGW+SMBH       & $ 0.05 $ & $ 0.20 $ & $ 0.06 $ & $ -0.01 $ \\
\hline\hline

\multirow{16}{=}{\centering Sub-\\solar} &

\multirow{4}{=}{\centering CE} &
\multirow{2}{=}{\centering free amplitude} &
SIGW            & $ -0.01 $ & {\cellcolor{ForestGreen!20} $ 0.77 $} & $ 0.07 $ & {\cellcolor{Maroon!50} $ -2.20 $} \\
\cline{4-8}
& & & 
SIGW+SMBH       & $ 0.34 $ & {\cellcolor{ForestGreen!20} $ 0.81 $} & $ 0.32 $ & $ 0.05 $ \\
\cline{3-8}
& &
\multirow{2}{=}{\centering $\tt{GWOnly\text{-}Ext}$} &
SIGW            & {\cellcolor{ForestGreen!20} $ 0.55 $} & {\cellcolor{ForestGreen!40} $ 1.33 $} & {\cellcolor{ForestGreen!20} $ 0.66 $} & {\cellcolor{Maroon!50} $ -1.64 $} \\
\cline{4-8}
& & & 
SIGW+SMBH       & {\cellcolor{ForestGreen!20} $ 0.72 $} & {\cellcolor{ForestGreen!40} $ 1.31 $} & {\cellcolor{ForestGreen!20} $ 0.74 $} & $ 0.17 $ \\
\cline{2-8}

& \multirow{4}{=}{\centering ET} &
\multirow{2}{=}{\centering free amplitude} &
SIGW            & $ 0.16 $ & {\cellcolor{ForestGreen!20} $ 0.95 $} & $ 0.23 $ & {\cellcolor{Maroon!50} $ -2.05 $} \\
\cline{4-8}
& & & 
SIGW+SMBH       & $ 0.45 $ & {\cellcolor{ForestGreen!20} $ 0.96 $} & $ 0.42 $ & $ 0.09 $ \\
\cline{3-8}
& &
\multirow{2}{=}{\centering $\tt{GWOnly\text{-}Ext}$} &
SIGW            & {\cellcolor{ForestGreen!20} $ 0.72 $} & {\cellcolor{ForestGreen!40} $ 1.51 $} & {\cellcolor{ForestGreen!20} $ 0.82 $} & {\cellcolor{Maroon!50} $ -1.49 $} \\
\cline{4-8}
& & & 
SIGW+SMBH       & {\cellcolor{ForestGreen!20} $ 0.86 $} & {\cellcolor{ForestGreen!40} $ 1.48 $} & {\cellcolor{ForestGreen!20} $ 0.87 $} & $ 0.23 $ \\
\cline{2-8}

& \multirow{4}{=}{\centering LVK~O5} &
\multirow{2}{=}{\centering free amplitude} &
SIGW            & {\cellcolor{ForestGreen!20} $ 0.62 $} & {\cellcolor{ForestGreen!40} $ 1.40 $} & {\cellcolor{ForestGreen!20} $ 0.69 $} & {\cellcolor{Maroon!50} $ -1.50 $} \\
\cline{4-8}
& & & 
SIGW+SMBH       & {\cellcolor{ForestGreen!20} $ 0.74 $} & {\cellcolor{ForestGreen!40} $ 1.36 $} & {\cellcolor{ForestGreen!20} $ 0.72 $} & $ 0.20 $ \\
\cline{3-8}
& &
\multirow{2}{=}{\centering $\tt{GWOnly\text{-}Ext}$} &
SIGW            & {\cellcolor{ForestGreen!40} $ 1.18 $} & {\cellcolor{ForestGreen!40} $ 1.96 $} & {\cellcolor{ForestGreen!40} $ 1.28 $} & {\cellcolor{Maroon!20} $ -0.94 $} \\
\cline{4-8}
& & & 
SIGW+SMBH       & {\cellcolor{ForestGreen!40} $ 1.21 $} & {\cellcolor{ForestGreen!40} $ 1.90 $} & {\cellcolor{ForestGreen!40} $ 1.25 $} & $ 0.37 $ \\
\cline{2-8}

& \multirow{4}{=}{\centering LVK~O4} &
\multirow{2}{=}{\centering free amplitude} &
SIGW            & {\cellcolor{ForestGreen!20} $ 0.66 $} & {\cellcolor{ForestGreen!40} $ 1.59 $} & {\cellcolor{ForestGreen!20} $ 0.79 $} & {\cellcolor{Maroon!50} $ -1.24 $} \\
\cline{4-8}
& & & 
SIGW+SMBH       & {\cellcolor{ForestGreen!20} $ 0.77 $} & {\cellcolor{ForestGreen!40} $ 1.52 $} & {\cellcolor{ForestGreen!20} $ 0.79 $} & $ 0.24 $ \\
\cline{3-8}
& &
\multirow{2}{=}{\centering $\tt{GWOnly\text{-}Ext}$} &
SIGW            & {\cellcolor{ForestGreen!40} $ 1.22 $} & {\cellcolor{ForestGreen!40} $ 2.15 $} & {\cellcolor{ForestGreen!40} $ 1.37 $} & {\cellcolor{Maroon!20} $ -0.68 $} \\
\cline{4-8}
& & & 
SIGW+SMBH       & {\cellcolor{ForestGreen!40} $ 1.27 $} & {\cellcolor{ForestGreen!40} $ 2.06 $} & {\cellcolor{ForestGreen!40} $ 1.33 $} & $ 0.40 $ \\
\hline
\end{tabular}

\caption{\label{tab:BF} Bayes factors obtained after running MCMC sampling of the NG15 dataset. We compare the SIGW and SIGW+SMBH interpretations of the PTA signal with the astrophysical one in terms of SMBHs alone. Cells are shaded light/dark green or red when model $Y$ is substantially/strongly favored or disfavored with respect to model $X$, i.e when $|\textrm{Log}_{10}\mathcal{B}_{Y,X}|~>~0.5 ~/ ~1.0$.}
\end{table}

\begin{table*}[ht]
\centering
\renewcommand{\arraystretch}{1.2}
\begin{tabular}{|
    >{\centering\arraybackslash}p{2.5cm}|
    >{\centering\arraybackslash}p{1.7cm}|
    >{\centering\arraybackslash}p{2.6cm}|
    >{\centering\arraybackslash}p{2.5cm}|
    >{\centering\arraybackslash}p{2.5cm}|
}
\hline
\multicolumn{1}{|c|}{\multirow{4}{*}{\makecell[c]{\textbf{Prior} \\ \textbf{on PBHs} }}}
& \textbf{Model X} 
& \multicolumn{1}{c|}{\textbf{Model Y}} 
& \multicolumn{2}{c|}{\multirow{2}{*}{\centering
 ${\bf\textbf{Log}_{\bf 10}~\bm{\mathcal{B}}_{Y,X}}~(\pm 0.01)$ 
}}  \\
\cline{2-3}
&  \multicolumn{1}{c|}{
  \multirow{1}{*}{\multirow{2}{*}{\makecell[c]{Gaussian}}}}
& \multicolumn{1}{c|}{\multirow{2}{*}{\centering Non-Gaussian}}  & \multicolumn{2}{c|}{}  \\
\cline{4-5}
& & &
\multicolumn{1}{c|}{
  \multirow{2}{*}{\makecell[c]{SIGW}}} &\multicolumn{1}{c|}{
  \multirow{2}{*}{\makecell[c]{SIGW+SMBH \\ ($\tt{GWOnly\text{-}Ext}$)}}} \\ \cline{2-3}
& \multicolumn{2}{c|}{\multirow{1}{*}{\shortstack{Threshold statistics}}} 
& & \\
\hline\hline

\multirow{6}{=}{\centering No prior } 
& \multirow{6}{=}{\centering $ f_{\rm NL}=0 $} 
& $ f_{\rm NL} \in [-10,10] $ & {$ 0.14  $} & { $  0.13  $}\\
\cline{3-5}
& & $ f_{\rm NL} = 1 $ & { $ 0.02  $} & { $  0.01  $}\\
\cline{3-5}
& & $ f_{\rm NL} = -1 $ & {$ 0.00  $} & { $  0.00 $}\\
\cline{3-5}
& & $ f_{\rm NL} = -2 $ & {$ 0.01  $} & { $  0.01 $}\\
\cline{3-5}
& & $ f_{\rm NL} = -3 $ & {$ 0.03  $} & { $  0.03 $}\\
\cline{3-5}
& & $ f_{\rm NL} = -4 $ & {$ 0.04  $} & { $  0.04 $}\\
\hline\hline

\multirow{6}{=}{\centering Astrophysical } 
& \multirow{6}{=}{\centering $ f_{\rm NL}=0 $} 
& $ f_{\rm NL} \in [-10,10] $ & {$ 0.29  $} & { $  0.05  $}\\
\cline{3-5}
& & $ f_{\rm NL} = 1 $ & {\cellcolor{Maroon!40} $ -1.17  $} & { $  -0.10  $}\\
\cline{3-5}
& & $ f_{\rm NL} = -1 $ & {$ 0.47  $} & { $ 0.07  $}\\
\cline{3-5}
& & $ f_{\rm NL} = -2 $ & {\cellcolor{ForestGreen!20} $ 0.72  $} & { $  0.13 $}\\
\cline{3-5}
& & $ f_{\rm NL} = -3 $ & {\cellcolor{ForestGreen!20} $ 0.51  $} & { $  0.03 $}\\
\cline{3-5}
& & $ f_{\rm NL} = -4 $ & {$ 0.34  $} & { $  0.05  $}\\
\hline\hline

\multirow{6}{=}{\centering Sub-solar in CE } 
& \multirow{6}{=}{\centering $ f_{\rm NL}=0 $} 
& $ f_{\rm NL} \in [-10,10] $ & {$ 0.30  $} & { $  0.21 $}\\
\cline{3-5}
& & $ f_{\rm NL} = 1 $ & {\cellcolor{Maroon!40} $ -1.18 $} & {\cellcolor{Maroon!40} $ -0.56  $} \\
\cline{3-5}
& & $ f_{\rm NL} = -1 $ & {$ 0.47  $} & { $  0.34  $}\\
\cline{3-5}
& & $ f_{\rm NL} = -2 $ & {\cellcolor{ForestGreen!20} $ 0.73  $} & {\cellcolor{ForestGreen!20} $ 0.54  $}\\
\cline{3-5}
& & $ f_{\rm NL} = -3 $ & {\cellcolor{ForestGreen!20} $ 0.52  $} &  { $  0.36  $} \\
\cline{3-5}
& & $ f_{\rm NL} = -4 $ & {$ 0.35  $} & { $  0.21 $} \\
\hline
\end{tabular}
\caption{Same as Table \ref{tab:BF}, but here the comparison is the between the Gaussian limit and a range of values for the NG parameter $f_{\rm NL}$ of the SIGW and SIGW+SMBH interpretations of the PTA signal.}
\label{tab:BF_NG}
\end{table*}

In this section, we combine the information from transient gravitational-wave events detected by ground-based laser interferometers with the SGWB observed by PTAs. Specifically, we confront hypotheses for interpreting the PTA data by setting up two distinct comparisons. The first asks whether the PTA signal can be explained entirely by SIGW rather than by SMBH binaries alone. The second asks whether a \emph{combination} of SIGW and SMBH binaries (which we denote as SIGW+SMBH) provides a better explanation than SMBH binaries by themselves. Even though SMBH binaries constitute a genuine astrophysical background, the size of their contribution to the PTA signal is not known with high accuracy~\cite{NANOGrav:2023hvm,NANOGrav:2023hfp,Antoniadis:2023zhi}. In this way, our aim is to probe whether the PTA observations would necessarily point to new physics once a sub-solar PBH population is discovered in future kHz GW observatories.

For the SIGW and SIGW+SMBH hypotheses, we examine three prior choices for the treatment of PBH abundance, as outlined in Sec.~\ref{sec:GW_curvature}: 
(i) a \emph{no PBH prior}  case, which does not penalize any PBH abundance and serves as a reference baseline; 
(ii) the \emph{astrophysical PBH prior}, which incorporates the fact that the scalar density fluctuations needed to fit NG15 would otherwise lead to an excessive PBH population --- overclosing the Universe, producing microlensing signals, or yielding merger events already excluded by current kHz GW observatories such as LVK. The corresponding excluded region in the $f_{\rm PBH}$–$M_{\rm PBH}$ plane is shown in Fig.~\ref{fig:future_reach}, above the thick black line and denoted $f_{\rm PBH} < f_{\rm astro}$ (see Eq.~\eqref{eq:PBH_prior_1} for further details); 
(iii) the \emph{sub-solar PBH prior in LVK, CE, and ET}, which enforces that the PBH population should be detectable in LVK, CE, and ET (see Eq.~\eqref{eq:PBH_prior_2}).

As emphasized in Sec.~\ref{sec:PBH_formation}, the main theoretical uncertainty affecting the inferences drawn in this paper lies in the computation of the PBH abundance, due to the existence of two distinct methods: threshold statistics and peak theory. In what follows, our comments are based on the results calculated with threshold statistics unless explicitly noted otherwise. This choice is motivated by the fact that, as shown in Table~\ref{tab:BF}, the calculation using peak theory cannot at present convincingly distinguish new physics scenarios from the astrophysical background, once phenomenologically motivated priors are taken into account. This is mainly due to the fact that we rely on a formulation of peak theory that does not incorporate NGs --- an ingredient that plays a crucial role in obtaining a phenomenologically viable best fit. As a result, the comparison is somewhat asymmetric, and our conclusions should be understood with this caveat.

\paragraph{Impact of priors.}

We present the posteriors of our Bayesian analysis for the SIGW and SIGW+SMBH hypotheses in Figs.~\ref{fig:SIGW_posteriors} and~\ref{fig:SIGW+SMBH_posteriors}, respectively, and present the $68\%$, $95\%$, and $99\%$ credible regions obtained without imposing PBH priors (choice~i). A central question is how well the PTA data can be described by these interpretations compared to the astrophysical SGWB in terms of SMBH binaries. As reflected by the Bayes factors in Table~\ref{tab:BF}, we find that in this case (case i), the SIGW hypothesis provides a much better fit to the PTA data than SMBHs irrespective of how their SGWB is modeled. SIGW is also slightly preferred over SIGW+SMBH. The latter gains flexibility from its additional parameters, as illustrated by its broader preferred regions when compared directly with SIGW in Figs.~\ref{fig:SIGW+SMBH_posteriors}, \ref{fig:2D_Posteriors_no_priors}, \ref{fig:2D_Posteriors_PBH_priors}, but this freedom is penalized in the Bayesian analysis.

However, we notice that the above solution is already almost entirely excluded by the combination of LVK O3 and microlensing probes, and therefore drawing conclusions from such comparisons while ignoring the respective physical priors is misleading. We visualize this in the left panel of Fig.~\ref{fig:2D_Posteriors_no_priors} by projecting the posteriors onto the $\mathcal{A}_\zeta$–$k_\star$ plane, as well as onto the corresponding PBH parameter space $f_{\rm PBH}$–$M_{\rm PBH}$, while fixing the other parameters close to their mean posterior values. This highlights the tension between interpreting the PTA data solely in terms of SIGW and the independent searches for PBHs. Consequently, we repeat the analysis including the astrophysical bounds (choice~ii). In this case, the credible regions for SIGW accumulate below the constraints and shrink as can be seen in the top panel of Fig. \ref{fig:2D_Posteriors_PBH_priors}, while the SMBH+SIGW regions are enlarged and stretch to very low $f_{\rm PBH}$ reflecting the greater freedom of the model. Importantly, both the SIGW and SIGW+SMBH interpretations are no longer favored over the SMBHs one in this case.\footnote{We note that, the tension on the SIGW interpretation due to the astrophysical bound on PBH can be relaxed by increasing the stiffness of the universe during collaspe $1/3<w < 1$~\cite{Balaji:2023ehk}.}

We then include as a prior the condition that sub-solar PBH populations lie within the detectability region of future experiments (choice~iii). This addition overturns the preference introduced by astrophysical priors, restoring a clear advantage for SIGW over SMBHs. In fact, the preferred regions for SIGW and SIGW+SMBH nearly overlap, rendering the SMBH contribution effectively redundant, as shown in the bottom panel of Fig.~\ref{fig:2D_Posteriors_PBH_priors}. Moreover, SIGW performs slightly better, since the relevant normalization parameter $Z$ (see Eq.~\eqref{eq:Z_factor_2}) is smaller, leading to a larger Bayes factor. The effect persists for sub-solar detections across all experiments, although it becomes less pronounced for those planned further in the future, as their enlarged sensitivity further decreases $Z$. Conversely, this also implies that if a discovery were made in the LVK O5 run, the SIGW hypothesis would become \emph{strongly} favored over the SMBH interpretation. 

We find that the SIGW posterior computed with the sub-solar PBH prior shows no significant deviation from the one obtained with the purely astrophysical PBH prior, see e.g. Fig.~\ref{fig:SIGW_posteriors}. This simply reflects the fact that the parameter space compatible with a sub-solar PBH signal coincides with the region that yields the optimal PTA fit in terms of GWs.

As expected the relative strength of these preferences depends on the modeling of the astrophysical SGWB. When we impose the more constraining $\tt{GWOnly\text{-}Ext}$ prior, the preference for SIGW or SIGW+SMBH over SMBHs is always stronger. At the same time, there is little difference between the SIGW and SIGW+SMBH interpretations themselves given a specific SMBH prior. Finally, it is worth noticing that if we use the $\tt{GWOnly\text{-}Ext}$ prior, even the Gaussian peak theory calculation lies on the verge of showing a \emph{substantial} preference in favor of SIGW+SMBH over SMBHs in the case of LVK O4 prior, i.e. $|\textrm{Log}_{10}\mathcal{B}|=0.4 \pm 0.1 \sim 0.5$.  

In Fig.~\ref{fig:PTA_violins}, we illustrate the SIGW and SIGW+SMBH spectra obtained using the posterior mean values listed in Table~\ref{tab:meanPoseterior_SIGW} for the three prior choices (resp. orange, red, green), as well as the GW spectrum from SMBH mergers only (dashed), obtained under the assumption of the stricter $\tt{GWOnly\text{-}Ext}$ prior. The visual comparison confirms the previous observation, namely that SIGW and SIGW+SMBH interpretations either with the unphysical case of no priors but also with the sub-solar detectability prior provide better fits than SMBHs to the NG15 violin plots. Furthermore, regarding the astrophysical prior, the reduced fit quality of the SIGW explanation relative to SMBHs is directly visible in the spectra, whereas in the SIGW+SMBH it is not, because the lower preference compared to SMBH alone originates mainly from the Bayesian penalty associated with the additional parameters.

Additionally, as discussed in Sec.~\ref{sec:GW_SMBH}, other exotic mechanisms could in principle lead to sub-solar black holes which originate from large curvature perturbations. We can evaluate how future observations could exclude these scenarios using different detection criteria based on specific SNR and redshift cuts. Comparing these criteria as illustrated in Fig.~\ref{fig:future_reach}, with the posteriors in the lower left panel of Fig.~\ref{fig:2D_Posteriors_PBH_priors}, we find that under the most permissive assumptions the $68\%$ credible region lies partly within the reach of LVK O5 and ET and fully within CE’s reach. Stricter criteria, however, progressively restrict this sensitivity, and in the most conservative case only CE remains capable of probing the entire $68\%$ credible region. This implies that even if LVK O5 were to report sub-solar black hole merger events, confirming their primordial origin would still require the improved sensitivity of future observatories such as ET and CE.

 \paragraph{Impact of non-Gaussianities.} 

Table~\ref{tab:BF_NG} presents the impact of introducing NGs by comparing the Bayes factors of the SIGW intepretation with and without NGs. Without imposing astrophysical priors (choice~i), the Bayesian analysis does not show a significant preference between the two hypotheses, because the NG15 data can be fitted equally well.
However, as already discussed in the previous subsection and shown clearly in the top panel of Fig.~\ref{fig:2D_Posteriors_no_priors}, the  $99\%$ and $95\%$ preferred regions for the SIGW and SIGW+SMBH scenarios, respectively, are excluded by the overclosure constraint in the Gaussian limit. Yet, the bottom panel of this figure provides a first indication of the usefulness of negative NGs: the $99\%$ credible level of the SIGW interpretations extends into the detectable sub-solar regime allowed by current constraints.

We then impose the astrophysical prior (choice~ii), and find that the NG hypothesis with positive $f_{\rm NL}$ is always strongly \emph{disfavored} relative to the Gaussian case, whereas negative NGs are instead \emph{favored} if $f_{\rm NL}\sim-2$, consistent with the findings of Refs.~\cite{Franciolini:2023pbf,Iovino:2024tyg}. The reason is that negative $f_{\rm NL}$ suppresses PBH formation while leaving the SIGW spectrum largely unaffected, thereby alleviating the tension with the PBH bounds. The preference for $f_{\rm NL}\sim -2$ reflects the behavior of the overclosure bound on $\mathcal{A}_\zeta$ as a function of $f_{\rm NL}$, which peaks near this value (upper right panel of Fig.~\ref{fig:fPBH_log_normal_DM}, see also the discussion below Eq.~\eqref{eq:f_PBH_final} and Fig.~\ref{fig:why_fnl_m2}). This trend persists when we further impose the prior that the PBH population lies within the detection range of future experiments.

\section{Conclusions}
\label{sec:conclusion}

Gravitational waves provide unprecedented access to physics in both the early and late universe, although their origin remains uncertain. A promising strategy to disentangle the nature of the underlying source(s) is to scrutinize models that yield correlated predictions across different frequency bands, allowing independent experiments to test the same underlying physics from complementary perspectives. A particularly compelling example is the scenario in which large curvature perturbations simultaneously generate sub-solar PBHs and a stochastic SIGW background at nHz frequencies, with the PBHs later forming binaries whose mergers give rise to transient signals in the Hz–kHz band, thereby linking the two frequency regimes. 

A reliable way to distinguish between such models is through a Bayesian framework that enables quantitative model identification. 
In this work we have applied such an analysis by confronting PTA data with the sensitivity of upcoming LVK runs and future gravitational-wave observatories such as the Einstein Telescope and Cosmic Explorer to signals from sub-solar PBH mergers (as shown in Fig.~\ref{fig:future_reach}). The obtained unified assessment of the scenario, against the standard astrophysical explanation in terms of SMBH binaries, is summarized in Table~\ref{tab:BF}. Our results show that the preferred interpretation depends crucially on the choice of priors. 
For instance, if priors are ignored, the PTA data appear to be well described by SIGW; yet this result is misleading, since the corresponding best-fit region lies in parameter space already excluded by existing astrophysical bounds. Including these bounds as priors removes the apparent preference for a primordial origin, and places the purely astrophysical explanation in terms of SMBH binaries, or a mixed scenario where both SIGW and SMBH binaries contribute, on comparable footing. This contrasts with the nominal Bayesian analysis in Ref.~\cite{NANOGrav:2023hvm}, which assumed no priors on the PBH abundance.

However, requiring that sub-solar PBHs fall within the reach of future observatories restores a predictive and testable advantage for the primordial-only scenario. In Bayesian terms, this happens because the requirement significantly reduces the viable parameter space while still yielding a rather good fit, thereby increasing the evidence for the model relative to the purely astrophysical hypothesis. Importantly, negative non-Gaussianities play a central role in this picture: by suppressing PBH abundances without altering the SIGW spectrum, they reconcile PTA data with independent constraints and extend the parameter space into the detectable regime.
Taken together, these results show that the observation of sub-solar PBH mergers in future experiments would provide decisive evidence for a primordial origin of the PTA signal. 
In turn, the continued absence of such signals will progressively restrict the viable parameter space of the mechanism and imply that, if it exists at all, the SIGW contribution to the PTA background must be subdominant with respect to the SMBH component. 

Additionally, we stress that the quantitative conclusions remain subject to significant theoretical systematics associated with the choice of PBH formation prescription. As discussed in Sec. \ref{sec:PBH_formation} and \ref{sec:results}, peak theory and threshold statistics yield systematically different PBH abundances—with peak theory typically predicting larger abundances. Consequently, some parameter regions that appear allowed under a Press–Schechter estimate become excluded once peak theory is used. As future PTA and interferometer data become more precise, a correspondingly improved theoretical treatment will be required, in particular one that incorporates primordial non-Gaussianities consistently within peak theory.

We also note that, following Ref.~\cite{NANOGrav:2023hvm}, we have modeled the GW spectrum from SMBH binaries assuming circular, GW-driven evolution. In practice, additional orbital energy losses due to environmental interactions (e.g. stellar scattering, gas drag, or triple interactions), as well as highly eccentric binaries, can suppress the GW signal at low frequencies \cite{Sesana:2013wja,Kelley:2016gse,Burke-Spolaor:2018bvk,Taylor:2021yjx,Ellis:2023dgf,Raidal:2024tui}. Such effects would strengthen the SMBH-binary interpretation of the PTA signal by increasing its Bayes factor relative to primordial scenarios.

Looking forward, from the theoretical side several candidate sources for the PTA signal already face strong phenomenological constraints, e.g. inflationary blue-tilted spectra are disfavored by CMB and large-scale structure bounds~\cite{Vagnozzi:2023lwo},  global cosmic strings are excluded by limits from Big-Bang nucleosynthesis~\cite{Servant:2023mwt} and the SGWB due to the merger of supermassive PBH are disfavored by the combination of large-scale structure and CMB bound~\cite{Depta:2023qst,Gouttenoire:2023nzr,Clesse:2024epo,Ivanov:2025pbu,Gerlach:2025vco}. 
Beyond the curvature-peak mechanism studied in this work, other new-physics scenarios remain viable and in some cases highly predictive. 
These include first-order phase transitions~\cite{Addazi:2023jvg,Bringmann:2023opz,Gouttenoire:2023bqy,Ellis:2023oxs,Lewicki:2024ghw}, collapse of domain wall networks~\cite{Ferreira:2022zzo,Kitajima:2023cek,Gouttenoire:2023ftk,Gouttenoire:2025ofv,Lu:2024ngi,Lu:2024szr}, local cosmic strings~\cite{Blasi:2020mfx,Ellis:2023tsl,Ellis:2020ena,Buchmuller:2020lbh,Buchmuller:2023aus,Kume:2024adn}, and axion–gauge field dynamics~\cite{Ratzinger:2020koh,Madge:2023dxc,Geller:2023shn,Murai:2023gkv,Unal:2023srk}.
Notably, the first two scenarios above can also link the PTA signal to the production of solar- or sub-solar PBHs, placing them within the reach of current and future kHz GW observatories as well (see Refs.~\cite{Gouttenoire:2023bqy,Ellis:2023oxs,Lewicki:2024ghw} and Refs.~\cite{Ferreira:2022zzo,Kitajima:2023cek,Gouttenoire:2023ftk}, respectively). In this sense, multi-band searches for sub-solar black holes will serve not only as a broader discriminator among competing theoretical explanations of the PTA background. Studying the Bayesian comparison of these models under phenomenologically motivated priors will therefore be essential to establish which of them, if any, provides the most consistent explanation of the PTA signal.

Finally, the relatively high collapse threshold for PBH formation ($\delta_c \gtrsim 0.5$) implies that not every enhancement in scalar perturbations will efficiently produce PBHs. By contrast, much smaller overdensities ($\delta \gtrsim 10^{-4}$) can seed ultra-compact dark-matter minihalos~\cite{Ricotti:2009bs}, adding a clumpy substructure component to the otherwise smooth halo population expected in standard $\Lambda$CDM. This consideration has been used to argue that several primordial interpretations of the PTA background are challenging to reconcile with dark matter as a thermal relic~\cite{Liu:2023tmv,Gouttenoire:2025wxc}. Strengthening astrophysical probes of these objects --- via stellar dynamics~\cite{Buschmann:2017ams}, CMB measurements~\cite{Kawasaki:2021yek}, gravitational lensing~\cite{Delos:2023fpm}, and potential direct signatures in PTAs~\cite{Lee:2020wfn} --- would be crucial for breaking degeneracies between new physics explanations and an SMBH origin of the PTA signal.

\paragraph*{Added note}

Since the completion of this work, the LVK collaboration has released 
the O4 results~\cite{LIGOScientific:2025slb}, which reported no evidence for sub-solar mass PBHs. We have 
not incorporated the O4 exclusion region into our astrophysical prior in Eq. \eqref{eq:PBH_prior_1}, 
as a careful reanalysis of the newly released data is ongoing. Nevertheless, given the modest increase in sensitivity relative 
to O3 in the relevant mass range, we do not expect our main conclusions to be 
significantly affected. 

\section*{Acknowledgements}

We are particularly grateful to Sam Young for extensive discussions on various subtle technical points. It is a pleasure also to thank Virgile Dandoy, Gabriele Franciolini, Qing-Guo Huang, Antonio J. Iovino, David Kaiser, Anupam Ray, Ville Vaskonen, Shao-Jiang Wang, and Noah E. Wolfe for clarifying discussions. We further thank Antonio J. Iovino for insightful comments on the manuscript and  Tomer Volansky for facilitating the use of the computational cluster from his research group in Tel Aviv University. Y.G. acknowledges support by the Cluster of Excellence ``PRISMA+'' funded by the German Research Foundation (DFG) within the German Excellence Strategy (Project No. 390831469). 
S.T. was supported by the Office of High Energy Physics of the US Department of Energy (DOE) under Grant No.~DE-SC0012567, and by the DOE QuantISED program through the theory consortium “Intersections of QIS and Theoretical Particle Physics” at Fermilab (FNAL 20-17).
S.T. is additionally supported by the Swiss National Science Foundation project number P5R5PT\_222350, and acknowledges CERN TH Department for hospitality while this research was being carried out.
M.V. is supported
by the "Strategic Research Program High-Energy Physics” of the Vrije Universiteit Brussel, by the
iBOF “Un-locking the Dark Universe with Gravitational Wave Observations: from Quantum Optics to Quantum Gravity” of the Vlaamse Interuniversitaire Raad,
and by the 
the ``Excellence of Science - EOS" - be.h project n.30820817. M.V. also recognizes partial support by the DOE Grant No. DE-SC0020223. 

\appendix

\section{Mean posterior values of model parameters}
\label{app:meanPoseterior}

In this appendix, we report the value of the mean posterior values for each parameter and each hypothesis we have made. The mean posterior value of a quantity $X$ is defined as 
\bea
\langle X\rangle = \frac{\int d\theta \,X(\theta) \mathcal{P}(\theta|\vec{R}_{\rm obs})}{ \int d\theta \,\mathcal{P}(\theta|\vec{R}_{\rm obs}) } \qquad \qquad \text{(mean posterior value of $X$)} \, ,
\eea
where $ \mathcal{P}(\theta|\vec{R}_{\rm obs}) $ is the posterior distribution of model parameters  $\theta$ given the dataset $\vec{R}_{\rm obs}$. Notice however that this quantity does not correspond to the best fit value which instead is defined by 
\bea 
\label{eq:best_fit_def}
X_{\text{best fit}} = \text{Max}_X[\mathcal{P}(\theta|\vec{R}_{\rm obs})] \, . 
\eea 
However the mean posterior value captures the ballpark where the parameter is favored by the posteriors.
We present the mean posterior values of each parameter for each hypothesis in Tables \ref{tab:meanPoseterior_SMBH} and \ref{tab:meanPoseterior_SIGW}.

\begin{table}[ht!]
\renewcommand{\arraystretch}{1.5}
  \begin{center}
    \begin{tblr}{|Q[c,1.8cm]|Q[c,2.8cm]|Q[c,2.5cm]|Q[c,3cm]|Q[c,3cm]|}
      \hline
      \SetCell[r=2]{c}{{{$\textbf{Model}$ }}}&\SetCell[r=2,c=2]{c}{{{$\textbf{Prior}$ }}}& &\SetCell[r=2]{c}{{{\centering\parbox{1.9cm}{\textbf{Parameters}\\
     $(\log_{10}{X})$}}}}
    & 
\SetCell[r=2, c=1]{c}{{{$\textbf{Posterior mean}$}}}\\
&&&&\\
      \hline \hline 
           \SetCell[r=3]{c}{{{SMBH}}}& \SetCell[r=1,c=2]{c}{{{free amplitude}}}& &$A_{\rm SMBH} $ & $-14.61\pm 0.06$    \\
         \hline 
           & \SetCell[r=2,c=2]{c}{{{\tt{GWOnly-Ext}
           }}} && $A_{\rm SMBH}$&  $-14.56\pm 0.13$     \\
      \hline
      & && $\gamma_{\rm SMBH}$&  $4.15\pm 0.28$     \\
      \hline
    \end{tblr}
    \caption{\label{tab:meanPoseterior_SMBH} Mean posterior values of model parameters and their $68\%$ confidence intervals for the interpretation of the PTA signal based only on astrophysical SMBH binaries.}
  \end{center}
\end{table}

\begin{table}[ht!]
\centering
\renewcommand{\arraystretch}{1.2}
\begin{tabular}{|
    >{\centering\arraybackslash}p{1.5cm}|
    >{\centering\arraybackslash}p{0.8cm}|
    >{\centering\arraybackslash}p{1.9cm}|
    >{\centering\arraybackslash}p{2.cm}|
    >{\centering\arraybackslash}p{2.cm}|
    >{\centering\arraybackslash}p{2.cm}|
    >{\centering\arraybackslash}p{2.cm}|
}
\hline
\multicolumn{2}{|c|}{\multirow{4}{*}{\textbf{Prior on PBHs}}} & 
\multirow{4}{*}{\parbox{1.4cm}{\centering \textbf{Param.}}} & 
\multicolumn{4}{c|}{\textbf{Posterior mean}} \\ \cline{4-7}
\multicolumn{2}{|c|}{} & & \multicolumn{2}{c|}{\multirow{2}{*}{SIGW}} & \multicolumn{2}{c|}{\multirow{2}{*}{\centering \makecell[c]{SIGW+SMBH \\($\tt{GWOnly\text{-}Ext}$)}}} \\ 
\multicolumn{2}{|c|}{} & & \multicolumn{2}{c|}{} & \multicolumn{2}{c|}{} \\ \cline{4-7}
\multicolumn{2}{|c|}{} & & G & NG & G & NG \\
\hline\hline

\multicolumn{2}{|c}{\multirow{8}{*}{\centering No prior} }
& \multicolumn{1}{|c|}{$\text{log}_{10 }\mathcal{A}_\zeta$} & $ -0.91^{+0.37}_{-0.18}$ & $-0.91^{+0.37}_{-0.18}$ & $-0.90^{+0.38}_{-0.17}$ & $-0.91^{+0.38}_{-0.18}$ \\ \cline{3-7}
\multicolumn{2}{|c|}{} & $\text{log}_{10 }k_{\star}$ & $ 7.76\pm 0.32$ &  $7.75\pm 0.32$ & $ 7.77\pm 0.33$ & $7.97^{+0.43}_{-0.56}$ \\ \cline{3-7}
\multicolumn{2}{|c|}{} & $\Delta$ & $0.92^{+0.32}_{-0.78}$ & $ 0.91^{+0.31}_{-0.77}$ & $0.89^{+0.30}_{-0.77}$ & $0.88^{+0.30}_{-0.76}$ \\ \cline{3-7}
\multicolumn{2}{|c|}{} & $f_{\rm NL}$ & $0$ & $0.0\pm 6.0$ & $0$ & $0.0\pm 6.0$ \\ \cline{3-7}
\multicolumn{2}{|c|}{} & $\text{log}_{10 } f_{\rm PBH}$ & $7.7 ^{+1.9}_{-1.6}$ &  $9.15 ^{+0.33}_{-1.88}$ & $7.7 ^{+2.0}_{-1.6}$ & $9.14 ^{+0.34}_{-1.84}$ \\  \cline{3-7}
\multicolumn{2}{|c|}{} & $\text{log}_{10 }M_{\rm PBH}$ & $-2.4 \pm 0.0$ & $-2.62 ^{+0.001}_{-0.001}$ & $7.7 ^{+2.0}_{-1.6}$ & $-2.66 ^{+0.001}_{-0.001}$ \\ \cline{3-7}
\multicolumn{2}{|c|}{} & $\text{log}_{10 } A_{\rm SMBH}$ & --- & --- & $-15.7^{+0.51}_{-0.40}$ &  $-15.74^{+0.50}_{-0.40}$\\ \cline{3-7}
\multicolumn{2}{|c|}{} & $\gamma_{\rm SMBH}$ & --- & --- & $4.65\pm 0.34$ &  $4.70\pm 0.35$ \\\cline{3-7}
\hline\hline

\multicolumn{2}{|c}{\multirow{8}{*}{\centering \makecell[c]{~Astrophysical} }}
&  \multicolumn{1}{|c|}{$\text{log}_{10 }\mathcal{A}_\zeta$} & $-1.76\pm0.10$ & $-1.79\pm 0.07$ & $-2.42^{+0.24}_{-0.54}$ & $-2.38^{+0.51}_{-0.45}$ \\ \cline{3-7}
\multicolumn{2}{|c|}{} & $\text{log}_{10 } k_{\star}$ & $7.00\pm0.10$ & $6.99\pm 0.12$ & $7.48^{+0.99}_{-0.86}$ & $7.45^{+1.2}_{-0.89}$ \\ \cline{3-7}
\multicolumn{2}{|c|}{} & $\Delta$ & $0.23^{+0.080}_{-0.13}$ & $0.23^{+0.073}_{-0.15}$ & $1.34^{+0.90}_{-1.3}$ & $1.26^{+0.87}_{-1.3}$ \\ \cline{3-7}
\multicolumn{2}{|c|}{} & $f_{\rm NL}$ & $0$ & $-3.0^{+2.3}_{-2.9}$ & $0$ & $-1.1^{+3.5}_{-6.8}$ \\ \cline{3-7}
\multicolumn{2}{|c|}{} & $\text{log}_{10 }f_{\rm PBH}$ & $-2.0 ^{+1.4}_{-1.1}$ & $-3.2 ^{+2.2}_{-1.6}$ & $-51.5 ^{+26.2}_{-123}$ & $-40.0 ^{+29.0}_{-73.0}$ \\\cline{3-7}
\multicolumn{2}{|c|}{} & $\text{log}_{10 } M_{\rm PBH}$ & $-0.59 ^{+0.026}_{-0.027}$ & $ -0.49 ^{+0.010}_{-0.030}$ & $-1.6 ^{+0.017}_{-0.14}$ & $-1.6^{+0.12}_{-0.11}$\\ \cline{3-7}
\multicolumn{2}{|c|}{} & $\text{log}_{10 }A_{\rm SMBH}$ & --- &  --- & $-14.7^{+0.29}_{-0.061}$ &$-14.8^{+0.43}_{-0.019}$ \\ \cline{3-7}
\multicolumn{2}{|c|}{} & $\gamma_{\rm SMBH}$ & --- & --- &  $ 4.2^{+0.27}_{-0.36}$ & $4.3^{+0.29}_{-0.39}$ \\
\hline\hline

\multirow{16}{*}{\centering  \makecell[c]{Sub-solar} }
& \multirow{8}{*}{\centering  \makecell[c]{LVK \\ O5}} 
& $\text{log}_{10 }\mathcal{A}_\zeta$ & $-1.74^{+0.061}_{-0.041}$ & $-1.77^{+0.095}_{-0.053}$ & $-1.77^{+0.10}_{-0.027}$ & $-1.79^{+0.15}_{-0.041}$ \\ \cline{3-7}
& & $\text{log}_{10 }k_{\star}$ & $6.97^{+0.071}_{-0.037}$ & $6.97^{+0.059}_{-0.074}$ & $6.96^{+0.091}_{-0.036}$ & $6.96\pm 0.093$ \\ \cline{3-7}
& & $\Delta$ & $0.25^{+0.080}_{-0.10}$ & $0.22^{+0.084}_{-0.095}$ & $0.35^{+0.039}_{-0.23}$ & $0.32^{+0.044}_{-0.21}$ \\ \cline{3-7}
& & $f_{\rm NL}$ & $0$ & $-3.3^{+2.6}_{-2.1}$ & $0$ & $-3.2^{+2.8}_{-3.6}$ \\ \cline{3-7}
 & & $\text{log}_{10 }f_{\rm PBH}$ & $ -1.4 ^{+1.0}_{-0.76}$ & $-2.6 ^{+2.0}_{-1.5}$ & $-2.0 ^{+1.8}_{-0.59}$ & $-4.0 ^{+3.4}_{-1.1}$ \\ \cline{3-7}
 & & $\text{log}_{10 }M_{\rm PBH}$ & $ -0.52 ^{+0.023}_{-0.030}$ & $-0.51 ^{+0.0080}_{-0.048}$ & $-0.45 ^{+0.0070}_{-0.056}$ & $-0.49 ^{+0.010}_{-0.10}$ \\ \cline{3-7}
& & $\text{log}_{10 }A_{\rm SMBH}$ & --- & --- & $-15.6^{+0.52}_{-0.62}$ & $-15.6\pm 0.55$ \\ \cline{3-7}
& & $\gamma_{\rm SMBH}$ & --- & --- & $4.6\pm 0.36$ & $4.6^{+0.37}_{-0.32}$ \\  \cline{2-7}

& \multirow{8}{*}{~CE} 
& $\text{log}_{10 }\mathcal{A}_\zeta$ & $-1.76^{+0.071}_{-0.051}$ & $-1.79^{+0.10}_{-0.063}$ & $-1.85^{+0.20}_{-0.030}$ & $-1.85^{+0.16}_{-0.044}$ \\ \cline{3-7}
& & $\text{log}_{10 } k_{\star}$ & $7.01^{+0.070}_{-0.079}$ & $7.00^{+0.066}_{-0.076}$ & $7.09^{+0.070}_{-0.23}$ & $7.06^{+0.068}_{-0.17}$ \\ \cline{3-7}
& & $\Delta$ & $0.22^{+0.083}_{-0.11}$ & $0.21^{+0.075}_{-0.11}$ & $ 0.54^{+0.044}_{-0.56}$ & $0.39^{+0.0023}_{-0.34}$ \\ \cline{3-7}
& & $f_{\rm NL}$ & $0$ & $-3.3^{+2.6}_{-1.5}$ & $0$ & $-2.6^{+2.4}_{-2.8}$ \\ \cline{3-7}
 & & $\text{log}_{10 }f_{\rm PBH}$ & $-1.9 ^{+1.2}_{-1.1}$ & $-3.1 ^{+2.2}_{-1.7}$ & $-4.5 ^{+4.4}_{-0.93}$ &  $-6.2 ^{+4.3}_{-1.6}$ \\ \cline{3-7}
& & $\text{log}_{10 }M_{\rm PBH}$ & $-0.62 ^{+0.029}_{-0.024}$ & $ -0.49 ^{+0.015}_{-0.054}$ & $-0.72 ^{+0.000}_{-0.11}$ & $ -0.62 ^{+0.0010}_{-0.049}$\\ \cline{3-7}
 & & $\text{log}_{10 }A_{\rm SMBH}$ & --- & --- & $-15.3^{+0.91}_{-0.65}$ & $-15.5^{+0.93}_{-0.65}$ \\ \cline{3-7}
& & $\gamma_{\rm SMBH}$ & --- & --- & $4.5\pm 0.38$ & $4.6^{+0.43}_{-0.37}$ \\

\hline
\end{tabular}
\caption{\label{tab:meanPoseterior_SIGW} Mean posterior values of model parameters with $68\%$ confidence intervals across multiple priors for the interpretation of the PTA signal which includes SIGW. On this table the units of $k_{\star}$ is $\text{Mpc}^{-1}$, and $M_{\rm PBH}$ is $M_{\odot}$, while the rest of the entries are dimensionless. For computing those values we used threshold statistics for the abundance of PBHs.}
\end{table}
\FloatBarrier

\bibliographystyle{JHEP}
{\footnotesize
\bibliography{main}}
\end{document}